\documentclass[12pt]{article}
\usepackage{formatting}
\usepackage{shortcuts}

\begin{document}
\begin{titlepage}
\begin{center}
\phantom{ }
\vspace{1.25cm}

{\bf \Large{An SYK wormhole under the microscope}}

\vskip 1.4cm
\large{
Valérie Bettaque${}^{1}$, Martin Sasieta${}^{2}$, Brian Swingle${}^{3}$, and Alejandro Vilar L\'opez${}^{4}$
}

\vskip 1cm

\small{${}^{1}$ \textit{Quantum Horizons Alberta, Department of Physics and Astronomy,}
\vskip -.05cm
\textit{University of Calgary, 2500 University Drive NW, Calgary, Alberta T2N 1N4, Canada}}
\vskip 0.3cm
\small{${}^{2}$ \textit{Leinweber Institute for Theoretical Physics and Department of Physics,}
\vskip -.05cm
\textit{University of California, Berkeley, California 94720, USA}}
\vskip 0.3cm
\small{${}^{3}$ \textit{Martin Fisher School of Physics, Brandeis University}
\vskip -.05cm
\textit{Waltham, Massachusetts 02453, USA}}
\vskip 0.3cm
\small{${}^{4}$ \textit{Department of Physics and Astronomy, University of British Columbia,}}
\vskip -.05cm
{\textit{6224 Agricultural Road, Vancouver, B.C. V6T 1Z1, Canada}}
\vskip 0.7cm

\small{\texttt{E-mail}: \href{mailto:valerie.bettaque@ucalgary.ca}{valerie.bettaque@ucalgary.ca},
 \href{mailto:msasieta@berkeley.edu}{msasieta@berkeley.edu},}\\[.1cm]
  \small{\href{mailto:bswingle@brandeis.edu}{bswingle@brandeis.edu},
  \href{mailto:alejandro.vilarlopez@ubc.ca}{alejandro.vilarlopez@ubc.ca}}\\[.5cm]

\end{center}

\vspace{.4cm}

\begin{abstract}
We derive a large-$N$ wormhole saddle in the Sachdev–Ye–Kitaev (SYK) model for the variance of the thermal one-point function of a heavy operator across realizations of the microscopic couplings. We propose that this saddle describes a double trumpet geometry. In the double-scaling limit, we express the wormhole as a transition amplitude in an emergent bulk Hilbert space of chords connecting the two boundaries. In the Schwarzian limit, the chords proliferate and the wormhole defines a quantum transition amplitude in the open-universe Hilbert space of Jackiw–Teitelboim (JT) gravity. Comparing the SYK wormhole amplitude with its counterpart in JT gravity coupled to matter provides a direct test of the heavy-operator randomness hypothesis in SYK.

The microscopic SYK couplings specify the $\alpha$-parameters of the wormhole. For a fixed realization of the couplings, we compute the observable in the corresponding $\alpha$-state. We provide the bulk wavefunction of the associated half-wormhole, which exhibits erratic correlations in the length basis that do not self-average. We extend the bulk description to the Hilbert space of a closed universe. We point out that the correlation chords from which the wormhole geometry emerges structurally resemble local observer worldlines.
\end{abstract}

\vfill

\end{titlepage}

\setcounter{tocdepth}{3}
\tableofcontents
\noindent\makebox[\linewidth]{\rule{\textwidth}{0.4 pt}}

\section{Introduction}

Spacetime wormholes provide a window onto the microstructure of quantum gravity, but only in a coarse-grained or statistical form. In holography, where this is most precisely realized, wormholes describe the statistical phase correlations of chaotic microscopic boundary data, as first proposed in \cite{Saad:2018bqo,Saad:2019lba} for the spectrum of the black hole Hamiltonian. This phenomenon recalls the effect of wormholes anticipated by Coleman, Giddings and Strominger \cite{Coleman:1988cy,Giddings:1988cx}, but with important differences. Here, wormholes are macroscopic geometries connecting asymptotic boundaries of spacetime, and the resulting statistics appears in non-perturbative, microscopic boundary data rather than as a direct probability distribution over the low-energy couplings of the bulk theory.

On the other hand, from a purely bulk perspective, macroscopic wormholes have a more direct physical implication: they prepare semiclassical states of closed ``baby universes'' \cite{Maldacena:2004rf, Saad:2019pqd,Marolf:2020xie,Chen:2020tes,VanRaamsdonk:2020tlr,Saad:2021uzi,Antonini:2022blk,Antonini:2023hdh,Usatyuk:2024mzs,Belin:2025ako,Sasieta:2025vck,VanRaamsdonk:2026tnv,Sontag:2026iiu,Chen:2025fwp}. From this perspective, the wormhole is not merely a probe of microscopic boundary statistics, but part of the mechanism by which a closed-universe wavefunction emerges from the microscopic theory. The hope is that understanding this mechanism well enough may provide fundamental lessons for quantum cosmology (cf. \cite{Marolf:2020xie,Abdalla:2025gzn,Harlow:2025pvj,Akers:2025ahe,Antonini:2024mci,Higginbotham:2025dvf,Antonini:2025ioh,Chen:2025fwp,Engelhardt:2025azi, Liu:2025cml,Kudler-Flam:2025cki,Klinger:2026kqj,Harlow:2026hky,Zhao:2026mpl,Engelhardt:2026blp,Kudler-Flam:2026nzz,Miyata:2026myd,Bintanja:2026vkr}). Schematically, the wormhole and the associated wavefunction of the universe emerge from the holographic system as
\be\label{eq:whbaby} 
\langle Z^*_{\rm micro}\,Z_{\rm micro} \rangle_c\;\sim\; Z_{\rm wormhole}
= \bra{\Psi_{\rm cl}}\ket{\Psi_{\rm cl}}\, .
\ee 
Here $\langle \cdot \rangle_c$ denotes the connected statistical average over microscopic boundary data, and $Z_{\rm micro}$ is the boundary partition function. The quantity $Z_{\rm wormhole}$ is the contribution of the semiclassical wormhole, which we assume to be an isolated saddle of the gravitational path integral. This saddle may be viewed as computing the norm of a wavefunction $\Psi_{\rm cl}$ for the baby universe. This wavefunction is determined in a WKB approximation by a semiclassical solution of the gravitational equations, and it depends on the closed universe data, such as the induced metric and matter fields on the time-symmetric slice. The relation $\sim$ denotes equality at leading order in the large-$N$ limit of the holographic system, or equivalently in the $G_{\rm N}\to 0$ limit in the bulk.

\begin{figure}[h]
    \centering
    \includegraphics[width= \linewidth]{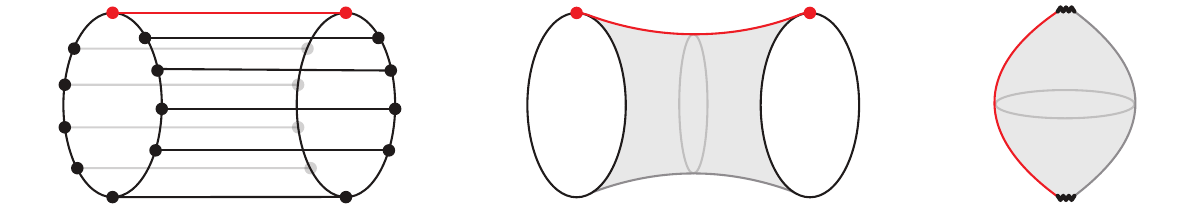}
    \caption{Left: $\langle Z_{\rm micro}^2 \rangle$ for $Z_{\rm micro} =\Tr (e^{-\beta H}\mathcal{O})$ and the average over couplings is dominated by configurations with large correlations in the form of chords. Middle: the matter worldline wormhole in JT gravity, which prepares a semiclassical state $\Psi_{\rm cl}$ of the closed universe at the gray slice. Right: time evolution of its initial data describes a big bang big  crunch cosmology.}
    \label{fig:introwh}
\end{figure}

In this paper we explicitly derive a version of \eqref{eq:whbaby} in the Sachdev-Ye-Kitaev (SYK) model \cite{Kitaev:2015SYKTalks,Maldacena:2016hyu}. We use the natural disorder of the coupling constants of the Hamiltonian to define the statistical variance of the thermal one-point function of a heavy operator. This builds on \cite{Bettaque:2026vpl}, which derived and numerically solved the relevant Schwinger-Dyson equations, exhibited a connected wormhole-like saddle for the SYK path integral, and showed that a simple truncation of the bulk dual of SYK consisting of $N$ light fermions propagating on a wormhole geometry could give a good quantitative account of the numerical results. Our main results for this observable are twofold:
\begin{itemize}
\item In Section~\ref{sec:GS} we find an analytic large-$N$ replica-connected saddle in the path integral description of SYK, where the operator acts as a defect that changes the thermal spin structure of the fermions it contains. We interpret the solution as the correlator in a double trumpet geometry. The length of the bottleneck of the wormhole is completely determined by the size of the operator. We compare this solution to the wormhole of JT gravity and a heavy particle of \cite{Stanford:2020wkf} (see also \cite{Saad:2019pqd,Usatyuk:2024mzs}), represented in Fig.~\ref{fig:introwh}, and to numerical solutions that extend the results of  \cite{Bettaque:2026vpl}.

\item In Section~\ref{sec:thermalonept} we derive a ``bulk'' Hilbert space expression for $Z_{\rm wormhole}$ in the double-scaling limit of SYK. This bulk Hilbert space is spanned by the number of correlation chords connecting the two boundaries. In the Schwarzian limit, these chords proliferate, and the bulk Hilbert space becomes that of JT gravity. This allows a direct comparison of the resulting wormhole amplitude, \eqref{eq:SYKwormhole}, with the JT gravity wormhole amplitude of \cite{Stanford:2020wkf}. We find that the SYK wormhole is exponentially suppressed by an “entropy” associated with the operator, whereas the JT wormhole is exponentially suppressed by the topological entropy $2S_0$. Moreover, the SYK wormhole kernel is diagonal in the length basis, whereas the JT wormhole kernel is diagonal in the energy basis. We display the full comparison in Table \ref{tab:SYK-JT-kernels}. For an operator of fixed size in the fermion basis, the SYK kernel projects onto a single length eigenstate determined by the operator’s size. For a smeared operator over different sizes, the kernel can reproduce the JT gravity wormhole kernel in the limit of a very heavy particle.
\end{itemize}

Ref.~\cite{Stanford:2020wkf} observed that the JT gravity wormhole supported by a matter particle captures what has come to be called “heavy operator randomness,” i.e., the statistics of the thermal one-point function of an ETH-type banded random operator for the matter particle (cf.~\cite{Saad:2019pqd,Jafferis:2022wez}). That is, the wormhole amplitude is fixed by the operator’s smooth microcanonical envelope, which can be extracted from its thermal two-point function, a single-boundary observable. In higher dimensions, similar banded random matrix ansatze reproduce the effect of wormholes supported by homogeneous dust-shell operators \cite{Sasieta:2022ksu}. In AdS$_3$, the dust-shell wormholes admit an alternative effective description in terms of expectation values of line defects in Liouville CFT \cite{Chandra:2024vhm,Abdalla:2026wdx,Chandra:2026hfj}.

While such a random-operator model may capture the gravitational wormhole, the wormhole contribution studied in this paper is derived directly from a microscopic holographic system.\footnote{The full bulk dual of SYK remains only partially understood, but heavy matter
is precisely the regime in which a classical description is expected, so we will not rely on precise details about its particle content.} The operators we study are not random but are fixed, structured operators in the fermion basis. Likewise, the Hamiltonian is not a random matrix, but a structured few-body Hamiltonian.

For the thermal one-point wormhole, the question therefore becomes how well the microscopic theory satisfies the heavy-operator-randomness hypothesis. Such behavior is far from obvious. In particular, connected multi-trace observables are usually dominated by system-dependent contributions that are not naturally interpreted as geometric wormholes and yield signals parametrically larger than gravitational wormhole contributions. Examples in SYK include the ``non-geometric wormhole'' contributions discussed in \cite{Berkooz:2020fvm}, which dominate the connected spectral form factor or light correlators at early times, before the usual ramp regime \cite{Altland:2017eao}. Their large signal reflects the fact that the SYK Hamiltonian depends on polynomially many couplings in $N$, which are parametrically fewer random parameters than those in a random matrix. These contributions are absent in the limit in which the SYK Hamiltonian is scaled to be fully non-local \cite{Raz:2025wjw}. Likewise, our results show that sufficiently heavy operators isolate a leading contribution with a geometric wormhole interpretation.

\paragraph{Couplings as $\alpha$-parameters \& factorization.}
In Section~\ref{sec:alphacouplings} we turn to viewing our wormhole in the $\alpha$-parameter formalism of Marolf and Maxfield \cite{Marolf:2020xie}. We naturally identify the SYK coupling constants with the $\alpha$-parameters of the wormhole. The GNS construction turns the disorder
average into an inner product on functions of the microscopic
couplings. The resulting Hilbert space is our formal
``baby-universe Hilbert space'' $\mathcal H_{\rm BU}$. The $\alpha$-states are coupling eigenstates $\ket{\alpha}:=\ket{\mathbf J}$, where $\mathbf J$ denotes the collection of couplings.\footnote{See \cite{Almheiri:2021jwq,Qi:2021oni} for other situations in which SYK couplings have been considered as $\alpha$-parameters.}

In this formalism, the interpretation of \eqref{eq:whbaby} is as follows. One first promotes the microscopic observable to an operator $\hat Z_{\rm micro}$ in $\mathcal{H}_{\rm BU}$ whose role is to ``create'' asymptotic boundaries. The $\alpha$-states are eigenstates of this operator $\hat Z_{\rm micro}\ket{\alpha}=Z_{\rm micro}^{\alpha}\ket{\alpha}$, where $Z_{\rm micro}^{\alpha}$ is the value of the observable with couplings fixed by $\alpha$. Starting from a reference ``Hartle-Hawking no-boundary'' state $\ket{{\rm HH}}\in\mathcal{H}_{\rm BU}$, this operator creates the baby-universe state
\be\label{eq:MM}
\hat Z_{\rm micro}\ket{{\rm HH}}
=\int {\rm d}\alpha\,\Psi_{\rm HH}(\alpha)
Z_{\rm micro}^{\alpha}\ket{\alpha}\,.
\ee
Here $\Psi_{\rm HH}(\alpha)=\langle\alpha|{\rm HH}\rangle$. The probability density $|\Psi_{\rm HH}(\alpha)|^2$ is the Gaussian disorder density of the SYK couplings. The wormhole \eqref{eq:whbaby} computes the norm of the part of \eqref{eq:MM} orthogonal to the no-boundary state. The parallel part vanishes when $\langle Z_{\rm micro}\rangle=0$, as it does for the observable considered in this paper.

Repeated applications of $\hat Z_{\rm micro}$ prepare states associated with several asymptotic boundaries. Even for the state \eqref{eq:MM}, the gravitational path integral potentially allows components with different numbers of closed universes, or superpositions of universes with macroscopically different features. In this paper, however, we remain within the saddle-point approximation and assume that the simplest topology dominates, so that $\Psi_{\rm cl}$ is a WKB wavefunction of a single macroscopic universe. In other words, we focus on two-boundary wormhole saddles with topology $\mathbb{R}\times\Sigma$, where $\Sigma$ is the closed spatial slice of the baby universe. This is the kind of wormhole contribution that often arises in higher-dimensional gravity, and as such, we would like to understand it better in this model.

\begin{figure}[h]
    \centering
    \includegraphics[width= .85\linewidth]{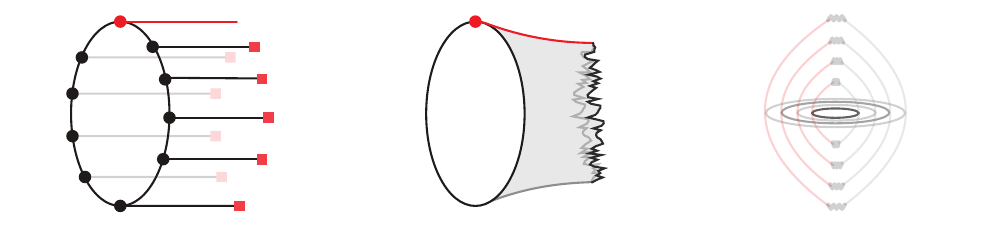}
    \caption{Left: $Z^\alpha_{\rm micro} =\Tr\!\left(e^{-\beta H(\mathbf{J})}\mathcal{O}\right)$ for an individual $\alpha$-state is dominated by configurations with many ``half-chords'', where each half-chord carries information about a microscopic coupling. Middle: the limit in which the number of half chords proliferate gives a bulk Hilbert space description of the ``half-wormhole'' in terms of a white-noise correlated wavefunction $\ket{\Psi_{\bf J}}$ in the length basis of JT. Right: the corresponding wavefunction of the closed universe is a very erratic superposition of geometries of different sizes.}
    \label{fig:introhalfwh}
\end{figure}

One of our main results in this section is a bulk description of individual $\alpha$-states. At fixed couplings, the two-boundary observable factorizes into a product of one-point functions, but each one-point function remains highly non-self-averaging (cf. \cite{Swingle:2026ecy}). We find that the one-point function can be expressed in the bulk Hilbert space of JT gravity as an overlap between a self-averaging thermal Hartle-Hawking state $\ket{\Psi_\beta}$ with semiclassical interpretation, and a coupling-dependent bulk state $\ket{\Psi_{\mathbf J}}$,
\be\label{eq:hwhintro} 
Z_{\rm micro}^{\mathbf{J}} \sim  \bra{\Psi_{\bf J}}\ket{\Psi_\beta}\,.
\ee 
We provide this description in both an “open” and a “closed” universe bulk Hilbert space.

This provides a concrete realization of the ``half-wormhole'' phenomenon of \cite{Saad:2021rcu,Mukhametzhanov:2021nea,Blommaert:2020seb,Blommaert:2021fob,Saad:2021uzi}. In the present setting, the state $\ket{\Psi_{\bf J}}$ represents the bulk wavefunction of the half-wormhole associated with the $\alpha$-state.\footnote{See \cite{Blommaert:2021fob} for a similar description of the $\alpha$-states of spectral wormholes in the JT gravity matrix integral \cite{Saad:2019lba}.} We show that, for a Majorana string, $\ket{\Psi_{\bf J}}$ is a length eigenstate whose eigenvalue is determined by the size of the string and whose amplitude is non-self-averaging. For an operator smeared over strings of different sizes, the wavefunction $\ket{\Psi_{\bf J}}$ exhibits white-noise correlations in length. Averaging $\ket{\Psi_{\bf J}}\bra{\Psi_{\bf J}}$ over the couplings (or, equivalently, taking expectation values in the no-boundary state $\ket{\rm HH}$ in the baby-universe formalism) glues two such half-wormholes together and yields the smooth wormhole inner product.

\paragraph{Closed universes, observers, and related work.}

Our setup provides a concrete microscopic formulation of questions about closed universes in holography that have attracted interest in recent years~\cite{Maldacena:2004rf, Saad:2019pqd,Marolf:2020xie,Chen:2020tes,VanRaamsdonk:2020tlr,Saad:2021uzi,Antonini:2022blk,Antonini:2023hdh,Usatyuk:2024mzs,Belin:2025ako,Sasieta:2025vck,VanRaamsdonk:2026tnv,Sontag:2026iiu,Chen:2025fwp,Abdalla:2025gzn,Harlow:2025pvj,Akers:2025ahe,Antonini:2024mci,Higginbotham:2025dvf,Antonini:2025ioh,Engelhardt:2025azi, Liu:2025cml,Kudler-Flam:2025cki,Engelhardt:2026blp,Klinger:2026kqj,Harlow:2026hky,Kudler-Flam:2026nzz,Zhao:2026mpl,Miyata:2026myd,Bintanja:2026vkr}. A closed universe has no asymptotic boundary, so its bulk description cannot be encoded in a boundary Hilbert space in the conventional holographic way. The baby-universe Hilbert space provides a candidate description~\cite{Marolf:2020xie}. Observers have emerged as a way to interpret this description physically, by relating closed-universe states and their overlaps to relationally-defined Hilbert spaces~\cite{Abdalla:2025gzn,Harlow:2025pvj,Akers:2025ahe,Chen:2025fwp,Harlow:2026hky,Zhao:2026mpl}.

We provide an interpretation of the correlation chords connecting the two boundaries as a delocalized observer over space. An interesting fact of this picture is that the same chords also underlie the emergence of space in the wormhole: upon appropriate scaling, their number becomes the geodesic length of a bulk slice of the wormhole. The reference system that keeps track of the $\alpha$-parameters is thus directly connected to the microscopic origin of the geometry. We develop this interpretation further in the discussion.

Our analysis uses the chord techniques developed in Refs.~\cite{Berkooz:2018jqr,Lin:2022rbf} for the double-scaled limit of the SYK model; see also \cite{Erdos:2014zgc,Cotler:2016fpe,Berkooz:2018qkz,Goel:2023svz,Lin:2023trc} and the comprehensive review \cite{Berkooz:2024lgq}. A wormhole contribution to the square of the thermal one-point function in double-scaled SYK has also been proposed in Refs.~\cite{Okuyama:2023yat,Cui:2025sgy}. These works construct cylinder amplitudes using gluing rules for disk amplitudes, incorporating, implicitly or explicitly, a version of the heavy operator randomness hypothesis. This relates the wormhole amplitude to a disk two-point function, much as in \cite{Stanford:2020wkf}, but now with the $q$-deformed disk two-point function. The connection to SYK microscopics, however, does not follow directly from these gluing prescriptions: as discussed above, connected observables in SYK do not generally have the same structure as wormhole amplitudes in JT gravity.

\paragraph{Organization of the paper.}

The paper is organized as follows. In Sec.~\ref{sec:GS}, we introduce the thermal one-point observable, derive its replica-connected saddle in the large-$N$, large-$p$ limit of $p$-body SYK, and discuss its geometric interpretation. In Sec.~\ref{sec:thermalonept}, we derive the wormhole amplitude from chord diagrams in the double-scaling limit of SYK and identify the emergent bulk Hilbert space and gluing kernel. We then take the Schwarzian limit, compare the amplitude with JT gravity, and study the effect of smearing over operator sizes. In Sec.~\ref{sec:alphacouplings}, we identify the SYK couplings as the $\alpha$-parameters of the wormhole and describe the corresponding half-wormholes in both the open and closed universe Hilbert spaces. We also explain how these bulk descriptions are related to the microscopic baby-universe Hilbert space. We conclude in Sec.~\ref{sec:discussion} with a discussion of the results and open questions. The technical details are collected in seven appendices: Appendix ~\ref{app:SD-onepoint} contains details of the large-$p$ analytic solution of the Schwinger-Dyson equations, Appendix ~\ref{sec:extensive-wormhole-windings} discusses extensive operators, Appendix ~\ref{sec:finitepnum} performs the numerical analysis of the Schwinger-Dyson equations at finite $p$, Appendix ~\ref{sec:appKrawtchouk} derives the endpoint probability formula used in the double-scaled analysis, Appendix ~\ref{app:ConditionedEnsemble} contains details of the double-scaled analysis, Appendix ~\ref{app:averaged-light} discusses light probes on the wormhole, and Appendix ~\ref{app:higher-moments} considers higher statistical moments.

\section{The wormhole from the path integral}
\label{sec:GS}

We consider the $p$-body SYK model of $N$ Majorana fermions
$\psi_1,\ldots,\psi_N$, satisfying
$\{\psi_i,\psi_j\}=2\delta_{ij}$, with Hamiltonian \cite{Kitaev:2015SYKTalks,Maldacena:2016hyu}
\be
H=\sum_I J_I\,\psi_I,
\qquad
\psi_I=\mathrm{i}^{p/2}\psi_{i_1}\cdots\psi_{i_p},
\ee
where $I=(i_1<\cdots<i_p)$. The couplings $J_I$ are independent Gaussian
random variables with
\be
\langle J_I\rangle_J=0,
\qquad
\langle J_IJ_{I'}\rangle_J
=
\frac{\mathcal J^2N}{2p^2\binom{N}{p}}\,
\delta_{I,I'},
\label{eq:syk-disorder}
\ee
where $\langle\cdot\rangle_J$ denotes the disorder average.

For an index set $W=(i_1<\cdots<i_w)$ of even size $w>0$, we define
the Majorana string
\be
\psi_W=\mathrm{i}^{w/2}\psi_{i_1}\cdots\psi_{i_w}.
\label{eq:string}
\ee
Its unnormalized thermal one-point function is
$\Tr(e^{-\beta H}\psi_W)$. Throughout, we normalize the trace so that ${\rm Tr}(\mathbf{1}) = 1$. By the $O(N)$ invariance of the ensemble of couplings,
the average thermal one-point function vanishes,
\be
\left\langle\Tr\!\left(e^{-\beta H}\psi_W\right)\right\rangle_J=0.
\label{eq:disconvanish}
\ee
We therefore study its covariance,
\be
\mathcal C_\beta(W)
:=
\left\langle
\left[\Tr\!\left(e^{-\beta H}\psi_W\right)\right]^2
\right\rangle_J.
\label{eq:twopt}
\ee
Since the one-point function has zero mean, $\mathcal{C}_\beta(W)$ is
entirely connected. The same $O(N)$ invariance implies that it depends on
$W$ only through its size $w$, so the choice of a particular string
within the fixed-$w$ sector entails no loss of generality.

\paragraph{Heavy operator.}
We take $N\to\infty$ first, followed by $p\to\infty$, at fixed
$\beta\mathcal J$. We focus on a heavy Majorana string whose length scales as\footnote{For definiteness, we restrict to
$p,w\in4\mathbb Z$, avoiding antiunitary selection rules that can otherwise force the one-point function to vanish in each disorder realization.}
\be
w=\frac{\kappa N}{p},
\qquad
\kappa>0\ \text{fixed}.
\ee
Thus the operator is extensive at fixed $p$, while the fraction of fermions
it contains, $w/N=\kappa/p$, vanishes in the subsequent large-$p$ limit.

We construct the replica-off-diagonal saddle directly from the path integral
and extract the associated wormhole geometry. This gives an analytic large-$p$
description of the saddle identified in Ref.~\cite{Bettaque:2026vpl}, which we
then compare with the worldline wormhole of JT gravity~\cite{Stanford:2020wkf}.

\subsection{The operator insertion and the two spin structures}
\label{eq:secopspin}

Consider the Majorana string $\psi_W$ defined in \eqref{eq:string}. For even $w$, conjugation by this string gives
\be
\psi_W\psi_i\psi_W^{-1}
=
\begin{cases}
-\psi_i,&i\in W,\\
+\psi_i,&i\notin W.
\end{cases}
\label{eq:string-conjugation}
\ee
Combining \eqref{eq:string-conjugation} with the usual minus sign from
the fermion trace shows that the $w$ fermions in $W$ are periodic around
the thermal circle, whereas the remaining $N-w$ fermions are
antiperiodic.  The operator affects the collective description only
through this split in spin structure.

We use two replicas, $a,b=L,R$, for the two traces in
$\mathcal{C}_\beta(W)$, and introduce one bilocal for each sector:
\be
G_{\rm AP}^{ab}(\tau_1,\tau_2)
=
\frac{1}{N-w}\sum_{i\notin W}
\psi_i^a(\tau_1)\psi_i^b(\tau_2),
\qquad
G_{\rm P}^{ab}(\tau_1,\tau_2)
=
\frac{1}{w}\sum_{i\in W}
\psi_i^a(\tau_1)\psi_i^b(\tau_2).
\label{eq:app-bilocals}
\ee
They satisfy $G_s^{ab}(\tau_1,\tau_2)
=-G_s^{ba}(\tau_2,\tau_1)$, where $s={\rm AP,P}$. Shifting either
argument by $\beta$ multiplies $G_{\rm AP}$ by $-1$ and leaves $G_{\rm P}$
unchanged. The SYK interaction depends on their weighted average
\be
G^{ab}
=(1-\nu)G_{\rm AP}^{ab}+\nu G_{\rm P}^{ab},
\qquad
\nu:=\frac{w}{N}=\frac{\kappa}{p}.
\label{eq:app-mixed-G}
\ee
Because the two terms in \eqref{eq:app-mixed-G} have different spin
structures, $G^{ab}$ has no single boundary condition of its own.  We define
it on $0\leq\tau<\beta$ and extend the AP and P terms separately when
needed.

\paragraph{Collective action and Schwinger--Dyson equations.}

With our Majorana normalization, the free kinetic operator is
$\frac12\partial_\tau$. Its inverse in the antiperiodic sector is
$\sign(\tau)$, with antiperiodic continuation around the thermal circle understood. The collective representation of
the observable is
\be
\mathcal{C}_\beta(W)
=
\mathcal N_W
\int DG_{\rm AP}\,DG_{\rm P}\,D\Sigma_{\rm AP}\,D\Sigma_{\rm P}\,
e^{-NI[G,\Sigma]},
\label{eq:app-collective-integral}
\ee
where
\be
\begin{split}
I[G,\Sigma]={}&
-(1-\nu)\log{\rm Pf}_{\rm AP}
\left(\frac12\partial_\tau-\Sigma_{\rm AP}\right)
-\nu\log{\rm Pf}_{\rm P}
\left(\frac12\partial_\tau-\Sigma_{\rm P}\right)
\\
&+\frac12\sum_{a,b=L,R}
\int_0^\beta {\rm d}\tau_1\,{\rm d}\tau_2
\left[
(1-\nu)\Sigma_{\rm AP}^{ab}G_{\rm AP}^{ab}
+\nu\Sigma_{\rm P}^{ab}G_{\rm P}^{ab}
-\frac{\mathcal J^2}{2p^2}(G^{ab})^p
\right].
\end{split}
\label{eq:app-correct-collective-action}
\ee
The normalization $\mathcal N_W$ is independent of the collective fields
and will be fixed in Appendix \ref{app:SD-onepoint} by matching it to the microscopic high-temperature
expansion.

Varying \eqref{eq:app-correct-collective-action} with respect to the
self-energies gives
\be
G_{\rm AP}
=
\left(
\frac12\partial_\tau-\Sigma_{\rm AP}
\right)_{\rm AP}^{-1},
\qquad
G_{\rm P}
=
\left(
\frac12\partial_\tau-\Sigma_{\rm P}
\right)_{\rm P}^{-1}.
\label{eq:app-resolvents}
\ee
The subscripts on the inverses specify the corresponding antiperiodic or
periodic function space. Although the self-energies below coincide on the
fundamental interval, the two propagators remain distinct because their
inverses are taken with different spin structures.

Variation with respect to $G_{\rm AP}$ and $G_{\rm P}$ gives
\be
\Sigma_{\rm AP}^{ab}(\tau_1,\tau_2)
=
\Sigma_{\rm P}^{ab}(\tau_1,\tau_2)
=
\frac{\mathcal J^2}{2p}
\left[
G^{ab}(\tau_1,\tau_2)
\right]^{p-1}.
\label{eq:app-self-energy}
\ee
The two variations give the same result because the interaction depends on
$G_{\rm AP}$ and $G_{\rm P}$ only through the weighted combination
\eqref{eq:app-mixed-G}.

\subsection{Zero-mode pairing}

\begin{figure}[h]
\centering
\includegraphics[width=.5\linewidth]{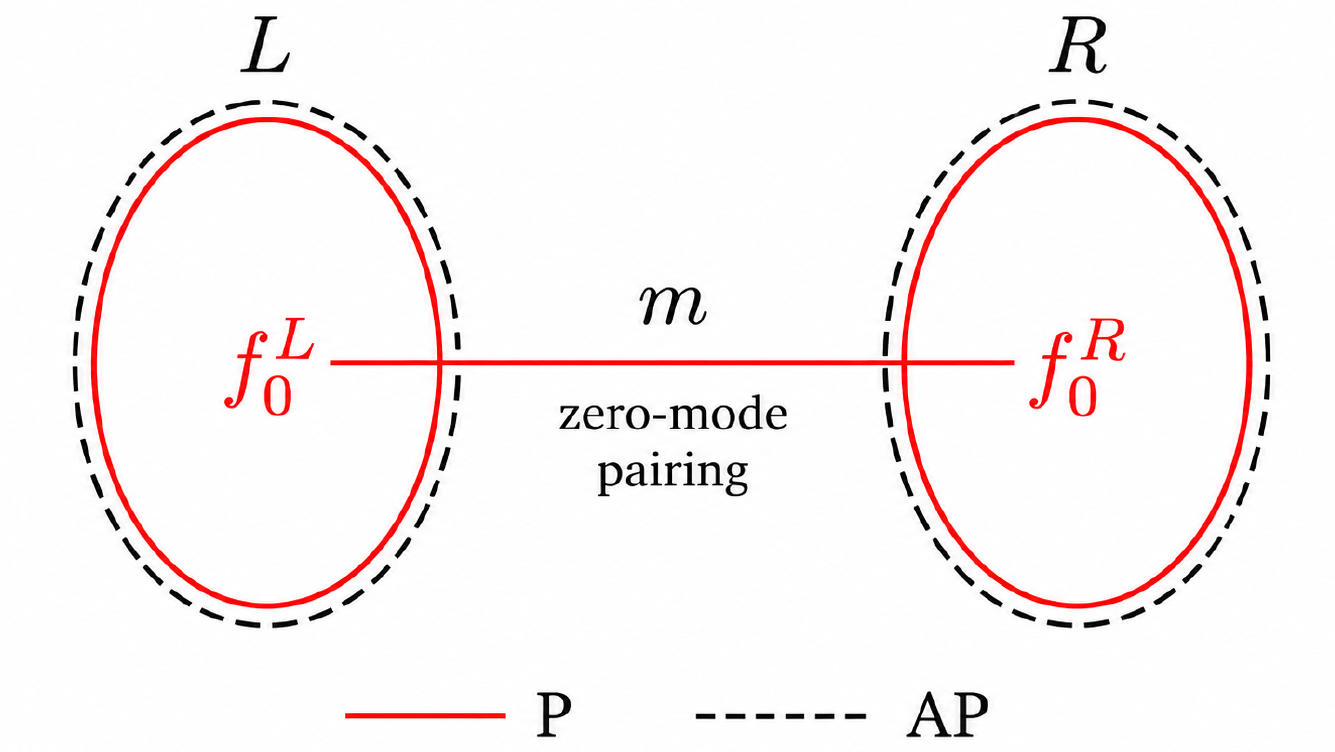}
\caption{Each thermal replica contains a periodic sector (solid red) and an
antiperiodic sector (dashed black). The constant modes $f_0^L$ and $f_0^R$
extend around their respective thermal circles. The off-diagonal
self-energy pairs them through the matrix element $m$ defined in
\eqref{eq:app-zero-mode-matrix-element}. This pairing produces the leading
$LR$ propagator in the periodic sector and supports the connected saddle.}
\label{fig:thermal-one-point-wormhole}
\end{figure}

The periodic kinetic operator has the normalized constant zero mode
\be
f_0(\tau)=\beta^{-1/2}.
\ee
Consequently, the periodic Pfaffian vanishes unless interactions lift this
mode. A replica-diagonal
self-energy has a vanishing projection onto this mode by antisymmetry.
An off-diagonal self-energy can instead pair the zero modes of the
two replicas, allowing a nonzero contribution to the observable, as
illustrated in Fig.~\ref{fig:thermal-one-point-wormhole}.
This pairing also explains why the periodic sector remains important
even though it contains only a fraction $\nu=\kappa/p$ of the fermions.
The periodic propagator is inversely proportional to the coupling
between the zero modes: a weak pairing, therefore, produces a large
propagator, which can compensate for the small weight of this sector.

We seek a regular saddle at fixed $\kappa$ and $\beta\mathcal J$.
Writing $x=\tau_1$ and $y=\tau_2$, we take
\begin{gather}
G^{LR}(x,y)
=1+\frac{h(x,y)}{p}+O(p^{-2}).
\label{eq:sec-offdiagonal-ansatz}\\
\Sigma^{LR}(x,y)
=
\frac{\mathcal J^2}{2p}e^{h(x,y)}
\left[1+O(p^{-1})\right].
\label{eq:sec-offdiagonal-largep}
\end{gather}
Here $h(x,y)$ is of order one, and we make no assumption of translation
invariance.

Because the antiperiodic kinetic operator has no zero mode, its contribution to $G^{LR}(x,y)$ remains perturbative, $O(p^{-1})$. The order-one term in
\eqref{eq:sec-offdiagonal-ansatz} must therefore come from the
periodic zero modes. To evaluate this contribution, project the
off-diagonal self-energy onto the normalized constant modes:
\be
m
=
\frac1\beta
\int_0^\beta {\rm d}x\,{\rm d}y\,\Sigma^{LR}(x,y).
\label{eq:app-zero-mode-matrix-element}
\ee
Using \eqref{eq:app-resolvents}, the zero-mode contribution to the mixed propagator is therefore
\be
G^{LR}(x,y)
=
\nu \,\frac{f_0(x)f_0(y)}{m}+O(p^{-1})
=
\frac{\kappa}{p\beta m}+O(p^{-1}).
\label{eq:app-zero-mode-matching}
\ee
This contribution is independent of $x$ and $y$ because, at leading order,
zero-mode wavefunctions are constant. Matching it to the unit
leading term in \eqref{eq:sec-offdiagonal-ansatz} and using \eqref{eq:sec-offdiagonal-largep} gives the consistency condition
\be
\int_0^\beta {\rm d}x\,{\rm d}y\,e^{h(x,y)}
=\frac{2\kappa}{\mathcal J^2}.
\label{eq:app-zero-mode-normalization}
\ee
Thus the string size $\kappa$ fixes the strength of the $LR$ correlations through this constraint.

\subsection{Solution and local geometry}
\label{sec:solwinding}

At leading nontrivial order in $1/p$, the Schwinger-Dyson equations yield the Liouville equation
\be
\partial_x\partial_y h(x,y)=-2\mathcal J^2e^{h(x,y)},
\qquad 0<x,y<\beta.
\label{eq:sec-offdiagonal-liouville}
\ee
The boundary conditions of $h(x,y)$ are particularly involved because this field contains the two spin structures and is moreover affected by the zero mode. In particular, these boundary conditions couple this field to the replica diagonal fields. We write and solve this coupled system in Appendix~\ref{app:SD-onepoint}. Here we focus on providing the geometric description of the solution and its interpretation in JT gravity.

Consider reversing the orientation of the right boundary and introducing the dimensionless time variables $u=\mathcal Jx$ and $v=\mathcal J(\beta-y)$. On the domain $0< u,v<\beta \mathcal J$, the solution takes the form
\be
e^{h(x,y)}=
\frac{\dot\sigma(u)\dot\sigma(v)}
{4\cosh^2[(\sigma(u)-\sigma(v))/2]},
\qquad \sigma(0)=-\frac{b}{2},\quad \sigma(\beta \mathcal{J})=\frac{b}{2},
\label{eq:duniformization}
\ee
where a dot denotes a derivative with respect to dimensionless time.
The increasing function $\sigma(u)$ reparametrizes the boundary time on the hyperbolic disk.
Its range for one thermal proper time follows directly from the zero-mode normalization
\eqref{eq:app-zero-mode-normalization}:
\be
2\kappa=
\int_{-b/2}^{b/2}\!{\rm d}s\int_{-b/2}^{b/2}\!{\rm d}t\,
\frac{1}{4\cosh^2[(s-t)/2]}
=2\log\cosh(b/2),
\qquad b=2\operatorname{arcosh}(e^\kappa).
\label{eq:dspan}
\ee
Thus the operator fixes the range of this coordinate, while temperature determines the trajectory of the boundary in the hyperbolic disk. More explicitly, the full saddle gives
\be
\begin{gathered}
\dot\sigma=A_0+A_1\cosh\sigma>0,
\qquad A_1>0,\\
e^{2\kappa}A_1^2-A_0^2=4,
\qquad
\beta\mathcal J=\int_{-b/2}^{b/2}\frac{{\rm d}s}{A_0+A_1\cosh s}.
\end{gathered}
\label{eq:dclock-equations}
\ee
The last two equations determine $A_0$ and $A_1$ as functions of
$\beta\mathcal J$ and $\kappa$. Together with $\sigma(0)=-b/2$,
the first equation then uniquely determines $\sigma(u)$.

\begin{figure}[h]
\centering
\includegraphics[width=\linewidth]{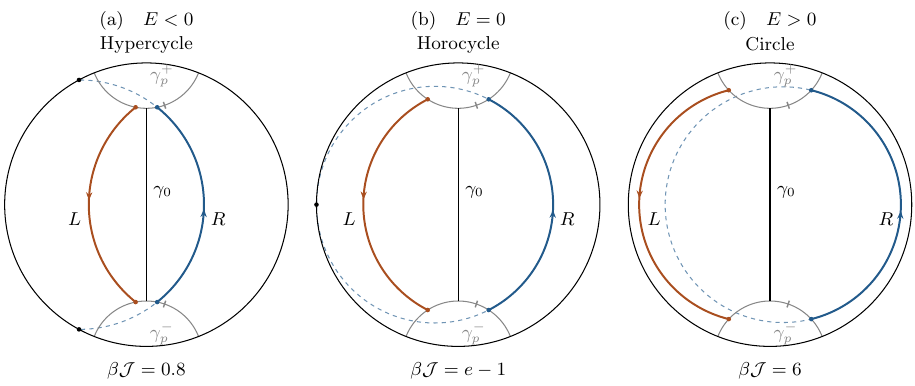}
\caption{Geometric representation of the LR solution. From left to right, the negative-, zero-, and positive-Schwarzian
branches give a hypercycle, a horocycle, and a circle boundary trajectory on the hyperbolic disk. Gray geodesics $\gamma_p^\pm$
mark the cuts used to construct the cylinder below, and the black
segment $\gamma_0$ becomes the bottleneck. Its length $b$ is solely determined by the operator. All panels have $\kappa=1$.
At fixed $\kappa$, the Schwarzian energy changes sign at
$\beta\mathcal J=e^\kappa-1$.}
\label{fig:dtrajectories}
\end{figure}

To read off the local geometry, use the reparametrizations
$f_R=e^{\sigma(u)}$ and $f_L=-e^{\sigma(v)}$ in the upper half-plane
representation.
Eq.~\eqref{eq:duniformization} is then the renormalized
geodesic kernel between them. To understand the cutoff boundary trajectories, note that \eqref{eq:dclock-equations} implies that these are constant Schwarzian trajectories
\be
E:=\{e^\sigma,u\}=\frac{A_1^2-A_0^2}{2},
\qquad \partial_uE=0.
\label{eq:dSchwarzian}
\ee
Here $\{f,u\}$ denotes the Schwarzian derivative of the reparametrization $f$. As $\beta \mathcal{J}$ increases at fixed $\kappa$, $E$ changes
from negative to positive at $\beta \mathcal{J}=e^\kappa-1$. The corresponding boundary trajectory
orbit changes from a hypercycle through a horocycle to a circle in the
disk, as shown in Fig.~\ref{fig:dtrajectories}. The saddle remains regular through this change.

\subsection{Wormhole topology and winding}
\label{sec:fixed-alpha-winding-match}

To close the boundaries in this hyperbolic representation, we take the insertion to glue the two geodesic cuts by the upper half-plane isometry $f\mapsto e^b f$, equivalently $\sigma\mapsto\sigma+b$. This identification joins the endpoints of each boundary trajectory, with a kink at the insertion. The resulting metric quotient is a double trumpet:
\be
{\rm d}s^2={\rm d}r^2+\cosh^2r\,{\rm d}\sigma^2,
\qquad \sigma\sim\sigma+b,
\qquad b=2\operatorname{arcosh}(e^\kappa).
\label{eq:dcylinder-reconstruction}
\ee
The metric can be glued in this way for every $\kappa>0$.
With this prescription, the operator fixes the bottleneck
circumference, while the temperature modifies the boundary trajectories.
The dilaton and matter junctions needed to interpret this solution as a solution to JT gravity are discussed in the next subsection.

The continuation of $h(x,y)$ is determined by
the microscopic propagators: we continue $G_{\rm P}$ periodically
and $G_{\rm AP}$ antiperiodically, and then combine them in
$G=(1-\kappa/p)G_{\rm AP}+(\kappa/p)G_{\rm P}$.
The mixed kernel $h(x,y)$ therefore does not itself obey a single
periodic or antiperiodic condition.

The question, following \cite{Bettaque:2026vpl}, is whether the mixed correlator in the solution
\eqref{eq:duniformization} can, in some approximation, be
understood as a weighted combination of free-particle propagators
on this geometry, for probes of boundary scaling dimension
$\Delta_\psi=1/p$. The individual propagators are given by the usual
sums over winding images:
\be
\begin{gathered}
K_n(u,v)=\left[
\frac{\dot\sigma(u)\dot\sigma(v)}
{4\cosh^2[(\sigma(u)-\sigma(v)+nb_{\rm w})/2]}
\right]^{1/p},\\[.2cm]
\mathcal{G}_{\rm P}^{\rm w}=\sum_{n\in\mathbb Z}K_n,
\qquad
\mathcal{G}_{\rm AP}^{\rm w}=\sum_{n\in\mathbb Z}(-1)^nK_n.
\end{gathered}
\label{eq:dwinding-sums}
\ee
Here $\sigma$ is the boundary trajectory on the original interval,
with a shift by $nb_{\rm w}$ on its $n$th copy.  Taking an endpoint once
around the cylinder shifts the winding label by one, so the two
sums have precisely the required thermal signs. We denote the winding bottleneck by $b_{\rm w}$ to distinguish it from $b$, with which it will be compared below.

The periodic zero mode gives a simple test of this interpretation.
Summing all windings before expanding at large $p$, we find
\be
\mathcal{G}_{\rm P}^{\rm w}=\frac{2p}{b_{\rm w}}+O(1),
\qquad
G_{\rm P}^{LR}=\frac p\kappa+O(1).
\label{eq:dwinding-zero-mode}
\ee
The winding sum produces the periodic enhancement, which is absent
from the unwound term alone. Its coefficient also reveals a
mismatch at small $\kappa$: exact agreement would require an image circumference
$b_{\rm w}=2\kappa+O(p^{-1})$, whereas the exact SYK solution gives
$b=2\operatorname{arcosh}(e^\kappa)$.

For large $\kappa$, the difference is small, and both correlators agree at leading order. A more non-trivial check is the matching of the $O(p^{-1})$ term in the propagator. To reproduce the SYK solution $h(x,y)$ via sum over winding images in \eqref{eq:dwinding-sums}, one must shift the effective winding bottleneck by an $O(p^{-1})$ amount,
\be
b_{\rm w}=2\kappa+
\frac{2\kappa\log(2\sinh\kappa)}p+O(p^{-2}).
\label{eq:dmatched-image-circumference}
\ee
Recall that this is a negligible shift in the geometry at large-$p$, but because the periodic zero mode eigenvalue is enhanced, it will produce an $O(p^{-1})$ effect in the propagator. For the winding sum, we take the boundary reparametrization to agree with $\sigma$
away from the operator insertion, in the large-$\kappa$ interior.
We denote the matched images by $K_n^{\rm w}$ and their sums by
$\mathcal{G}_{\rm P,AP}^{\rm w}$.

The single-geodesic form of the interaction kernel can now be
compared with the complete winding sums. The two spin sectors enter
with different weights:
\be
\begin{aligned}
\mathcal{G}^{\rm w}
&=\left(1-\frac{\kappa}{p}\right)\mathcal{G}_{\rm AP}^{\rm w}
+\frac{\kappa}{p}\mathcal{G}_{\rm P}^{\rm w}\\
&=\sum_{n\in\mathbb Z}
\left[\frac{\kappa}{p}
+\left(1-\frac{\kappa}{p}\right)(-1)^n\right]K_n.
\end{aligned}
\label{eq:dmixed-winding-sum}
\ee
Even windings have weight $1$, while odd windings have weight
$-1+2\kappa/p$. To evaluate their cancellation, we first take
$p\to\infty$ at fixed $\kappa$, and then consider $\kappa\gg1$.
Suppose we evaluate the correlator at points that remain a distance $L\gg1$ from the insertion
in the covering coordinate:
$|\sigma(u)|,|\sigma(v)|\leq b/2-L$.
Writing $d=\sigma(u)-\sigma(v)$ and
$\mathcal V=\dot\sigma(u)\dot\sigma(v)$,
the resummed propagators give
\be
\begin{aligned}
\frac{\kappa}{p}\mathcal{G}_{\rm P}^{\rm w}
&=1+\frac{\log\mathcal V-\log(2\sinh\kappa)}p
+O\!\left(\frac{e^{-L}}p\right)+O(p^{-2}),\\
\left(1-\frac{\kappa}{p}\right)\mathcal{G}_{\rm AP}^{\rm w}
&=\frac{\kappa-\log[4\cosh^2(d/2)]}{p}
+O\!\left(\frac{e^{-L}}p\right)+O(p^{-2}).
\end{aligned}
\label{eq:dweighted-sector-expansion}
\ee
At $O(p^{-1})$, the $+\kappa$ in the antiperiodic contribution
cancels the leading part of $-\log(2\sinh\kappa)$ in the periodic
contribution:
\be
\kappa-\log(2\sinh\kappa)
=-\log(1-e^{-2\kappa})
=O(e^{-2\kappa}).
\label{eq:dsector-cancellation}
\ee
The remaining logarithm,
$\log[\mathcal V/(4\cosh^2(d/2))]$, is precisely $h(x,y)$ in
\eqref{eq:duniformization}. Thus, the complete weighted sum
reproduces the mixed correlator in this regime:
\be
\mathcal{G}^{\rm w}(u,v)
=1+\frac{h(x,y)}p
+O\!\left(\frac{e^{-L}}p\right)+O(p^{-2}).
\label{eq:dmixed-winding-match}
\ee
Winding therefore remains essential to the separate sector propagators
even when their weighted sum has a single-geodesic form.
The periodic zero mode and the unwound image should not be
identified: the former supplies the leading constant through the
complete winding sum, while the latter describes the local
geodesic dependence that emerges after the sectors are combined.

At small $\kappa$, the geometric neck of the solution shrinks as
$b\sim2\sqrt{2\kappa}$, while the circumference inferred from the
sum over winding scales as $b_{\rm w} \sim 2\kappa$. There is no longer a large
region in which the insertion can be neglected, and the exact
propagators depend on where the endpoints lie relative to it.
The metric gluing remains well-defined, but the free image sums
do not reproduce these propagators. A bulk interpretation must
then account for the insertion in the winding amplitudes themselves.
The thermal signs specify how the fermions return around the
cylinder; they do not determine those amplitudes.

A similar matching between microscopic propagators and double-trumpet winding sums can be established when the operator contains a fixed fraction $\nu=w/N\in(0,1/2)$ of the fermions as $p\to\infty$. In this regime, the periodic sector has finite weight, so its full propagator must be retained instead of the zero-mode approximation used above. We derive the corresponding saddle in Appendix \ref{sec:extensive-wormhole-windings}, starting from the Schwinger–Dyson equations of Sec. \ref{eq:secopspin}. At leading order, the winding sums reproduce the propagators in each spin sector separately, while the nonzero winding contributions cancel in their weighted sum.

\subsection{Effective JT gravity description}
\label{sec:relative-geometry-dilaton}

We now compare the reconstructed geometry with the wormhole
in JT gravity \cite{Stanford:2020wkf}, asking whether an ordinary massive particle can
support the same boundary trajectories and dilaton.

With the AdS radius set to one and omitting the topological term, the Euclidean JT action coupled to a massive particle is
\be\label{eq:JTaction}
I=-\frac{1}{16\pi G_{\rm N}}\int_M {\rm d}^2x\,\sqrt g\,\phi(R+2)
-\frac{1}{8\pi G_{\rm N}}\int_{\partial M}{\rm d}s\,\phi(K-1)
+m_p\int_{\gamma_p}{\rm d}s.
\ee
The metric remains locally hyperbolic, $R = -2$, while matter backreaction is encoded in the dilaton and on the trajectory of the cutoff boundary. Away from the worldline, the dilaton obeys
\be
\nabla_\mu\nabla_\nu\phi=g_{\mu\nu}\phi.
\label{eq:dilatoneq}
\ee
The particle worldline joins vacuum regions across which $\phi$ is
continuous, but its normal derivative jumps with a coefficient proportional to $m_p$.

\begin{figure}[h]
\centering
\includegraphics[width=\linewidth]{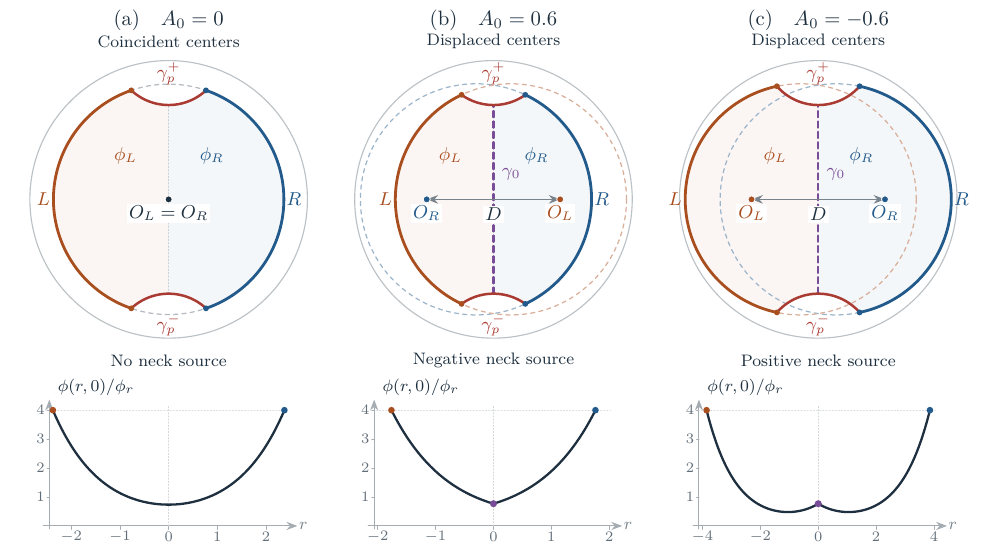}
\caption{The red geodesics $\gamma_p^\pm$ are identified and
become a single insertion worldline. In panel (a), $A_0=0$: the centers
coincide and the dilaton is smooth across the neck. In panel (b),
$A_0=0.6$: the centers are displaced and the prescribed gluing requires
an additional neck source (purple, $\gamma_0$). The lower panels show
$\phi(r,0)/\phi_r$ on the same axis scales, highlighting the smooth
profile in (a) and the derivative jump in (b). All examples have
positive Schwarzian energy, with $\kappa=1$.}
\label{fig:relative-geometry-dilaton}
\end{figure}

Consider the SYK wormhole solution. First, the $LR$ solution fixes the relative placement of the two boundary
arcs, in addition to their Schwarzian energy. On the positive-energy
branch, the hyperbolic distance $D$ between the centers of the arcs satisfies
\be
\sinh\frac D2=\frac{|A_0|}{\sqrt{A_1^2-A_0^2}}.
\label{eq:dcenter-separation}
\ee
Thus $A_0$ controls the displacement of the centers, which coincide only if $A_0=0$. Fig. ~\ref{fig:relative-geometry-dilaton} illustrates the three possible configurations.

In the metric \eqref{eq:dcylinder-reconstruction}, imposing the
same asymptotic dilaton coefficient $\phi_r$ on both boundaries fixes
the solution to \eqref{eq:dilatoneq} to be
\be
\phi_{L,R}=\phi_r\left[
A_1\cosh r\cosh\sigma\mp A_0\sinh r\right].
\label{eq:dtwo-dilatons}
\ee
For $A_0\ne0$, these fields coincide only at the neck, $r=0$.
Using $\phi_R$ for $r>0$ and $\phi_L$ for $r<0$ therefore gives the
continuous field
\be
\phi=\phi_r\left[
A_1\cosh r\cosh\sigma+A_0|\sinh r|\right],
\qquad
[\partial_r\phi]_{r=0}=2\phi_r A_0.
\label{eq:dneck-dilaton}
\ee
This profile satisfies the vacuum equations on either side of the
neck, but the jump in its normal derivative requires a localized
source there. Reproducing the general SYK solution within this
reconstruction therefore requires a bottleneck source with coefficient
$\mu_0$.

The second junction lies along the worldline $\gamma_p$. Using a signed normal
coordinate and the particle junction convention
$[\partial_n\phi]=-8\pi G_{\rm N}\mu$, the two source coefficients
are\footnote{The same particle mass also determines the boundary kink
at the insertion through a jump of the $SL(2,\mathbb{R})$ charges; see, for example, Appendix B of
\cite{Goel:2018ubv}. In our coordinates,
the gluing gives
$[\ddot f/\dot f]_{\gamma_p}=-2A_1\sinh(b/2)$, and the boundary
junction condition is
$m_p=-C[\ddot f/\dot f]_{\gamma_p}$.}
\be
m_p=2CA_1\sinh(b/2)>0,
\qquad
\mu_0=-2CA_0,
\qquad
C=\frac{\phi_r}{8\pi G_{\rm N}}.
\label{eq:drequired-sources}
\ee
The particle action $m_p\ell_p$, where $\ell_p$ is the renormalized
length of $\gamma_p$, supplies the transverse junction but cannot
account for the additional jump along the closed neck.  At low temperatures $\mu_0$ is exponentially suppressed in $\kappa$ with respect to $m_p$, so it might be a good approximation to neglect this source. However, $m_p$ generally varies with temperature,
 so a linear worldline action with fixed mass does not describe the
entire family.

For $A_0\ne0$, the bottleneck source can be supplied by a constraint term added to the JT gravity action \eqref{eq:JTaction} on the double trumpet topology
\be
\mu\left[b-2\operatorname{arcosh}(e^\kappa)\right],
\ee
whose multiplier takes the stationary value $\mu=\mu_0$ and may
have either sign.

\subsection{Diagonal correlators and low-temperature limit}
\label{sec:diagonal-correlators-low-temperature}

At finite temperature, the replica-diagonal correlators introduce a subtlety in this geometric picture. Viewed on their own, they contain a relative shift between the insertions, reminiscent of the shift that occurs in thermal correlators, and are geometrized by the fake disk \cite{Lin:2023trc}. This fake disk interpretation, however, does not extend straightforwardly to a fake double trumpet: a relative shift of the two boundaries would also affect the $LR$ correlator, while the solution \eqref{eq:duniformization} shows no sign of such a shift.

For $0<y<x<\beta$, we write the single-replica bilocals as $G^{LL}=G^{RR}=1+\Phi/p+O(p^{-2})$. On the positive-Schwarzian
branch, the solution in Appendix~\ref{app:SD-onepoint} gives
\be
e^{\Phi(x,y)}=
\frac{\epsilon^2}{4\sin^2\!\left[\frac\epsilon2
\bigl(\mathcal J(x-y)+a\bigr)\right]},
\qquad
\epsilon=\sqrt{2E},\qquad
a=\frac2\epsilon\arcsin\frac\epsilon2.
\label{eq:diagonal-shifted-kernel}
\ee
The translation invariance of $\Phi(x,y)$ applies only to the mixed replica-diagonal correlator, not to its individual periodic/antiperiodic components. The negative and zero Schwarzian energy branches can be obtained from the analytic continuation of this solution. The shift by $a$ keeps the diagonal correlator finite, with the coincidence-point limit $\lim_{x\to y}\Phi(x,y)=0$. As in the thermal fake disk~\cite{Lin:2023trc}, this kernel can be represented geodesically on a fake disk by using different endpoint maps for the bra and ket, separated by an auxiliary time interval $a/\mathcal J$.

The problem is that this prescription does not naturally extend to the full double-trumpet geometry without modifying the $LR$ solution. The full $LR$ correlator fixes the relative boundary trajectories, up to a common isometry and cylinder identifications. Once this relative positioning is fixed, the only remaining freedom in choosing their lifts is to act with the identification that defines the cylinder. The diagonal shift, however, requires an elliptic isometry about each boundary orbit’s hyperbolic center separately. Adding a conical defect at either center does not solve the problem. Such a modification would also change the geodesics connecting the two sides. This rules out a simple geometric picture for the full solution at finite temperature.

\paragraph{Low-temperature limit and finite $\kappa$.} The diagonal shift disappears in the low-temperature limit. After taking the large-$p$ limit, introduce
\be
t=\varepsilon x=\frac{\varepsilon}{\mathcal J}u,
\qquad
\beta_{\rm JT}=\varepsilon\beta,
\qquad \varepsilon\longrightarrow0,
\label{eq:diagonal-infrared-clock}
\ee
holding $\beta_{\rm JT}$ and $C_{\rm Schw}:= \varepsilon C$ fixed and
positive. The latter is the coefficient of the Schwarzian action
in the rescaled time coordinate $t$. At fixed $\kappa$, $a\to1$ while
$\epsilon/\varepsilon\to2\pi/(\mathcal{J}\beta_{\rm JT})$, so the rescaled shift
$\varepsilon a$ vanishes. For fixed separated interior times
$t_1=\varepsilon x$ and $t_2=\varepsilon y$, the correlators approach
\be
\varepsilon^{-2}e^{\Phi(x,y)}
\longrightarrow
\frac{\bigl(\pi/(\mathcal J\beta_{\rm JT})\bigr)^2}
{\sin^2[\pi(t_1-t_2)/\beta_{\rm JT}]},
\qquad
\varepsilon^{-2}e^{h(x,y)}\longrightarrow0.
\label{eq:diagonal-fixed-alpha-limit}
\ee
Thus $LL$ and $RR$ recover the ordinary thermal-disk kernel, while
the renormalized $LR$ separation diverges. The short-distance limit
is nonuniform: microscopic coincidence $x \to y$ remains finite if taken before the low-energy limit.

\begin{figure}[h]
\centering
\includegraphics[width=\linewidth]{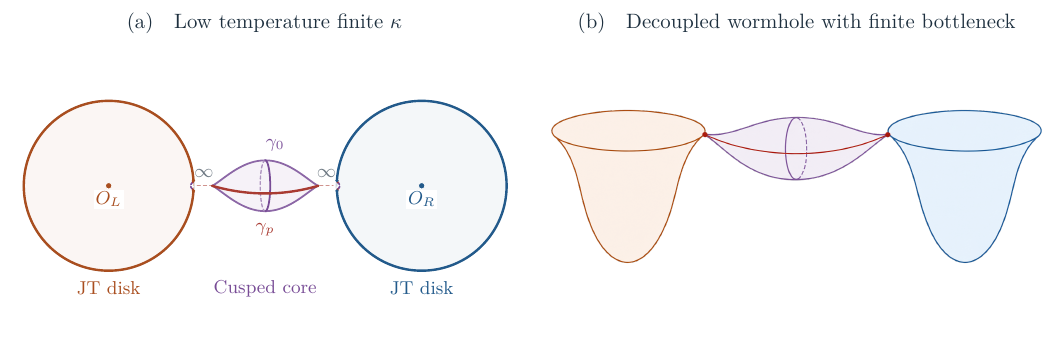}
\caption{Low temperature at fixed $\kappa$.
(a) The wormhole approaches two thermal JT disks glued by a residual cusped core which retains a finite closed neck $\gamma_0$.
The ideal boundary cusps attach to the disk ends at infinite
separation, indicated by dashed links.
(b) A schematic view of the wormhole topology.
The cusps do not represent a pinching of the closed neck.}
\label{fig:diagonal-fixed-alpha-geometry}
\end{figure}

Geometrically, taking this limit of the solution constructed in the preceding subsection yields two thermal disks connected by a cusped hyperbolic core whose closed geodesic has finite length $b=2\operatorname{arcosh}(e^\kappa)$.
Each side of the core develops an ideal boundary cusp, through which
the corresponding disk recedes to infinite distance. The geometry is illustrated in Fig.~\ref{fig:diagonal-fixed-alpha-geometry}.
These boundary cusps do not pinch the closed neck.
The required bottleneck source follows from
\eqref{eq:drequired-sources}:
\be
\mu_0\sim\frac{4C_{\rm Schw}}{\varepsilon \sinh(b/2)}
\longrightarrow \infty .
\label{eq:diagonal-fixed-alpha-source}
\ee
The core therefore retains a finite hyperbolic metric but requires
a diverging source for the dilaton junction along its neck.

\begin{figure}[h]
\centering
\includegraphics[width=.83\linewidth]{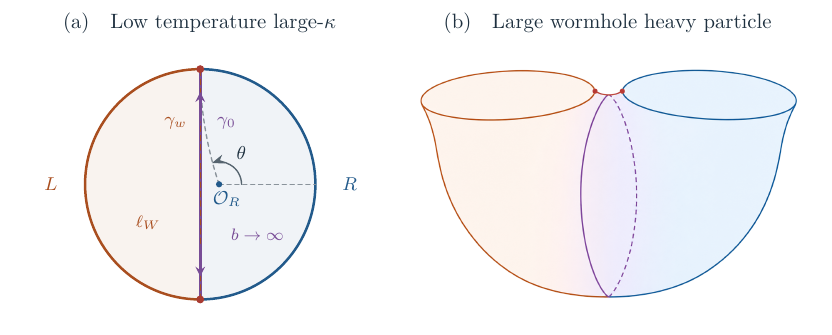}
\caption{Low temperature at fixed renormalized length $\ell_W$,
with $\kappa$ scaled as in \eqref{eq:diagonal-fixed-length-limit}.
(a) The closed neck grows without bound and opens locally into a
complete geodesic $\gamma_0$, approached by the winding geodesic
$\gamma_w$. The wormhole is supported by an infinitely heavy particle and a finite stress tensor at the bottleneck.
(b) A schematic view of the wormhole topology.}
\label{fig:diagonal-fixed-length-geometry}
\end{figure}

\paragraph{Scaling $\kappa$ and a large wormhole.}
To retain finite correlations between the two boundaries, we scale
the operator-size parameter $\kappa$ with the parameter $\varepsilon$. The required scaling will be logarithmic
\be
\kappa=\log\frac1\varepsilon+\frac{\ell_W}{2},
\qquad
b=2\log\frac2\varepsilon+\ell_W+o(1)
\longrightarrow\infty.
\label{eq:diagonal-fixed-length-limit}
\ee
Here, $\ell_W$ is fixed by the renormalized length of the geodesic that winds around the wormhole once and returns to the marked boundary point selected by the operator. In the next section, we will provide a more intuitive derivation of a similar scaling. As the bottleneck circumference $b$ diverges, the bottleneck geodesic approaches this winding geodesic. Thus, in this limit, $\ell_W$ also gives the renormalized length of the bottleneck, even if this bottleneck is not an open geodesic. As shown in Fig. \ref{fig:diagonal-fixed-length-geometry}, the limiting geometry can be represented as a hyperbolic disk with two marked points inherited from the original identification. These points separate the two boundaries, each of duration $\beta_{\rm JT}$. On the positive-Schwarzian branch, each boundary subtends an angle $2\theta$ of its circumference, where
\be
\mathcal J\beta_{\rm JT}\,e^{-\ell_W/2}
=\frac{\theta}{\sin\theta},
\qquad 0<\theta<\pi.
\ee
For fixed interior times $t_1=\varepsilon x$ and $t_2=\varepsilon y$, the $LR$ kernel has the explicit limit
\be
\varepsilon^{-2}e^{h(x,y)}
\longrightarrow
\frac{e^{-\ell_W}\sin^4\theta}
{\left[
\cos\!\frac{\theta(t_1-t_2)}{\beta_{\rm JT}}
-\cos\theta\,
\cos\!\frac{\theta(t_1+t_2-\beta_{\rm JT})}{\beta_{\rm JT}}
\right]^2}.
\ee
The diagonal correlators retain the same angle $\theta$:
\be
\varepsilon^{-2}e^{\Phi(x,y)}
\longrightarrow
\frac{\bigl[\theta/(\mathcal J\beta_{\rm JT})\bigr]^2}
{\sin^2[\theta(t_1-t_2)/\beta_{\rm JT}]}.
\label{eq:diagonal-fixed-length-kernel}
\ee

The worldline and bottleneck source coefficients in JT gravity that support such a geometry are
\be
m_p\sim\dfrac{4C_{\rm Schw}}{\varepsilon}\longrightarrow\infty,
\qquad
\mu_0\longrightarrow
-4 C_{\rm Schw} e^{-\ell_W/2}\cos\theta.
\label{eq:diagonal-limiting-sources}
\ee
Thus, in this limit, the wormhole is supported by an infinitely heavy particle. Moreover, the geometry carries a finite dilaton junction along
the bottleneck. At $\theta=\pi/2$, the bottleneck junction disappears, and
the two intervals form the smooth thermal disk of period
$2\beta_{\rm JT}$.

\subsection{Finite-$p$ numerics}
\label{sec:finitep-main}

The wormhole solution to \eqref{eq:app-correct-collective-action} can also be studied numerically at finite $p$, keeping
$N\to\infty$. We solve the Schwinger--Dyson equations
\eqref{eq:app-resolvents} and \eqref{eq:app-self-energy} using the
spectral method described in Appendix~\ref{sec:finitepnum}.
This involves expanding the periodic and antiperiodic propagators in their
respective Matsubara modes, and solving the equations iteratively
without assuming the saddles to be time-translation invariant. 

The periodic zero mode of the mixed cross-replica correlator plays an important role in choosing an initial guess for the iteration. At high temperatures, it is approximately constant, with magnitude
\be
\G_0^{LR}
\simeq
\left(\frac{2p\nu}{(\beta\mathcal J)^2}\right)^{1/p}
=
\left(\frac{2\kappa}{(\beta\mathcal J)^2}\right)^{1/p}.
\ee
This approximation agrees closely with the high-temperature
numerical results, which are shown in Appendix~\ref{sec:finitepnum}. At fixed $\kappa$ and $\beta\mathcal J$, it tends to one as $p\to\infty$, matching the leading term in \eqref{eq:sec-offdiagonal-ansatz}. At fixed $p$, however, it grows
without bound as $\beta\mathcal J\to0$. Accounting for this growth
in the initial guess is essential for numerical convergence at higher temperatures.

\begin{figure}[h!]
\centering
\includegraphics[width=\textwidth]{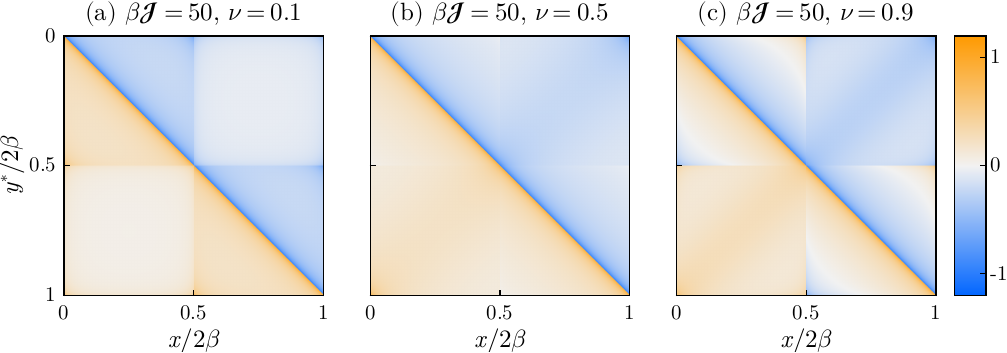}
\caption{(Block-) matrix heatmaps of numerical solutions for the two-point function $G$, with $p = 4$, $\beta \mathcal{J} = 50$ and varying values of $\nu$. $x$ and $y^* = 2\beta - y$ have been rescaled to run over both replicas. The off-diagonal blocks, which correspond to the cross-replica correlations giving rise to the wormhole geometry, exhibit roughly three distinct regimes of weak, intermediate and strong backreaction, depending on the weight of the inserted operator. $G^{LR}$ was chosen to be negative here to highlight the predicted emergence of an effective single-replica two-point function with inverse temperature $2 \beta$ around $\nu = 0.5$.}
\label{fig:G_LR_comparison_main_sec1}
\end{figure}

Fig. \ref{fig:G_LR_comparison_main_sec1} depicts numerical solutions for $G$ as a whole, with $p = 4$, $\beta \mathcal{J} = 50$ and representative choices of $\nu$. Most striking is the seeming emergence of an effective single-replica saddle with temperature $2\beta$ in the regime around $\nu = 0.5$ and $\beta\mathcal{J} \to \infty$. In fact, such a saddle emerges for any $\beta\mathcal{J}$ at a particular value of $\nu$. This is a manifestation of the identity $\sum_{\psi_W} \langle \Tr(e^{-\beta H} \psi_W)^2 \rangle_J \propto \langle \Tr(e^{-2 \beta H}) \rangle_J$, and is discussed in more detail in \cite{Bettaque:2026vpl}.

\begin{figure}[h!]
\centering
\includegraphics[width=\textwidth]{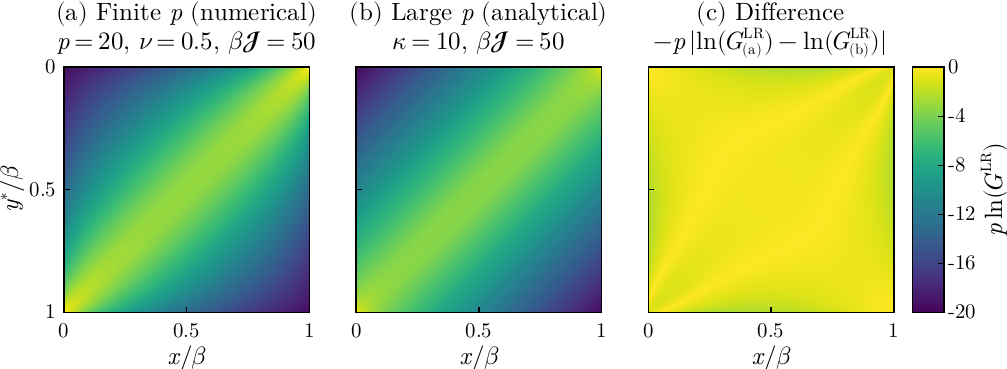}
\caption{Direct comparison between $p \ln(G^\mathrm{LR}(x, y))$ as obtained through finite-$p$ numerics, and the analytical large-$p$ solution $h(x, y)$ ($y^* = \beta - y$). Both are evaluated at the same $\beta \mathcal{J}$ and for the same value of $\kappa = p \, \nu$. Even though the correlations -- here describing a regime of strong backreaction -- are not expected to match over the whole configuration space, excellent agreement is observed for this choice of parameters, among others.}
\label{fig:G_LR_comparison_main_sec2}
\end{figure}

Furthermore, the structure of the off-diagonal entries in Fig. \ref{fig:G_LR_comparison_main_sec1} qualitatively matches that of the different regimes of backreaction for the large-$p$ solution as discussed in Appendix \ref{app:exact-coupled-saddle}. This similarity can be made more explicit by directly comparing finite-$p$ and large-$p$ solutions for the same $\beta\mathcal{J}$ and $\kappa = p \, \nu$. This is done in Fig. \ref{fig:G_LR_comparison_main_sec2}, which shows that both solutions agree to a very good degree, at least for certain choices of parameters. It is not expected that both limits produce similar results everywhere, which is a consequence of UV effects at finite $p$ and the order of limits becoming relevant

Further numerical results supporting these statements are presented in Appendix~\ref{sec:finitepnum}.

\section{The wormhole from chords}
\label{sec:thermalonept}

In this section, we take a different approach: we work in the double-scaling limit of SYK (cf. \cite{Berkooz:2024lgq}) and express the wormhole contribution as an amplitude in the ``bulk'' chord Hilbert space. We then focus on the Schwarzian regime of the observable, a limit where the chord Hilbert space naturally reduces to the Hilbert space of JT gravity. Unlike the classical limit of the previous section, this limit retains quantum Schwarzian fluctuations

\subsection{Chord expansion and endpoint condition}

\begin{figure}[h]
    \centering
    \includegraphics[width=0.42\linewidth]{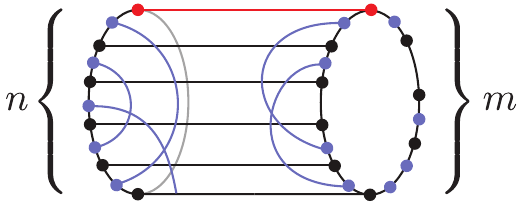}
    \caption{Chord diagram with $n$ Hamiltonian insertions on the first trace and $m $ on the second trace. Of these, $K$ cross-chords connect both traces (black), while $d_L = (n-K)/2$ and $d_R = (m-K)/2$ are disk-chords internal to each trace (blue).}
    \label{fig:disk-chords}
\end{figure}

We start by expanding the exponentials in terms of the moments
\be\label{eq:twopt2}
\mathcal{C}_\beta(W)
=
\left\langle
\Tr\!\left(e^{-\beta H}\psi_W\right)^2
\right\rangle_J
=
\sum_{n,m\ge 0} \frac{(-\beta)^{n+m}}{n!\,m!} \Gamma_{n,m} \, ,
\ee
where
\begin{equation}
\label{eq:gammaDefinition}
\Gamma_{n,m} := \left\langle
\Tr(H^n\psi_W)\Tr(H^m\psi_W)
\right\rangle_J \, .
\end{equation}
Within each moment, expanding the powers of the Hamiltonian into SYK interaction strings $\psi_{I}$ with $|I|=p$, the disorder average becomes a sum over pairings of the couplings $J_{I}$. As usual, each pairing is represented by a chord. Chords whose endpoints lie on the same trace will be called \emph{disk-chords}, while chords whose endpoints lie on different traces will be called \emph{cross-chords}. When drawing a diagram, we represent Wick pairings by chords that do not cross the matter line. Thus no chord winds around the cylinder, and the diagram can equivalently be viewed on a strip; see Fig. \ref{fig:disk-chords} for an example.

We will organize the expansion by the number of chords of each type: a given diagram has $K$ cross-chords, $d_L = (n-K)/2$ disk-chords in the left trace and $d_R = (m-K)/2$ disk-chords in the right trace. As an example, we could have a diagram with $n = 4$, $m=6$ and $K=2$ whose traces reduce to
\begin{equation}
\Tr\!\left( \psi_{I_1} \psi_{J_1} \psi_{I_2} \psi_{J_1} \psi_W\right) \Tr\!\left( \psi_{L_1} \psi_{I_2} \psi_{L_2} \psi_{I_1} \psi_{L_2} \psi_{L_1} \psi_W\right) \, .
\end{equation}

The insertion of $\psi_W$ constrains the allowed cross-chord backgrounds. For each trace not to vanish, the total product of Majorana strings on that trace must reduce to a multiple of the identity. Since $\psi_i^2=1$, fermions appearing in disk-chords cancel in pairs and only cross-chords are relevant; let $I_1,\dots,I_K$ be their index sets. Having a non-vanishing trace requires that, for every fermion label $i$, the total number of times $i$ appears among the sets $I_1,\dots,I_K$ has the same parity as its appearance in $W$. Equivalently, each $i\in W$ must appear an odd number of times among the cross-chords, while each $i\notin W$ must appear an even number of times. This is the condition
\be
I_1\oplus I_2\oplus\dots\oplus I_K=W,
\label{eq:xorcondition}
\ee
where $\oplus$ denotes the symmetric difference of sets, defined by $A \oplus B := (A \cup B)\setminus (A \cap B)$. This immediately implies the basic constraints $w \leq Kp$ and $w\equiv Kp \pmod 2$. Since $p$ is even, the second condition requires even $w$, which is the usual fermion-parity selection rule.\footnote{For $p\equiv0\!\!\pmod 4$, there is an additional antiunitary selection rule that further excludes $w\equiv 2\!\!\pmod 4$.\label{footnote2}} Much like in Sec. \ref{sec:GS}, in what follows we take $p, w \in 4 \mathbb{Z}_{>0}$ to automatically satisfy all selection rules.

The endpoint constraint admits an interpretation as a singlet
selection rule in the Majorana operator algebra. The trace retains only the $O(N)$-invariant component of the complete product formed by the external insertion and the Hamiltonian Majorana strings. Thus, only the identity channel ``propagates'' from one boundary to another. The following diagram
illustrates this projection:
\begin{center}
\begin{tikzpicture}[
  x=1.15cm,y=1cm,
  line cap=round,line join=round,
  every node/.style={font=\normalsize,inner sep=2pt},
  endpoint/.style={circle,fill=black,inner sep=1.65pt},
  leg/.style={draw=black!75,line width=0.65pt},
  identity/.style={draw=black,line width=1pt},
  tracelabel/.style={font=\small,text=black!65,anchor=north}
]
\path[use as bounding box] (-4.8,-1.8) rectangle (4.8,1.55);
\coordinate (L) at (-1.85,0);
\coordinate (R) at (1.85,0);
\draw[leg,line width=1.2pt] (-3.4,.75) -- (L);
\draw[leg,line width=1.2pt] (R) -- (3.4,.75);
\foreach \y in {0.15,0.55,0.95,1.25} {
  \draw[leg] (-3.4,-\y) -- (L);
  \draw[leg] (R) -- (3.4,-\y);
}
\node[anchor=south east] at (-3.42,.6) {$\psi_W$};
\node[anchor=south west] at (3.42,.6) {$\psi_W$};
\node[anchor=north east] at (-3.42,-1.1) {$\psi_{I_1}$};
\node[anchor=north east] at (-3.42,-0.7) {$\psi_{I_2}$};
\node[anchor=north east] at (-3.42,0.25) {$\psi_{I_K}$};
\node[anchor=north east] at (-3.6,-0.1) {$\vdots$};
\node[anchor=north west] at (3.42,-1.1) {$\psi_{I_1}$};
\node[anchor=north west] at (3.42,-0.7) {$\psi_{I_2}$};
\node[anchor=north west] at (3.42,0.25) {$\psi_{I_K}$};
\node[anchor=north west] at (3.6,-0.1) {$\vdots$};

\draw[identity] (L) -- (R);
\node[font=\large,anchor=south] at (0,0.13) {$\mathbf 1$};
\node[font=\small,text=black!65,anchor=north] at (0,-0.22)
  {identity channel};
\node[endpoint] at (L) {};
\node[endpoint] at (R) {};
\node[tracelabel] at (-1.8,-0.18) {$\Tr_L$};
\node[tracelabel] at (1.8,-0.18) {$\Tr_R$};

\end{tikzpicture}
\end{center}

For a fixed diagram, the disorder average produces one normalized sum
\begin{equation}
\binom{N}{p}^{-1} \sum_{|I|=p}
\end{equation}
for each chord, as follows from the coupling variance \eqref{eq:syk-disorder}. In particular, before imposing the endpoint constraint, the $K$ cross-chord labels are summed uniformly over all ordered $K$-tuples of $p$-subsets. The traces set to zero all configurations that fail to satisfy \eqref{eq:xorcondition}. The fraction of cross-chord assignments that survive is therefore
\begin{equation}
\label{eq:PKWdefinition}
P_K(W) \:= {\rm Prob}\!\left(I_1\oplus I_2\oplus\cdots\oplus I_K=W\right) \, ,
\end{equation}
where the $I_a$ are chosen independently and uniformly among all $p$-subsets of $\{1, \dots, N\}$. Thus every diagram with $K$ cross-chords is weighted by a factor $P_K(W)$. After extracting this factor, the remaining sum is a normalized average over the cross-chord configurations conditioned on \eqref{eq:xorcondition} (together with the unrestricted disk-chord labels) of the fermionic signs associated with the diagram. We make this conditional average explicit in Appendix \ref{app:ConditionedEnsemble}.

As we show in Appendix \ref{sec:appKrawtchouk}, for a fixed string of size $|W|=w$, the exact finite-$N$ value of $P_K(W)$ is\footnote{Viewing subsets of $\{1,\ldots,N\}$ as vertices of the hypercube $\mathbb Z_2^N$, this is the probability that the $p$-flip random walk, which at each step adds a uniformly random subset of size $p$ modulo $2$, reaches the vertex $W$ after $K$ steps.}
\be
P_K(W)
=\frac{1}{2^N}\sum_{k=0}^{N}
\left(\frac{K_p(k;N)}{\binom{N}{p}}\right)^K
K_k(w;N),
\label{eq:PW-final}
\ee
where $K_k(w;N)$ are the Krawtchouk polynomials, defined in \eqref{eq:Kt-def}. Although this formula is exact at finite $N$, it will be more useful below to replace it by a smooth large-$N$ approximation.

It is possible to see that \eqref{eq:PW-final} vanishes whenever the $K$ index sets cannot produce $W$ as their symmetric difference. In particular, $P_K(W) = 0$ if $K < w/p$. It will be important in the following discussion to know what is the smallest value of $K$ for which cross-chords can combine to produce $W$, i.e., for a given $p$ and matter operator size $w$, we define
\be
K_{\min}(w,p)
:=
\min\Bigl\{K:\ P_K(W) > 0\Bigr\} .
\label{eq:Kmin}
\ee
This is the minimal allowed number of cross-chords: in the large-$N$ limit relevant below, $K_{\rm min} = \lceil w/p \rceil$ for $w \geq p$. The only elementary exception is $0 < w < p$, when a single cross-chord cannot match $W$, so $K_{\rm min} = 2$. The leading contributions in a $\beta\to 0$ expansion of \eqref{eq:twopt2} come from such minimal cross-chord diagrams, although we will not be particularly interested in the high-temperature regime of the thermal one-point function. More importantly, for sufficiently heavy matter operators and taking the double-scaling limit, we will argue that diagrams with $K_{\rm min}$ cross-chords also dominate the expansion.

\subsection{Double-scaling limit: heavy operators and cross-chord localization}

We now consider the double-scaling limit of the SYK model
\cite{Erdos:2014zgc,Cotler:2016fpe,Berkooz:2018qkz,Berkooz:2018jqr,Goel:2023svz,Lin:2022rbf,Lin:2023trc,Berkooz:2024lgq},
\be
N\to\infty,
\qquad
p\to\infty,
\qquad
\lambda:=\frac{2p^2}{N}
\quad\text{fixed}.
\label{eq:doublescalinglimit}
\ee
The operators relevant for us have size
\be
w=\Delta p,
\label{eq:schwh}
\ee
where $\Delta$ is independent of $N$. They are subextensive in the microscopic theory since ${w}/{N}=
{\Delta\lambda}/{2p}\to 0$. In the low-energy Schwarzian limit, the two-point function of operators with fixed $\Delta$ reduces to that of Schwarzian bilocals associated with matter primaries of dimension $\Delta$ \cite{Berkooz:2018jqr}.\footnote{For random
superpositions of strings of size $\Delta p$, this correspondence extends
to higher-point functions \cite{Berkooz:2018jqr}.}

We consider a {\it heavy operator} by taking
\be
\Delta\sim\frac{1}{\lambda},
\ee
up to logarithmic corrections.\footnote{Logarithmic corrections are essential for identifying a finite renormalized wormhole length, as we will see below, and motivate the logarithmic shift taken in \eqref{eq:diagonal-fixed-length-limit}.} The reason is that the Schwarzian coupling is $C_{\rm Schw} =1/2\lambda$, and thus the unshifted scaling keeps the ratio of the operator dimension to the Schwarzian coupling fixed, yielding finite gravitational backreaction as $\lambda \to 0$. 

The exact formula \eqref{eq:PW-final} for the endpoint probability simplifies considerably in this regime. For operators of the form \eqref{eq:schwh}, at fixed $\lambda$ and $\Delta$, the dominant values of $K$ are $O(1)$: the endpoint contains $\Delta p$ fermions, whereas $K$ chords contain $Kp$ fermion factors before cancellations, and large cancellations are unlikely when drawing index sets at random. In Appendix \ref{app:ConditionedEnsemble} we make these statements precise by analyzing in detail the ensemble of cross-chords conditioned by the endpoint constraint \eqref{eq:xorcondition}. The end result is that $P_K(W)$ is exponentially suppressed for each additional cross-chord that we add. Quantitatively, for $k > 0$ fixed, equation \eqref{ConditionedEnsemble:PKSuppression} gives
\begin{equation}
P_{K_{\rm min} + k}(W) = P_{K_{\rm min}}(W)\, \exp \left[- \frac{k}{2} p \log(p) + O(p)\right] \, .
\end{equation}

\begin{figure}[h]
    \centering
    \includegraphics[width=0.7\linewidth]{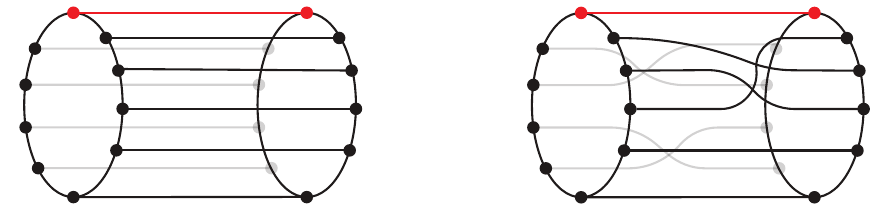}
    \caption{Left: simplest configuration with $K\geq K_{\rm min}$ cross-chords and no disk-chords. Right: configuration related by a relative permutation between the left and right insertions.}
    \label{fig:crosschords}
\end{figure}

Recall that $P_{K}(W)$ is the common factor weighting all diagrams with $K$ cross-chords. As we also show in Appendix \ref{app:ConditionedEnsemble}, the exponential suppression is so strong that it localizes the full squared thermal one-point function on the sector with the smallest allowed number of cross-chords, $K = K_{\rm min}$. After extracting $P_K(W)$, the remaining contribution of each chord diagram is a normalized average of fermionic phases arising from the traces, and is therefore $O(1)$, so it does not affect the exponential scaling with $p$. Moreover, at fixed $n$ and $m$, the number of diagrams with $K$ cross-chords is also independent of $p$, so summing over them does not change this scaling. Hence neither the weights nor the multiplicity of the diagrams can compensate for the parametric suppression of $P_K(W)$ as $p\to\infty$. We therefore conclude that $\mathcal{C}_{\beta}(W)$ localizes on diagrams with $K_{\rm min}$ cross-chords (with subleading terms exponentially suppressed in $p \log (p)$). We show some of these diagrams in Fig. \ref{fig:crosschords}, for the simplest case in which no extra disk-chords are present.

When $\Delta$ is an integer, the localization happens at $K_{\rm min} = \Delta$. In that case, a further simplification occurs. Since the number of fermions in the cross-chords is $\Delta p$, which coincides with the number of fermions in the matter operator $w$, the cross-chords must form a disjoint partition of $W$:
\begin{equation}
W = I_1 \sqcup  I_2 \sqcup \dots \sqcup I_{\Delta} \, .
\end{equation}
The endpoint probability \eqref{eq:PW-final} can be obtained in this case using simple combinatorics. We first extract a set of $p$ fermions from $W$ to form $I_1$, then successively select another group of $p$ fermions from the remaining $(\Delta - 1)p$ to construct $I_{2}$, and we continue iteratively. The result is
\begin{equation}
\label{eq:p_deltathresh}
P_\Delta(W) = \binom{N}{p}^{-\Delta} \frac{(\Delta p)!}{(p!)^{\Delta}} = \binom{N}{\Delta p}^{-1} \exp \left[- \frac{\lambda}{4} \Delta(\Delta-1) \right] \left( 1 + O(p^{-1}) \right) \, .
\end{equation}

For non-integer $\Delta$, the localization still occurs at $K_{\rm min}$, but the dominant cross-chords no longer form a disjoint partition of $W$: their conditioned overlaps remain nontrivial. We therefore restrict the discussion below to integer $\Delta$, referring to Appendix \ref{app:ConditionedEnsemble} for more details about the general case.

Before evaluating the chord diagrams, let us anticipate an important bulk consequence of the localization, which we will later develop in more detail. Through the chord–length dictionary, the chord number $K$ will become a bulk length variable. Localization at $K=\Delta$ therefore means that the matter operator size fixes the length of a geodesic in the wormhole, thus fixing its geometric modulus. A similar effect appears in the large-$p$ saddle discussed in Sec. \ref{sec:GS}, where the operator size also fixes the bottleneck circumference.

\subsection{Evaluation of chord diagrams}

Having localized the cross-chord number to $K=\Delta$, what remains is to evaluate the disk-chord dressing of this background; see Fig. \ref{fig:disk-chords}. Our way to proceed will be to first disentangle all of the disk-chords, commuting them past one another and past the cross-chord insertions. We will always commute them so that they do not pass across the operator. We take the double-scaling limit \eqref{eq:doublescalinglimit} to simplify the intersections, since in this limit the disk-disk-chord intersections are independent and weighted by a factor
\be
q = e^{-\lambda}\,.
\ee

We will use the same ordinary chord rule also for disk–cross intersections, but this requires some care because the cross-chord index sets are constrained by the matter insertion and are therefore correlated.\footnote{For a light operator $\psi_W$ these correlations are large. For
example, when $\psi_W=\mathbf 1$, the leading contributions to the connected moment $\langle \Tr(e^{-\beta H})^2\rangle_J|_{c}$ come from
pairs of cross-chords carrying the same index set \cite{Berkooz:2020fvm}.} For integer $\Delta$ and in the dominant threshold sector $K = \Delta$, the endpoint condition is especially simple: the cross-chords form a disjoint partition $W=I_1\sqcup\cdots\sqcup I_\Delta$. Consequently, a disk-chord crossing $s$ cross-chords intersects their disjoint union, of size $sp$. Its overlap with this union is asymptotically Poisson with mean $s \lambda / 2$ \cite{Berkooz:2018jqr}, so its averaged fermionic sign is
\be
e^{-s\lambda}=q^s \, .
\ee
Thus each disk–cross intersection carries the usual factor $q$. Away from threshold (or for non-integer $\Delta$) the cross-chords overlap, and this simple rule is modified. An analysis of this regime is presented in Appendix \ref{app:ConditionedEnsemble}, but in what follows we only need the simple result obtained for $\Delta \in \mathbb{Z}_{>0}$ at threshold.

Consider then the contribution to the moment \eqref{eq:gammaDefinition} with $d_L$ disk-chords on the left and $d_R$ on the right, corresponding to $n = \Delta + 2 d_L$ and $m = \Delta + 2 d_R$. As we showed, at threshold $K =\Delta$, the disk-chord weights depend only on the topology of the \textit{one-sided} chord diagrams. We thus get
\begin{equation}
\label{eq:gammadoublescaled}
\Gamma_{n,m} \sim \left(\frac{\mathcal J}{\sqrt\lambda}\right)^{n+m} c_{\Delta}(W) \left( \sum_{ D\in \mathcal{D}_{\Delta;d_L}} q^{\# {\rm int}(D)} \right) \left( \sum_{ D'\in \mathcal{D}_{\Delta;d_R}} q^{\# {\rm int}(D')} \right) \, ,
\end{equation}
where we use $\sim$ to denote equality to leading order in the double-scaled limit, $\mathcal{D}_{\Delta;d}$ is the set of chord diagrams with $\Delta$ open chords and $d$ closed disk-chords, and $\# {\rm int}(D)$ the total number (disk-disk and disk-open) of intersections in the diagram $D$. The coefficient $c_{\Delta}(W)$ contains the remaining sum over cross-chords and their relative pairing between the two traces,
\begin{equation}
\label{eq:c_K}
c_K(W) := \binom{N}{p}^{-K} \sum_{\sigma \in S_K}  \sum_{I_1, \dots, I_K} \Tr ( \psi_{I_{1}} \dots \psi_{I_{K}} \psi_W ) \Tr ( \psi_{I_{\sigma(1)}} \dots \psi_{I_{\sigma(K)}} \psi_W ) \, .
\end{equation}

At threshold the $I_a$ are mutually disjoint. Since $p$ is even, the corresponding Majorana strings commute, so changing their relative ordering between the two traces produces no fermionic sign. In the sector $p, w \in 4\mathbb{Z}_{>0}$, every nonzero product of traces in \eqref{eq:c_K} is equal to one. The sum over allowed index sets therefore gives $P_{\Delta}(W)$, while the relative permutation contributes $\Delta!$:
\begin{equation}
c_{\Delta}(W) = \Delta! P_{\Delta}(W) \, .
\end{equation}
We emphasize that with these definitions, the possible cross-chord pairings are contained in $c_{\Delta}(W)$ and should not be counted again in the sums over one-sided diagrams: open chords do not intersect there.

We conclude this section by resumming the disk-chord numbers $d_R$ and $d_L$, which gives the leading double-scaled result
\begin{equation}
\mathcal{C}_{\beta}(W) \sim c_{\Delta}(W) \left[ \sum_{d \geq 0} \frac{(-\beta \mathcal{J}/\sqrt{\lambda})^{\Delta + 2 d}}{(\Delta + 2d)!} \sum_{D \in \mathcal{D}_{\Delta;d}} q^{\# {\rm int}(D)} \right]^2 \, .
\end{equation}
The quantity in square brackets is precisely a sum over all one-sided chord diagrams with $\Delta$ open chords. It admits a natural interpretation as the wavefunction of a state in the ``bulk'' Hilbert space of chords, which we develop next.

\subsection{Emergence of the bulk Hilbert space}

The decomposition of the previous subsection has a natural bulk interpretation which is reminiscent of the construction of the wormhole in JT gravity. In fact, as we will later show, certain parts of the SYK wormhole directly reduce to gravitational objects in the Schwarzian limit.

\begin{figure}[h]
    \centering   \includegraphics[width=0.9\linewidth]{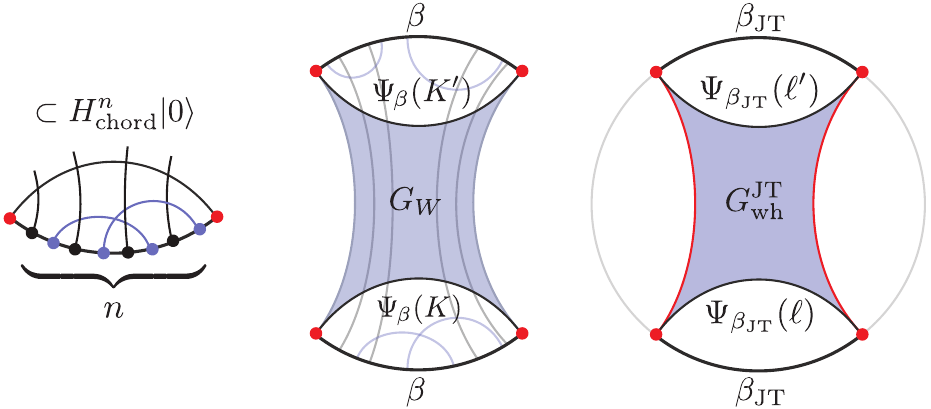}
    \caption{Left: contribution to $H_{\rm chord}^n|0\rangle$ from $n$ chord
    insertions. Middle: SYK thermal one-point wormhole written as the bulk amplitude
    $\langle \Psi_\beta|G_{W}|\Psi_\beta\rangle$. Right: JT gravity
    wormhole, $\langle \Psi_\beta|G^{\rm JT}_{\rm wh}|\Psi_\beta\rangle$,
    where the gravitational wormhole amplitude $G^{\rm JT}_{\rm wh}$ is obtained by tracing over the two red geodesic slices with a
    matter propagator inserted. In the Schwarzian limit, the Hartle-Hawking wavefunctions reduce to the JT ones. The SYK wormhole reproduces the JT wormhole only if the kernel $G_{\rm wh}$ likewise reduces to its JT counterpart $G^{\rm JT}_{\rm wh}$ near the saddle point.}
        \label{fig:decomposition}
\end{figure}

\paragraph{Chord Hilbert space and Hamiltonian.} We first recall some basic facts about the chord Hilbert space
\cite{Berkooz:2018jqr,Lin:2022rbf}. We use the following conventions for the $q$-numbers and $q$-factorials:
\be\label{eq:qfact} 
[n]_q := \dfrac{1-q^n}{1-q}\,,\qquad [K]_q! := \prod_{n=1}^K\,[n]_q\,.
\ee 

In the double-scaled SYK chord
description, a bulk slice is labeled by the number of chords crossing it.
The chord number eigenstates on this slice form an orthonormal basis
\be
\lbrace |K\rangle\,; K=0,1,2,\dots\rbrace,
\qquad
\langle K|K'\rangle=\delta_{K,K'} .
\ee

The Euclidean path integral prepares a state in this Hilbert space by
evolving the zero chord state $\ket{0}$.  It is useful, however, to first recall the chord Hamiltonian evolution.  Suppose a slice is crossed by $K$ chords.  An elementary step can either open a chord,
increasing $K$ by one, or close a chord, decreasing $K$ by one.  In the direct
counting normalization, opening a chord has weight one, while closing a
chord carries the $q$-number weight $[K]_q$ defined in \eqref{eq:qfact}, which counts the weighted intersections with the chords already crossing
the slice.  Thus the direct chord transfer matrix acts as (from now on, we set $\mathcal{J} =1$)
\be
T|K\rangle
=
{1\over \sqrt{\lambda}}
\left(
|K+1\rangle+[K]_q |K-1\rangle
\right)\,,
\qquad T|0\rangle={1\over\sqrt{\lambda}}|1\rangle \,.
\label{eq:Tcount}
\ee
This form makes the chord combinatorics manifest, but $T$ is not symmetric.

The standard double-scaled SYK Hamiltonian is instead the symmetric
representative obtained by the similarity transformation $H_{\rm chord}=S T S^{-1}$ where $S|K\rangle=\sqrt{[K]_q!}\,|K\rangle$ and the $q$-factorial $[K]_q!$ is defined in \eqref{eq:qfact}. Altogether, this leads to the ``hopping'' chord Hamiltonian
\be\label{eq:chordH}
H_{\rm chord}|K\rangle
=
{1\over\sqrt{\lambda}}
\left(
\sqrt{[K+1]_q}\,|K+1\rangle
+
\sqrt{[K]_q}\,|K-1\rangle
\right)\,.
\ee
Formally, this chord Hamiltonian can be written as the operator $\sqrt{\lambda}\, H_{\rm chord}
=
a+a^\dagger$ of a $q$-deformed oscillator algebra $[a,a^\dagger]_q:=a a^\dagger-q\,a^\dagger a=\mathbf{1}$. The spectrum of $H_{\rm chord}$ is well known \cite{Berkooz:2018qkz,Berkooz:2018jqr,Lin:2022rbf,Berkooz:2024lgq}. Its continuum eigenstates $\ket{\theta}$ are labeled by $\theta\in[0,\pi]$.\footnote{The parameter $\theta$ is the asymptotic lattice momentum conjugate to the chord number $K$. For large $K$, the hopping coefficients approach constants, and $H_{\rm chord}$ becomes translation-invariant on the $K$-lattice, so its eigenstates behave as plane waves. Since $K$ is a discrete variable, this momentum is defined modulo $2\pi$; the boundary at $K=0$ lets one choose the range $0\leq \theta\leq \pi$.} Their energies are
\be\label{eq:chordspec}
E(\theta)=\frac{2\cos\theta}{\sqrt{\lambda(1-q)}} .
\ee
and their wavefunctions are proportional to the $q$-Hermite polynomials,
\be\label{eq:energyeig}
\Phi_\theta(K) :=\langle K|\theta\rangle
=
\frac{H_K(\cos\theta|q)}{\sqrt{(q;q)_K}}\,,
\ee
for $(q;q)_K = \prod_{n=1}^{K}(1-q^n)$. These eigenfunctions obey the completeness relation for the spectral measure
\be\label{eq:qhermitemeasure}
\mathbf 1=\int_0^\pi d\theta\,\rho_q(\theta)
|\theta\rangle\langle\theta|\,,
\qquad
\rho_q(\theta)
=
\frac{(q,e^{\pm 2i\theta};q)_\infty}{2\pi}\,,
\ee
where $\pm$ represents that the product of both signs is taken and $(x,y;q)_\infty
=
\prod_{r=0}^{\infty}(1-x q^r)(1-yq^r)$.

\paragraph{Hartle-Hawking state.} Since $S|0\rangle=|0\rangle$, the thermal evolution generated by the transfer
matrix can be expressed in terms of the symmetric chord Hamiltonian as $e^{-\beta T}|0\rangle
=
S^{-1}|\Psi_\beta\rangle$, for the Hartle-Hawking state
\be
|\Psi_\beta\rangle
:=
e^{-\beta H_{\rm chord}}|0\rangle\,.
\label{eq:HHstate}
\ee

Using the spectral decomposition of $H_{\rm chord}$, the Hartle-Hawking
wavefunction is
\be\label{eq:HHchordwf}
\Psi_\beta(K) = \langle K|\Psi_\beta\rangle
=
\int_0^\pi {\rm d}\theta\,
\rho_q(\theta)\,
e^{-\beta E(\theta)}
\Phi_\theta(K)\,.
\ee
The cross-chord wavefunction prepared by the original transfer matrix is then
\be
\langle K|e^{-\beta T}|0\rangle
=
\langle K|S^{-1}|\Psi_\beta\rangle
=
\frac{1}{\sqrt{[K]_q!}}\,
\Psi_\beta(K).
\label{eq:HHwavefunction}
\ee

\subsubsection{Wormhole amplitude}

At this point, the Euclidean evolution on each half of the cylinder has prepared a Hartle-Hawking wavefunction $\Psi_\beta(K)$ in the chord
Hilbert space. What remains is to specify how the two halves are glued
together. The insertion of $\psi_W$ defines this gluing amplitude: the
cross-chords ending on the two sides of the closed slice must combine
to give the fixed Majorana string $\psi_W$.

We denote the corresponding leading gluing kernel by $G_W$. In the
double-scaling limit considered above, only the threshold background
contributes, so that
\be
G_W=
g_W\,
\ket{\Delta}\bra{\Delta},\qquad g_W:=\frac{\Delta!\,P_\Delta(W)}{[\Delta]_q!}.
\label{eq:Gkerneldiag}
\ee
The weight $\Delta!\,P_\Delta(W)$ includes the sum over the cross-chord
indices and their matchings between the two boundaries. The factor
$[\Delta]_q!$ in the denominator accounts for the normalization of the
chord-number basis.

We can now evaluate the squared thermal one-point function explicitly.
We can write the moments \eqref{eq:gammadoublescaled} as
\be
\Gamma_{n,m}
\sim
g_W\,
\bra{0}H_{\rm chord}^n\ket{\Delta}\,
\bra{\Delta}H_{\rm chord}^m\ket{0}.
\label{eq:momentbulkdecomp2}
\ee
Resumming the two exponentials gives the wormhole amplitude
\be
\mathcal{C}_\beta(W)
\sim
g_W\,
\left|\Psi_\beta(\Delta)\right|^2.
\label{eq:GWnormfinal}
\ee
Thus we arrive at
\be
\boxed{
\mathcal{C}_\beta(W)
\sim
Z_{\rm wormhole}^{\rm SYK}:=\bra{\Psi_\beta}G_W\ket{\Psi_\beta}.
}
\label{eq:GWfinal}
\ee
The interpretation of this expression is that the operator leaves the
Hartle-Hawking evolution on each boundary unchanged but defines a
wormhole amplitude $G_W$ which characterizes the way in which the two
boundaries are glued together. This is represented in
Fig.~\ref{fig:decomposition}.

\subsubsection{Other observables \& Gaussian statistics}
\label{sec:otherobsgauss}

A similar wormhole contribution can be derived with different boundary moduli for the product of thermal one-point functions at different temperatures

\be\label{eq:covariance}
\mathcal{C}_{\beta,\beta'}(W) :=\left\langle
{\rm Tr}(e^{-\beta H}\psi_W)\,{\rm Tr}(e^{-\beta' H}\psi_W)
\right\rangle_J\,.
\ee
In this case one has that in the double-scaled limit
\be
\mathcal{C}_{\beta,\beta'}(W)\sim  
\bra{\Psi_\beta}G_{W}\ket{\Psi_{\beta'}}\succeq  0,
\ee
where the positive semi-definiteness as a function of $\beta,\beta'$ follows from the fact that $G_{W}\succeq  0$, from its definition above.\footnote{This may be interpreted as part of the ``closed universe positivity'' recently discussed in \cite{DiUbaldo:2026rly,DiUbaldo:2026tkx}. Although the bulk slice here is an ``open'' slice, the positive semi-definiteness of the kernel implies the positive semi-definiteness of the closed universe kernel that we define in Sec. \ref{sec:trumpet-bottleneck}.}

\begin{figure}[h]
\centering
\includegraphics[width = .58\textwidth]{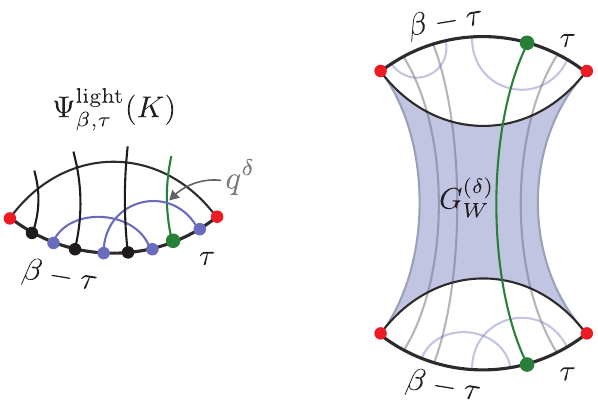}
\caption{Wormhole amplitude with a light probe of dimension $\delta$. Left: the boundary state $\Psi^{\mathrm{light}}_{\beta,\tau}(K)$; crossings between the light chord (green) and disk-chords (purple) contribute $q^\delta$. Right: two such states are glued through $G_W^{(\delta)}$, with the light chord connecting the insertions.}
\label{fig:finite-light-gluing}
\end{figure}

\paragraph{Light probes.} We can also probe the wormhole with light matter.  For definiteness, let
$\psi_{\mathrm{light}}$ be a Hermitian $p$-body operator with independent
Gaussian random coefficients, normalized so that
$\langle\Tr\psi_{\mathrm{light}}^2\rangle_{\mathrm{light}}=1$. Consider\footnote{We take $p,w\in4\mathbb Z$, for which the traces
below are real. For general Hermitian insertions, a positive covariance
is obtained by reflecting the insertion on the second boundary, since complex conjugation sends $\tau$ to $\beta-\tau$.}
\be
\mathcal C_{\beta,\tau}^{\mathrm{light}}(W)
:=\left\langle
\left[\Tr\!\left(e^{-\beta H}
\psi_{\mathrm{light}}(\tau)\psi_W\right)\right]^2
\right\rangle_{J,\mathrm{light}},
\qquad 0\leq\tau\leq\beta .
\label{eq:other-observables-light}
\ee
Averaging the light coefficients joins the two insertions by a matter
chord.  Its index set $V$ enters the boundary trace constraint, $I_1\oplus\cdots\oplus I_K\oplus V=W$, and therefore participates in the gluing of the two boundaries. In the double-scaled limit, the correlator is obtained by gluing two thermal states, through a kernel that incorporates the propagation of the light particle, as illustrated in Fig.~\ref{fig:finite-light-gluing},
\be
\mathcal C^{\mathrm{light}}_{\beta,\tau}(W)
\sim
\langle\Psi^{\mathrm{light}}_{\beta,\tau}|
G_W^{(\delta)}
|\Psi^{\mathrm{light}}_{\beta,\tau}\rangle .
\ee
Here $\delta$ is the dimension of the light operator. For the $p$-body probe considered above, this dimension is one, and
\be
G_W^{(1)}=\frac{1}{\Delta}\,aG_Wa^\dagger .
\ee
The interpretation is that $a^\dagger$ creates the light particle, $G_W$ imposes the endpoint constraint defining the wormhole amplitude, and $a$ annihilates the particle. Appendix~\ref{app:averaged-light} derives this formula and its generalization to light operators of integer dimension $\delta$.

\paragraph{Gaussian multi-trace statistics.} Higher moments provide a further test of the wormhole interpretation.
In the ordered double-scaled limit with $w=\Delta p$, positive integer
$\Delta$, and fixed moment order and thermal evolutions, the thermal
one-point functions obey Gaussian Wick factorization:
\be
\left\langle\prod_{a=1}^{2n}
\Tr\!\left(e^{-\beta_a H}\psi_W\right)\right\rangle_J
\sim
\sum_{\pi\in\mathcal P_2(2n)}
\prod_{(a,b)\in\pi}\mathcal{C}_{\beta_a,\beta_b}(W),
\label{eq:other-observables-Gaussian}
\ee
while odd moments vanish exactly.  Thus the leading contributions are
two-boundary wormholes, and connected higher cumulants are exponentially subleading in the double-scaled limit.
Appendix~\ref{app:higher-moments} proves this statement, using the single-instance bulk representation derived in Sec.~\ref{sec:alphacouplings}.

\subsection{Schwarzian limit}
\label{sec:schwlim}

We now describe the limit in which the wormhole amplitude \eqref{eq:GWfinal} acquires a gravitational interpretation in the Hilbert space of JT gravity. The main point is that the low-energy edge of the chord Hamiltonian reduces to the Hamiltonian of JT gravity. As a result, the Hartle-Hawking wavefunction $\Psi_\beta(K)$ admits a continuum limit in which the discrete cross-chord number $K$ is replaced by a renormalized geodesic length $\ell$. This winding geodesic begins and ends at the marked operator insertion at a single boundary (see Fig. \ref{fig:decomposition}). By contrast, we do \emph{not} assume that the $W$-dependent wormhole amplitude $G_W$ reduces to the JT wormhole kernel \cite{Stanford:2020wkf}. Comparing them is the main point of this subsection.

\paragraph{Continuum Hilbert space and Hamiltonian.} We zoom into the low-energy edge of the spectrum by taking the Schwarzian limit \cite{Berkooz:2018jqr,Lin:2022rbf}
\be
\lambda\to0,
\qquad
\beta_{\rm JT}=2\lambda\beta
\quad\text{fixed}.
\label{eq:schwarzian-limit}
\ee

In this regime the chord number scales to the renormalized geodesic length variable
\be
K\to\infty,
\qquad
\ell:=\lambda K+2\log(2\lambda)
\qquad\text{fixed}.
\label{eq:ellfromK}
\ee
The logarithmic shift is necessary to reproduce a finite JT Hamiltonian in the
Schwarzian limit. In this limit, the discrete chord-number basis becomes a
continuum renormalized geodesic length basis. Since $\Delta\ell=\lambda\Delta K$, one has $\sum_K\rightarrow\lambda^{-1}\int d\ell$ and $\delta_{K,K'}\rightarrow\lambda\delta(\ell-\ell')$. The continuum Hilbert space is the Hilbert space of JT gravity, $L^2(\mathbb R)$, with delta-function normalized length eigenstates $|\ell\rangle=\lambda^{-1/2}|K\rangle$. After subtracting its lower spectral edge $E_0=-2/(\sqrt{\lambda(1-q)})$ from \eqref{eq:chordspec}, the chord Hamiltonian \eqref{eq:chordH} limits to the physical Hamiltonian of JT gravity
\cite{Bagrets:2016cdf,Harlow:2018tqv,Maldacena:2018lmt}
\be
H_{\rm chord}-E_0\longrightarrow
H_{\rm JT}
=\frac{1}{C_{\rm Schw}}\left(
-\frac12\partial_\ell^2+2e^{-\ell}\right),
\qquad C_{\rm Schw}=\frac{1}{2\lambda}.
\label{eq:JT-length-H}
\ee
The low-energy eigenfunctions and spectral measure approach those of the Schwarzian theory. Consequently, after subtracting the ground-state energy and removing an overall normalization, the chord Hartle–Hawking wavefunction becomes the JT wavefunction \cite{Harlow:2018tqv,Yang:2018gdb},
\be
\Psi_{\beta_{\rm JT}}(\ell)
=
\frac{2}{\pi}\int_0^\infty {\rm d}s\,
\varrho_{\rm Schw}(s)\,
e^{-\beta_{\rm JT}s^2/2}
K_{2{\rm i}s}\!\left(4e^{-\ell/2}\right),\qquad \varrho_{\rm Schw}(s)=s\sinh(2\pi s)\,.
\ee
Here the modified Bessel function corresponds to the JT energy wavefunction in the length basis, $\Phi_s(\ell)=\bra{\ell}\ket{E_s} =\frac{2}{\pi}
K_{2\iw s}\!\left(4e^{-\ell/2}\right)$, while the factor multiplying the Boltzmann weight is the Schwarzian spectral density. The norm of the Hartle–Hawking state gives the thermal partition function of the Schwarzian theory \cite{Stanford:2017thb}:
\be
\bra{\Psi_{\beta_{\rm JT}}}\ket{\Psi_{\beta_{\rm JT}}}
=
\int_0^\infty{\rm d}s\,
s\sinh(2\pi s)\,e^{-\beta_{\rm JT}s^2}
=
Z_{\rm Schw}(2\beta_{\rm JT}).
\ee
The doubled inverse temperature reflects the two boundary segments joined in this gluing.

\paragraph{Comparison with the JT gravity wormhole.} In the Schwarzian
limit, the SYK wormhole amplitude \eqref{eq:GWfinal} becomes
\be
\mathcal{C}_\beta(W)
\sim
A_\lambda e^{-2\beta E_0}{
\bra{\Psi_{\beta_{\rm JT}}}G_W
\ket{\Psi_{\beta_{\rm JT}}}\,.
} \label{eq:SYKwormhole}
\ee
where, explicitly, $\bra{\Psi_{\beta_{\rm JT}}}G_W
\ket{\Psi_{\beta_{\rm JT}}}
= \int_{-\infty}^{\infty}{\rm d}\ell\,{\rm d}\ell'\,
\Psi_{\beta_{\rm JT}}(\ell)
G_W(\ell,\ell')
\Psi_{\beta_{\rm JT}}(\ell')$.
The prefactor in \eqref{eq:SYKwormhole} restores the spectral edge and the
spectral normalization in defining the Hartle-Hawking wavefunctions. The specific constant is $A_\lambda=\lambda(1-q)^2(q;q)_\infty^3/\pi^2$.\footnote{In the microscopic conventions
used above, the spectral measure near the lower edge satisfies
${\rm d}\theta\,\rho_q(\theta)
\sim A_\lambda\,{\rm d}s\,\varrho_{\rm Schw}(s)$.
Together with the energy eigenfunctions \eqref{eq:energyeig}, this gives
$\Psi_\beta(K)\sim
\sqrt{\lambda A_\lambda}\,e^{-\beta E_0}
\Psi_{\beta_{\rm JT}}(\ell)$,
up to phase conventions for the length eigenstates. This is where the factor of $A_\lambda$ originates.}

For a fixed Majorana string, the microscopic kernel
\eqref{eq:Gkerneldiag} selects a single length:\footnote{
Since $\ket{\Delta}=\sqrt{\lambda}\ket{\ell_W}$,
the continuum kernel is
$\lambda g_W
\ket{\ell_W}\bra{\ell_W}$.
}
\be
G_W \,\rightarrow\, \lambda \,g_W
|\ell_W\rangle\langle\ell_W|.
\label{eq:GW-localized}
\ee
The kernel is diagonal in length because Gaussian
disorder contractions enforce the same cross-chord
number (and thus same length) on the two boundaries. For the heavy strings
considered here, the threshold contribution further
fixes this common length to
\be
\ell_W=\lambda\Delta+2\log(2\lambda).
\label{eq:ell-W}
\ee
Keeping $\ell_W$ finite therefore requires that the Majorana string is heavy\footnote{The logarithmic shift in \eqref{eq:finite-length-heavy-scaling} is necessary if $\ell_W$ is required to be a finite renormalized geodesic length \eqref{eq:ellfromK}. In terms of the scaling of operator size used in Sec.~\ref{sec:GS}, in particular in \eqref{eq:diagonal-fixed-length-limit}, this is
$\Delta=2[\kappa-\log(2\lambda)]/\lambda$
with $\ell_W=2\kappa$.
The unshifted scaling $\Delta=2\kappa/\lambda$ at fixed
$\kappa$ instead sends $\ell_W$ to $-\infty$, giving
a pinching limit with a cusped hyperbolic core,
as in Fig.~\ref{fig:diagonal-fixed-alpha-geometry}.
}
\be
\Delta=
\frac{\ell_W-2\log(2\lambda)}{\lambda},
\qquad \ell_W\ \text{fixed}.
\label{eq:finite-length-heavy-scaling}
\ee

\begin{figure}[h]
    \centering   \includegraphics[width=0.9\linewidth]{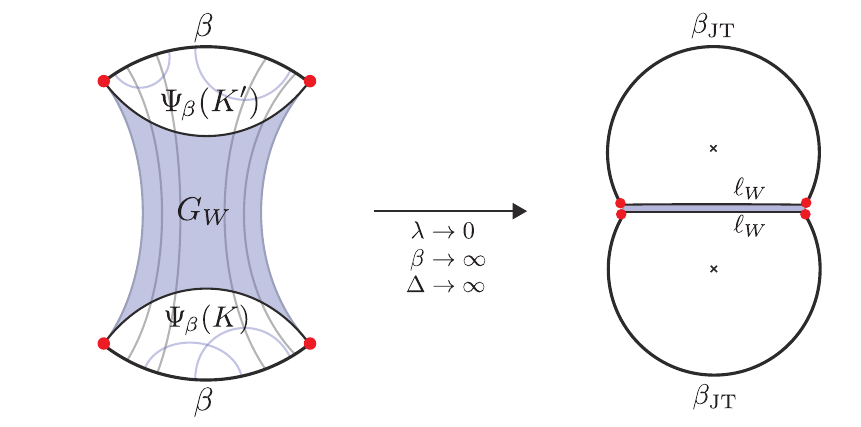}
    \caption{Schwarzian limit of the double-scaled SYK wormhole, the quantum
counterpart of Fig.~\ref{fig:diagonal-fixed-length-geometry}.
Left: the chord wavefunctions $\Psi_\beta(K)$ and $\Psi_\beta(K')$
are glued by $G_W$, which imposes $K=K'=\Delta$.
Right: as $\lambda\to0$, $\beta\to\infty$, and $\Delta\to\infty$
with $\beta_{\rm JT}$ and $\ell_W$ fixed, the kernel projects the
JT Hartle-Hawking wavefunctions onto the same length eigenstate
$|\ell_W\rangle$. Following Sec.~\ref{sec:GS}, $\ell_W$ can be
interpreted as the renormalized length of the wormhole bottleneck.
The shaded intermediate region is drawn with finite thickness
for illustration and should be understood as infinitesimal.}
        \label{fig:limitJT}
\end{figure}

Fig.~\ref{fig:limitJT} gives a geometric representation of the wormhole amplitude, which provides the quantum counterpart of the semiclassical wormhole geometry in Fig.~\ref{fig:diagonal-fixed-length-geometry}.
The two Hartle-Hawking wavefunctions meet at the
prescribed length $\ell_W$, interpreted as the
renormalized length of the wormhole bottleneck, since the latter is infinite in this limit. As described in Sec.~\ref{sec:GS}, the dilaton generally
jumps across this geodesic. The jump vanishes when
$\ell_W$ equals the renormalized length of the geodesic
bisecting the disk at the chosen temperature.

For a quantitative comparison with the JT gravity wormhole, we write the normalized thermal one-point wormhole amplitude in SYK,
\be
\frac{\mathcal{C}_\beta(W)}
{\langle Z(\beta)\rangle_J^2}
\sim
e^{-S_w-I_\Delta}\,
\frac{
\left|\Psi_{\beta_{\rm JT}}(\ell_W)\right|^2
}{
Z_{\rm Schw}(\beta_{\rm JT})^2
}.
\label{eq:SYKwormholenormalized}
\ee
Here $Z(\beta) = {\rm Tr}\,e^{-\beta H}$, and $\langle Z(\beta)\rangle_J\sim A_\lambda e^{-\beta E_0}
Z_{\rm Schw}(\beta_{\rm JT})$ is the disk partition function in the Schwarzian limit. The spectral edge factors cancel with \eqref{eq:SYKwormhole}, while one inverse factor of $A_\lambda$ remains.
We keep this factor explicit in the definition of $I_\Delta$ that we now provide.

The weight in \eqref{eq:SYKwormholenormalized} separates the ``entropic'' cost in $G_W$ of selecting a
fixed Majorana string of size $w$ from the remaining dependence on $\lambda$ and
$\Delta$, which remains finite in the double-scaled theory. For the former we define
\be
S_w:=w\log N-\log(w!)
\simeq w\left[\log\left(\frac Nw\right)+1\right]
-\frac12\log(2\pi w),
\label{eq:S0-def}
\ee
where the second step uses Stirling's formula. The remaining factor can be exactly defined at the level of the leading double-scaled answer
\be
e^{-I_\Delta}:=\frac{\lambda g_W e^{S_w}}{A_\lambda}=\frac{\pi^2\Delta!(1-q)^{\Delta-2}e^{\lambda\Delta/4}}
{(q;q)_\Delta(q;q)_\infty^3}
\qquad\text{at fixed }\lambda,\Delta.
\label{eq:I-Delta-exact-definition}
\ee
The difference between $S_w$ and $\log\binom Nw$ is included in $I_\Delta$. At fixed
$\ell_W$, its Schwarzian limit $\lambda \to 0$ gives
\be
I_\Delta
=
-\frac1\lambda\left[
\lambda\Delta\log(\lambda\Delta)
-\lambda\Delta+\frac{2\pi^2}{3}
\right]
+\frac{\lambda\Delta}{4}
-\frac12\log\left(\frac{\pi\Delta}{8}\right)
+O\!\left(\lambda\log\frac1\lambda\right),
\label{eq:Iw-finite-length}
\ee
where $\lambda\Delta=\ell_W-2\log(2\lambda)$. This term can be interpreted as a non-perturbative contribution in the Schwarzian theory.

The JT gravity matter wormhole has the same form as an amplitude between Hartle-Hawking wavefunctions, but for
a different kernel \cite{Stanford:2020wkf}. This kernel is built by identifying the two
opposite geodesic sides of a hyperbolic square and integrating over their length with a measure inherited from the
particle worldline action $e^{-\Delta\ell}$. This gives
\be
G^{\rm JT}_{\rm wh}(\ell,\ell')
=
\int_0^\infty {\rm d}s \,
\varrho_{\rm Schw}(s)\,
\Phi_s(\ell)\Phi_{s}(\ell')\,
G_\Delta(s,s)\,,
\qquad
G_\Delta(s,s') =
\frac{\Gamma(\Delta\pm \iw s\pm \iw s')}{4^{\Delta}\pi^{2}\,\Gamma(2\Delta)}\,,
\label{eq:GDeltaJT}
\ee
where $\Gamma(\Delta\pm \iw s\pm \iw s') = |\Gamma(\Delta + \iw (s +s'))|^2
|\Gamma(\Delta + \iw (s -s'))|^2$. The matrix elements $G_\Delta(s,s')$ are those of the Schwarzian
bilocal \cite{Mertens:2017mtv}. The wormhole kernel is diagonal in
energy, and therefore the wormhole amplitude is
\be
Z^{\rm JT}_{\rm wormhole}
=\bra{\Psi_{\beta_{\rm JT}}}G^{\rm JT}_{\rm wh}
\ket{\Psi_{\beta_{\rm JT}}}
=\int_0^\infty {\rm d}s\,
\varrho_{\rm Schw}(s)e^{-\beta_{\rm JT}s^2}G_\Delta(s,s).
\label{eq:JT-wormhole-energy}
\ee
Thus a fixed Majorana string selects the same prescribed length on
both sides, whereas the JT particle generally connects different
lengths. The cylinder itself has topological weight one, whereas
each disk carries $e^{S_0}$, so division by the two disk partition
functions gives $e^{-2S_0}$:
\be
\frac{Z^{\rm JT}_{\rm wormhole}}{(Z^{\rm JT}_{\rm disk})^2}
\sim e^{-2S_0}\,
\frac{\bra{\Psi_{\beta_{\rm JT}}}G^{\rm JT}_{\rm wh}
\ket{\Psi_{\beta_{\rm JT}}}}
{Z_{\rm Schw}(\beta_{\rm JT})^2}.
\label{eq:JTwormholenormalized}
\ee

The fixed-string SYK wormhole does not reproduce the
JT wormhole of a finite-mass particle. The SYK kernel
fixes both gluing lengths to $\ell_W$, whereas the
finite-$\Delta$ JT kernel generally connects different
lengths. This difference cannot be removed by changing
the normalization of the operator.

The appropriate comparison for our fixed-length
Schwarzian limit is instead with a heavy JT particle,
since \eqref{eq:finite-length-heavy-scaling} requires
$\Delta\to\infty$. We therefore take the heavy-mass
limit of the JT amplitude at fixed positive
$\beta_{\rm JT}$. In this limit,
\be
\frac{G_\Delta(s,s)}{G_\Delta(0,0)}
=
\left|
\frac{\Gamma(\Delta+2\iw s)}{\Gamma(\Delta)}
\right|^2
\longrightarrow 1
\qquad
(\Delta\to\infty,\ s\ \text{fixed}).
\label{eq:JT-heavy-identity}
\ee
After removing its scalar coefficient, the JT kernel
approaches the identity. Its length kernel is therefore
$\delta(\ell-\ell')$: the two lengths agree, but their
common value is integrated over. Consequently, $Z^{\rm JT}_{\rm wormhole}
\sim
G_\Delta(0,0)\,Z_{\rm Schw}(2\beta_{\rm JT})$. 

Table~\ref{tab:SYK-JT-kernels} compares the normalized
fixed-string SYK amplitude with the heavy-mass limit
of the JT amplitude. Even in this limit, their reduced
kernels differ: the SYK string selects one prescribed
length, while the heavy JT particle integrates over
the common gluing length.

The large-$N$ weights of the amplitudes have different origins.
In SYK, $e^{-S_w}$ measures the cost of selecting a
particular Majorana string of size $w$. In JT,
$e^{-2S_0}$ is the topological suppression of a cylinder
relative to two disks. The SYK selection cost is
therefore controlled by the number of possible strings
of the chosen size rather than by the number of
random couplings.

\begin{table}[h]
\centering
\small
\renewcommand{\arraystretch}{1.5}
\begin{tabular}{@{}lcc@{}}
\hline
 & SYK string & Heavy JT particle \\
\hline
Entropy factor
& $e^{-S_w}$
& $e^{-2S_0}$ \\
Remaining prefactor
& $e^{-I_\Delta}$
& $G_\Delta(0,0)$ \\
Reduced kernel
& $|\ell_W\rangle\langle\ell_W|$
& $\mathbf1$ \\
Thermal factor
& $|\Psi_{\beta_{\rm JT}}(\ell_W)|^2$
& $Z_{\rm Schw}(2\beta_{\rm JT})$ \\
\hline
\end{tabular}
\caption{Comparison of the thermally normalized
wormhole amplitudes. Both expressions have the common
denominator $Z_{\rm Schw}(\beta_{\rm JT})^2$.
The JT entries apply in the strict heavy-mass limit.}
\label{tab:SYK-JT-kernels}
\end{table}

To compare their scaling with $N$, we recall that
the double-scaled large-$N$ limit is taken at fixed
$\lambda$, before the Schwarzian limit
$\lambda\to0$. Combining $w=\Delta p$,
$\lambda=2p^2/N$, and
\eqref{eq:finite-length-heavy-scaling} gives $w/\sqrt{N}=
[\ell_W-2\log(2\lambda)]/\sqrt{2\lambda}$. At each fixed $\lambda$, the right hand side
is independent of $N$, so $w=O(\sqrt N)$.
Consequently,
\be
S_w+I_\Delta
\sim S_w
\sim
\frac{\ell_W-2\log(2\lambda)}{2\sqrt{2\lambda}}\,
\sqrt N\log N
=o(N).
\ee
Here $I_\Delta$ is independent of $N$ at fixed
$\lambda$ and $\ell_W$. Although both $I_\Delta$
and the coefficient multiplying $\sqrt N\log N$
diverge in the subsequent Schwarzian limit, the
suppression is subextensive in the prescribed
order of limits.\footnote{
The comparison with the conventional fixed-$p$
large-$N$ limit requires care. The unshifted scaling
$\lambda\Delta=2\kappa$, with $\kappa$ fixed, gives
$w=\kappa N/p$ and an extensive selection cost whose
coefficient decreases as $\log p/p$ at large $p$.
Holding $\ell_W$ fixed instead introduces a logarithmic
enhancement, giving $w/N\sim\log N/p$, which eventually
violates $w\leq N$ at fixed $p$. The fixed-length
Schwarzian limit is therefore nonuniform and does not admit a direct
fixed-$p$ extrapolation.
}

By contrast, the JT topological exponent $2S_0$
is $O(N)$ when matched to the SYK entropy.
The remaining JT prefactor depends on $\Delta$
but is independent of $N$ at fixed $\lambda$,
so it does not change this comparison.
The two suppressions also have different origins:
the SYK weight measures the cost of selecting
a particular string, while the JT weight comes
from the topology of the cylinder relative
to two disks.

The geometric distinction between the kernels is
independent of the previous scaling. The fixed
SYK string evaluates the Hartle-Hawking wavefunction
at $\ell_W$, while the heavy JT particle integrates
over all lengths. This motivates smearing over
string sizes to allow the SYK gluing length
to vary.

\subsection{Smearing operator size}
\label{sec:smearingsize}

Thus far, we have considered a fixed Majorana string.
We now superpose strings of different sizes. Writing $G_W=g_w\,|\Delta_w\rangle\langle\Delta_w|$ for integer $
\Delta_w=w/p$, we normalize
each string so that its gluing kernel has unit weight:
\be
\mathcal O
=
\sum_{|W|=O(p)} \dfrac{a_W}{\sqrt{g_w}}\,\psi_W.
\label{eq:smearedop}
\ee
The coefficients $a_W$ are independent of the SYK
disorder.\footnote{This operator is more similar to a banded random operator \cite{Stanford:2020wkf},
whose expansion in the Majorana-string basis has random coefficients
$a_W$. Here, however, the contributing strings have $|W|=O(p)$
in the double-scaled limit, whereas the Majorana-string expansion
of a banded random operator is dominated by extensive strings.}

The $O(N)$ invariance of the coupling ensemble ensures
that distinct strings do not interfere in the
disorder-averaged variance. Their gluing kernels therefore
add weights $g_{\mathcal{O}}(w) = \sum_{|W|=w}|a_W|^2$, giving
\be
\mathcal{C}_\beta(\mathcal O)
:=
\left\langle
\left|\Tr\!\left(e^{-\beta H}\mathcal O\right)\right|^2
\right\rangle_J
\sim
\langle\Psi_\beta|G_{\mathcal O}|\Psi_\beta\rangle,
\ee
where 
\be 
G_{\mathcal O} = \sum_K g_{\mathcal{O}}(K)|K\rangle \langle K|\,.
\ee 

In the Schwarzian limit \eqref{eq:schwarzian-limit},
with the retained sizes scaling as in
\eqref{eq:finite-length-heavy-scaling}, smearing over
sizes becomes smearing over lengths:
\be
G_{\mathcal O}
\longrightarrow
\int_{\mathcal I}{\rm d}\ell\,
g_{\mathcal{O}}(\ell)\,|\ell\rangle\langle\ell|.
\ee
Here $\mathcal I$ is the interval of lengths covered by
the smearing, and $g_{\mathcal{O}}(\ell)$ is the total weight assigned
to the corresponding size. For uniform size smearing, we take
$g_{\mathcal{O}}(w)=g_{\mathcal O}$ at every retained
size. Then $g_{\mathcal{O}}(w)=g_{\mathcal O}$, and the length
projectors give
$G_{\mathcal O}=g_{\mathcal O}\mathbf1_{\mathcal I}$.

The smeared kernel remains diagonal in length and thus
generally differs from the finite-mass JT kernel
\eqref{eq:GDeltaJT}. Uniform smearing nevertheless reproduces the JT gluing kernel on $\mathcal I$ in the
strict heavy-mass limit \eqref{eq:JT-heavy-identity},
up to an overall normalization.\footnote{
Recall that this does not mean that $\mathcal O$ satisfies the heavy
operator randomness hypothesis, since its thermal two-point
function combines contributions of different scaling
dimensions.
}

\section{Half-wormholes, $\alpha$-states and the baby universe}
\label{sec:alphacouplings}

In this section, we develop a bulk description of individual $\alpha$-states, labeled by the microscopic SYK couplings. Our central object is the half-wormhole associated with a fixed realization of these couplings. We construct its bulk wavefunction and show that its overlap with a thermal Hartle--Hawking preparation reproduces the thermal one-point function.

We first construct the half-wormhole in the open bulk Hilbert space $\mathcal H_{\rm op}$ introduced in the previous section, and explain how its linked contributions account for factorization at fixed couplings. We then give a closed-universe representation of the same amplitudes in $\mathcal H_{\rm cl}$ and relate its length profile to the classical bottleneck geometry. Finally, we consider the quantization of the microscopic couplings as the baby-universe Hilbert space $\mathcal H_{\rm BU}$ and construct the maps from the two bulk descriptions to this Hilbert space.

\subsection{Open bulk description of the half-wormhole}
\label{sec:bulkohwh}

An individual $\alpha$-sector is specified by a fixed realization of the microscopic SYK couplings $\mathbf J$. Its thermal one-point function is
\be
\begin{aligned}
Z_\beta(W;\mathbf J)
&=\Tr\!\left(e^{-\beta H(\mathbf J)}\psi_W\right)
=
\sum_{n=0}^{\infty}
\frac{(-\beta)^n}{n!}\, \Gamma_n(\mathbf{J})\,,\\
\Gamma_n(\mathbf{J}) &:=\sum_{I_1,\ldots,I_n}
J_{I_1}\dots J_{I_n}\,
\Tr\!\left(\psi_{I_1}\dots\psi_{I_n}\psi_W\right).
\end{aligned}
\label{eq:factorized-one-point}
\ee
Each moment $\Gamma_n(\mathbf{J})$ contains two distinct ingredients.  The first one is
the trace, which determines which Majorana strings can combine with
the operator $ \psi_W$.  The second one is the product of couplings
$J_{I_1}\dots J_{I_n}$, whose value depends on the particular microscopic
realization.

We denote the number of independent couplings by $N_c=\binom{N}{p}$ and their individual disorder variance by $\sigma_J^2=\mathcal J^2/(\lambda N_c)$. For a typical realization of the couplings, there are self-averaging and non-self-averaging terms in \eqref{eq:factorized-one-point}. We shall consider them by introducing {\it half-chords}:
\begin{center}
\includegraphics[width=0.18\linewidth]{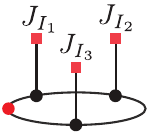}
\end{center}
Each half-chord carries an index associated with the value of the coupling of that particular $p$-string.

The self-averaging parts are the terms in which the coupling indices come in pairs, and correspondingly, the half-chords are paired up. After summing the indices, terms with more than two equal indices are suppressed by factors of $1/N_c$ with respect to the leading answer (recall that in the double-scaled theory this is an exponential suppression in $ \sqrt{N} \log N$). Thus, for an instance of the couplings, with probability approaching one in the large-$N$ limit, the moment \eqref{eq:factorized-one-point} contains terms of the form
\be
\Gamma_n(\mathbf{J})\supset\sum_{I_a \,{\rm distinct}}
J_{I_1}\cdots J_{I_K}
\sum_{L_1,\ldots,L_d}
J_{L_1}^2\cdots J_{L_{d}}^2\,
\Tr\!\left(\dots \right) \;\sim \,\sigma_J^{2d}\;\sum_{I_1,\ldots,I_K} J_{I_1}\cdots J_{I_K}\sum_{L_1,\ldots,L_d}
\Tr\!\left(\dots \right).
\label{eq:paired-and-unpaired}
\ee
with $K\leq n$ and $d = (n-K)/2$. Here $\Tr\!\left(\dots \right)$ contains the trace, the ordering of the operators, and the corresponding crossing signs. The diagrammatic rule is that two half-chords that carry the same index can be replaced by a disk-chord without any coupling-dependence, including a factor of $\sigma_J^2$.
\begin{center}
\includegraphics[width=0.47\linewidth]{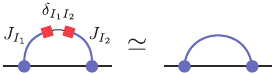}
\end{center}

The natural method is then to disentangle the disk-chords across the unpaired half-chords, getting factors of $q$ for each crossing. As for the number $K$ of unpaired half-chords, we impose the threshold condition from the outset. As in
Sec.~\ref{sec:thermalonept}, we take $w=\Delta p$ with positive integer
$\Delta$. The index sets of such half-chords form a disjoint partition of $W$. Configurations with more
unpaired half-chords require $O(p)$ repetitions of fermion labels and have an exponentially smaller mean-square contribution. Thus, we obtain an expression analogous to \eqref{eq:gammadoublescaled}, summing over chord diagrams  $\mathcal{D}_{\Delta;d}$ on the strip with $d = (n-\Delta)/2$ disk-chords and $\Delta$ unpaired half-chords
\be
\Gamma_n(\mathbf J)
\sim\left(\frac{\mathcal J}{\sqrt\lambda}\right)^{n-\Delta}\,\xi_W(\mathbf{J})
\sum_{D\in \mathcal{D}_{\Delta;d}}
q^{\#\mathrm{int}(D)}.
\label{eq:threshold-half-chord-moment}
\ee
Here $\#\mathrm{int}(D)$ counts the total number of intersections in diagram $D$ (corresponding to disk-disk and disk-half chord intersections).  Unlike in the previous section, the coefficient associated with the
unpaired half-chords depends on the realization of the couplings and
is highly non-self-averaging:
\be
\xi_W(K;\mathbf J)
=
\sum_{\substack{I_1,\ldots,I_K\\ I_a\neq I_b\;\text{for }a\neq b}}
J_{I_1}\cdots J_{I_K}\,
\Tr\!\left(\psi_{I_1}\cdots\psi_{I_K}\psi_W\right).
\ee
We write
$\xi_W(\mathbf J):=\xi_W(\Delta;\mathbf J)$ for the threshold
coefficient.

At threshold, the contributing index sets form a disjoint partition
of $W$, and the trace gives the sign required to put their fermions
in canonical order. Consequently,
\be
\xi_W(\mathbf J)
=\Delta!\,\operatorname{Pf}(\mathbf J|_W).
\ee
Here $\operatorname{Pf}$ denotes the hyperpfaffian of the coupling
tensor restricted to indices in $W$. More precisely, for $W=(i_1<\cdots<i_w)$ with $w=\Delta p$, this is
\be
\operatorname{Pf}(\mathbf J|_W)
:=
\frac{1}{\Delta!\,(p!)^\Delta}
\sum_{\sigma\in {\rm Sym}(w)}\operatorname{sgn}(\sigma)\,
J_{i_{\sigma(1)}\cdots i_{\sigma(p)}}\,
J_{i_{\sigma(p+1)}\cdots i_{\sigma(2p)}}\cdots
J_{i_{\sigma(w-p+1)}\cdots i_{\sigma(w)}}.
\ee
Here the couplings are extended to a completely antisymmetric
tensor, and the denominator removes the overcounting from
permutations within each block and permutations of the blocks.\footnote{A similar hyperpfaffian structure appears in one-time-point
SYK \cite{Saad:2021rcu,Mukhametzhanov:2021nea}, where the partition
function involves the full coupling tensor. For our observable,
the leading high-temperature coefficient admits an analogous
one-time-point Grassmann integral, restricted by the insertion
to the fermions in $W$. Our result, however, is not restricted to
high temperatures: in the double-scaled regime, the same
coupling-dependent hyperpfaffian is dressed by disk-chord
diagrams, retaining the full temperature dependence and applying
also at low temperatures.}

The sum in \eqref{eq:threshold-half-chord-moment} admits
the same bulk description as in Sec.~\ref{sec:thermalonept}.
Resumming the moments expresses the one-point function
in an individual realization as a bulk overlap in $\mathcal{H}_{\rm op}$:
\be
\boxed{
Z_\beta(W;\mathbf J)
\sim \langle\Psi_{\mathbf J,W}|\Psi_\beta\rangle_{\rm op}
}.
\ee
where, setting $\mathcal{J} = 1$,
\be
\ket{\Psi_{\mathbf J,W}}
:=\dfrac{\lambda^{\Delta/2}}{\sqrt{[\Delta]_q!}}\,\xi_W(\mathbf J)
\ket{\Delta}\,.
\label{eq:hwstateds}
\ee
Therefore, the self-averaging thermal dependence is fully encoded in
the Hartle-Hawking wavefunction $\ket{\Psi_\beta}$,
while the microscopic realization of the couplings enters through
the state $\ket{\Psi_{\mathbf J,W}}$.

For a fixed string, the bulk state
$\ket{\Psi_{\mathbf J,W}}$ is supported at a single
chord number and hence at a single length in the
Schwarzian limit:
\be
\ket{\Psi_{\mathbf J,W}}
\propto \xi_W(\mathbf J)\ket{\ell_W}.
\ee
The microscopic realization therefore changes only
the normalization of the state, not the selected length. The random amplitude has zero mean, while its
covariance reproduces the wormhole gluing kernel \eqref{eq:GW-localized}:
\be
\left\langle\Psi_{\mathbf J,W}(\ell)\right\rangle_J=0,
\qquad
\left\langle
\Psi_{\mathbf J,W}(\ell)
\Psi_{\mathbf J,W}(\ell')^*
\right\rangle_J
= g_W\delta(\ell-\ell_W)\delta(\ell'-\ell_W).
\ee

We exclude realizations in which the wavefunction has an anomalously small (exponential in $p$) value relative to its standard deviation, which could allow the subleading sectors to
compete with the threshold contribution. Such
realizations have vanishing probability in the
large-$N$ limit.\footnote{
The amplitude $\Psi_{\mathbf J,W}(\ell)$ is a degree-$\Delta$ polynomial of the
Gaussian couplings. The Carbery-Wright inequality
(Theorem $8$ in \cite{CarberyWright2001}) bounds the
probability of its magnitude being below
$\varepsilon$ times its root-mean-square value by
$C\Delta\,\varepsilon^{1/\Delta}$.
Taking $\varepsilon=e^{-cp}$ with fixed $c>0$, this probability vanishes as $p\to\infty$ at fixed $\Delta$.
}

\begin{figure}[h]
    \centering
    \includegraphics[width=0.9\linewidth]{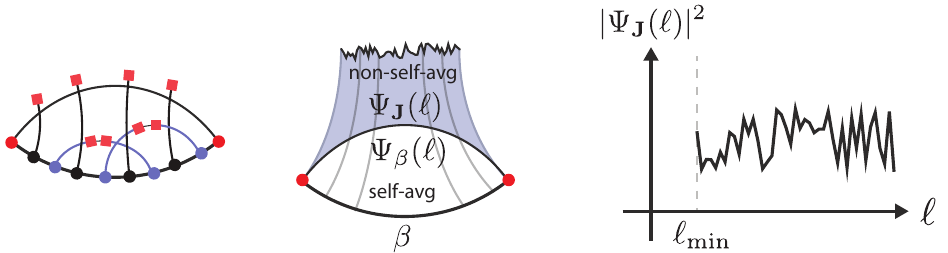}
    \caption{Left: paired half-chords form disk-chords, while
    unpaired half-chords retain the microscopic couplings.
    Middle: the one-point function is the overlap between the
    thermal Hartle-Hawking state and the bulk half-wormhole state.
    Right: for a size-smeared operator, the latter has an erratic,
    non-self-averaging wavefunction in the JT length basis.}
    \label{fig:halfchorddecomposition}
\end{figure}

\paragraph{Smeared operator.}
For the size-smeared operator \eqref{eq:smearedop},
linearity gives
\be
\boxed{Z_\beta(\mathcal O;\mathbf J)
:=\Tr\!\left(e^{-\beta H(\mathbf J)}\mathcal O\right)
\sim \langle\Psi_{\mathbf J}|\Psi_\beta\rangle_{\rm op}}.
\ee
The corresponding bulk state is
\be
\ket{\Psi_{\mathbf J}}
:=
\sum_W \frac{a_W^*}{\sqrt{g_W}}\,
\ket{\Psi_{\mathbf J,W}}\longrightarrow
\int_{\mathcal I}{\rm d}\ell\,
\Psi_{\mathbf J}(\ell)\ket{\ell},
\ee
where the sum runs over the retained strings and
$g_W$ is the weight of the chord kernel
\eqref{eq:Gkerneldiag}. The complex conjugation of
$a_W$ ensures that the overlap is linear in the
coefficients of $\mathcal O$. In the Schwarzian limit, the state becomes a white-noise correlated random state on the length basis within the interval $\mathcal I$
covered by the smearing. For uniform size weights
$\sum_{|W|=w}|a_W|^2=g_{\mathcal O}$ on the retained
sectors, its mean and covariance are
\be
\left\langle\Psi_{\mathbf J}(\ell)\right\rangle_J=0,
\qquad
\left\langle
\Psi_{\mathbf J}(\ell)
\Psi_{\mathbf J}(\ell')^*
\right\rangle_J
=
g_{\mathcal O}\,\delta(\ell-\ell'),
\qquad \ell,\ell'\in\mathcal I.
\ee

Fig.~\ref{fig:halfchorddecomposition} illustrates
the situation. The paired half-chords prepare
the smooth thermal Hartle-Hawking wavefunction,
while the unpaired half-chords determine the
coupling-dependent bulk state associated with the
microscopic $\alpha$-state. Their overlap gives
the one-point function in an individual realization.
A fixed string selects one length with a fluctuating
amplitude; smearing over sizes produces the extended,
erratic length profile.

\subsubsection{Linked half-wormholes and factorization}
\label{sec:linked-half-wormholes}

The half-chord wavefunctions also explain how the averaged wormhole
is compatible with factorization at fixed couplings. This is the
question addressed by the half-wormhole mechanism of
\cite{Saad:2021uzi}. Here we can identify the relevant contributions directly in the
open bulk description.

Consider the size-smeared state $\ket{\Psi_{\mathbf J}}$ above.
At fixed couplings, the product of two one-point functions is
computed by its projector. Subtracting the averaged kernel
defines the linked pair:
\be
\delta G_{\mathcal O}(\mathbf J)
:=\ket{\Psi_{\mathbf J}}\bra{\Psi_{\mathbf J}}-G_{\mathcal O},
\qquad
\left\langle\delta G_{\mathcal O}(\mathbf J)\right\rangle_J=0.
\label{eq:linked-half-wormhole-kernel}
\ee
This subtraction implements the linked-half-wormhole mechanism
at the level of the bulk amplitudes: it removes the averaged
contraction from the product of two individual half-wormholes
\cite{Saad:2021uzi}. For uniform size weights, its length kernel on $\mathcal I$ is
$\Psi_{\mathbf J}(\ell)\Psi_{\mathbf J}(\ell')^*
-g_{\mathcal O}\delta(\ell-\ell')$.

At fixed couplings, the product of the two boundary amplitudes
has the bulk decomposition
\be
\begin{aligned}
Z_\beta(\mathcal O;\mathbf J)^*
Z_{\beta'}(\mathcal O;\mathbf J)\quad\sim \underbrace{\bra{\Psi_\beta}\ket{\Psi_{\mathbf J}}\bra{\Psi_{\mathbf J}}\ket{\Psi_{\beta'}} }_{
\text{half-wormhole inner product at fixed }\mathbf J}
&\\[.2cm]
=\underbrace{\bra{\Psi_\beta}G_{\mathcal O}\ket{\Psi_{\beta'}}}_{
\text{wormhole inner product}}
&+\underbrace{\bra{\Psi_\beta}\delta G_{\mathcal O}(\mathbf J)
\ket{\Psi_{\beta'}}}_{
\text{linked-half-wormhole inner product}}.
\end{aligned}
\label{eq:fixed-J-two-boundary}
\ee
The last term retains the
realization-dependent information, and the sum gives the
factorized product in a single theory. Disorder averaging removes
the second term.

For a Hermitian smeared operator, with real coefficients $a_W$, the underlying half-wormhole amplitudes are real Gaussian. Consequently, at equal temperatures,
\be
\left\langle
\left|\bra{\Psi_\beta}\delta G_{\mathcal O}(\mathbf J)
\ket{\Psi_\beta}\right|^2
\right\rangle_J
\sim 2\,\bra{\Psi_\beta}G_{\mathcal O}\ket{\Psi_{\beta}}^2.
\label{eq:linked-half-wormhole-size}
\ee
The linked contribution, therefore, has the same order of magnitude
as the averaged wormhole.

\subsection{Closed universe description}
\label{sec:trumpet-bottleneck}

\begin{figure}[h]
    \centering
    \includegraphics[width=.88\linewidth]{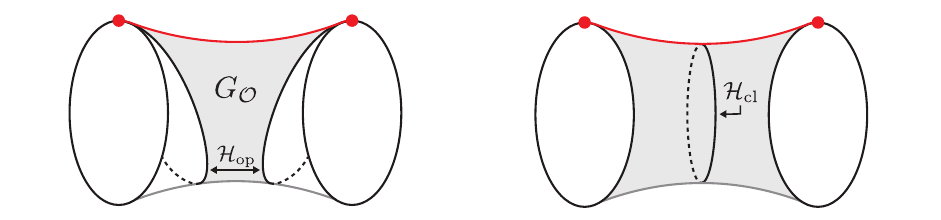}
    \caption{The wormhole is an amplitude $G_{\mathcal O}$ in the open
    bulk Hilbert space. Alternatively, it is the squared norm of a state in
    the bulk closed universe Hilbert space.}
    \label{fig:openvsclosed}
\end{figure}

Throughout this paper, we have described the wormhole and the
half-wormhole using the ``open'' bulk Hilbert space. This is the
Hilbert space naturally associated with the cross-chords, whose
number determines the renormalized length of the open winding
geodesic upon appropriate scaling. As we now show, the same amplitudes also
admit an equivalent quantum description in terms of the Hilbert space of the closed universe, as illustrated in Fig.~\ref{fig:openvsclosed}.

Let $ \ket a_{\rm cl}$ denote the closed geodesic states with $a=1,2,\ldots$. We denote by $\mathcal H_{\rm cl}$ the Hilbert space generated by these states. This basis describes the quantized values of the length of a closed geodesic. The inner product and resolution of the identity are
\be
{}_{\rm cl}\langle a\vert a'\rangle_{\rm cl}
=\frac{\delta_{aa'}}{a},
\qquad
\mathbf 1_{\rm cl}
=\sum_{a=1}^{\infty}
a\,\lvert a\rangle_{\rm cl}\,{}_{\rm cl}\langle a\rvert.
\ee

We denote the thermal and half-wormhole states in this Hilbert space by $\lvert\Psi_\beta^{\rm cl}\rangle$ and $\lvert\Psi_{\mathbf J,W}^{\rm cl}\rangle$. The thermal state is prepared by a trumpet with asymptotic boundary length $\beta$. Its wavefunction gives the trumpet amplitude \cite{Okuyama:2024BabyUniverse}:\footnote{The form of the trumpet wavefunction in double-scaled SYK can be motivated from sine dilaton gravity. When quantized on an open spatial slice, sine dilaton gravity reproduces the chord Hamiltonian \cite{Blommaert:2023wad,Blommaert:2024Hologram}. We adopt its independently
quantized closed sector, including its physical inner product
and trumpet wavefunctions, as a candidate geometric
representation \cite{Blommaert:2025Wormholes}. 
Lengths are measured in the Weyl-rescaled AdS metric of
\cite{Blommaert:2025Wormholes}. Their deformation
parameter is $\sqrt q$, so the length quantum is $\lambda$.
The zero-length mode is null and is omitted.}
\be
\Psi_\beta^{\rm cl}(a)
:={}_{\rm cl}\langle a\vert\Psi_\beta^{\rm cl}\rangle
=I_a(\beta |E_0|),
\ee
Here $I_a$ is the modified Bessel function of the first kind, and $|E_0| = 2/\sqrt{\lambda(1-q)}$ is the upper spectral edge of the chord Hamiltonian \eqref{eq:chordspec}.

The way to relate this description to the open Hilbert space is to express the open Hartle-Hawking state in terms
of these trumpet amplitudes. This will allow us to carry the
half-wormhole overlap into the closed description. As in
\cite{Okuyama:2024BabyUniverse}, the Chebyshev expansion of the
exponential gives, for $K\geq1$,
\be
\Psi_\beta(K)
=\sum_{a\geq1}\mathsf C_{Ka}\Psi_\beta^{\rm cl}(a).
\label{eq:chord-trumpet-transform}
\ee
Here $\mathsf C_{Ka}=2\bra K T_a(-H_{\rm chord}/|E_0|)\ket0$,
with $T_a(\cos\vartheta)=\cos(a\vartheta)$. All temperature
dependence on the right-hand side is carried by the trumpets;
the coefficients specify the combination of closed lengths
needed to reproduce a given open chord amplitude. The inner
product in each Hilbert space is specified separately.
Using the chord Hamiltonian of \eqref{eq:chordH}, the nonzero
coefficients are
\be
\mathsf C_{K,K+2r}
=\frac{(-1)^{K+r}q^{r(r-1)/2}
(1-q^{K+2r})(q;q)_{K+r-1}}
{(q;q)_r\sqrt{(q;q)_K}},
\qquad r\geq0,
\label{eq:explicit-chord-trumpet-coefficients}
\ee
where $(q;q)_n=\prod_{j=1}^n(1-q^j)$.\footnote{
We combine the positive and negative Fourier modes of
\cite{Okuyama:2024BabyUniverse}, with the sign appropriate to
$-H_{\rm chord}/|E_0|$.}
A fixed open chord number consequently contributes to the closed
lengths $a=K,K+2,\ldots$.

\paragraph{Closed representation of the half-wormhole.} For a Majorana string, the open half-wormhole state in
\eqref{eq:hwstateds} is proportional to
$\xi_W(\mathbf J)\ket\Delta$, so its thermal overlap selects
$\Psi_\beta(\Delta)$. We can reproduce this overlap in the closed
Hilbert space by substituting the trumpet expansion above.
The resulting closed half-wormhole wavefunction is
\be
\Psi_{\mathbf J,W}^{\rm cl}(a)
:=\langle a|\Psi_{\mathbf J}^{\rm cl}\rangle_{\rm cl}
=\frac{\lambda^{\Delta/2}\xi_W(\mathbf J)}
{\sqrt{[\Delta]_q!}}\,
\frac{\mathsf C_{\Delta a}^*}{a}.
\label{eq:closed-half-wormhole-wavefunction}
\ee
The factor $1/a$ compensates for the closed gluing measure,
so the microscopic one-point function becomes the closed overlap
\be
\boxed{
Z_\beta(W;\mathbf J)
\sim\langle\Psi_{\mathbf J,W}^{\rm cl}|\Psi_\beta^{\rm cl}\rangle_{\rm cl}
}.
\label{eq:closed-half-wormhole-overlap}
\ee
The dependence
on the microscopic couplings therefore remains entirely in
$\xi_W(\mathbf J)$. What changes is the length profile: the single
chord state is represented by a superposition of closed geodesic
states determined by $\mathsf C_{\Delta a}$.

The expression \eqref{eq:closed-half-wormhole-wavefunction} is also motivated by the microscopic observable.  It follows from the Chebyshev expansion and $\Tr\psi_W=0$, which yield
\be 
Z_\beta(W;\mathbf J)=2\sum_{a\geq1}I_a(\beta |E_0|)\,
\Tr[T_a(-H(\mathbf J)/|E_0|)\psi_W]\,.
\ee
In the closed overlap, each trumpet is multiplied by
$a\langle\Psi_{\mathbf J}^{\rm cl}|a\rangle_{\rm cl}$, where $a$
comes from the gluing measure. Matching these coefficients in
the threshold approximation identifies the half-wormhole amplitude:
\be
\frac{2}{a}\Tr\!\left[
T_a\!\left(-\frac{H(\mathbf J)}{|E_0|}\right)\psi_W\right]
\sim \langle\Psi_{\mathbf J,W}^{\rm cl}|a\rangle_{\rm cl}.
\label{eq:microscopic-BW-closed-formula}
\ee

\paragraph{The wormhole and the closed-universe state.}
At fixed couplings, the microscopic amplitude is an overlap
between a thermal state and a coupling-dependent half-wormhole
state. Averaging the product of two such amplitudes joins the
two thermal preparations through the operator
\be
G^{\rm cl}_{\mathcal O}
:=
\big\langle
|\Psi_{\mathbf J}^{\rm cl}\rangle
\langle\Psi_{\mathbf J}^{\rm cl}|
\big\rangle_J\,.
\ee
For a fixed string,
this is $G_W^{\rm cl}$; for the size-smeared operator, we first
take the corresponding superposition of half-wormhole states
and then average its projector to get $G_{\mathcal O}^{\rm cl}$.

Since $G^{\rm cl}_{\mathcal O}$ is the average of projectors, it is
positive semi-definite. We can therefore split it into two square roots
and absorb one into each thermal preparation. This gives
a {\it closed-universe state} whose squared norm is the
equal-temperature wormhole:
\be
|\Psi_{\rm cl}\rangle
:=\sqrt{G^{\rm cl}_{\mathcal O}}\,|\Psi_\beta^{\rm cl}\rangle\,,
\qquad
\boxed{\mathcal{C}_\beta(\mathcal{O})\sim
\langle\Psi_{\rm cl}|\Psi_{\rm cl} \rangle_{\rm cl}}\,.
\label{eq:closed-state}
\ee
For different temperatures, the wormhole instead computes
the overlap of the two corresponding states. This provides
the closed-universe interpretation anticipated in
\eqref{eq:whbaby} and Fig.~\ref{fig:openvsclosed}.

\paragraph{Continuous closed universe wavefunction.} We now study the $\lambda \to 0$ limit in which the spacing of $a$ becomes negligible. The closed universe wavefunction $\Psi_{\rm cl}(a)$ localizes in a bottleneck geodesic length exactly corresponding to the classical solution in Sec. \ref{sec:solwinding}.

For this comparison, we first identify the physical length
corresponding to $a$ by taking the Schwarzian limit of the thermal
trumpet. Keep $b=\lambda a$ and
$\beta_{\rm JT}=2\lambda\beta$ fixed as $\lambda\to0$.
The limiting wavefunction is the JT trumpet,
\be
\dfrac{e^{-\beta |E_0|}}{\lambda}\Psi_\beta^{\rm cl}(a)
\;\to \;Z_{\rm Schw}^{\rm trumpet}(b,\beta_{\rm JT}) = \dfrac{e^{-b^2/(2\beta_{\rm JT})}}{\sqrt{2\pi\beta_{\rm JT}}},
\ee
which identifies $b$ as the closed geodesic circumference.
After removing $e^{\beta |E_0|}$ from each thermal preparation,
the factors of $\lambda$ in the two wavefunctions convert the sum over $a$ into an integral with measure $b\,{\rm d}b$.

To locate the peak of the wavefunction, take $\lambda\Delta=2\kappa>0$ fixed. Write the allowed labels as
$a=\Delta+2r$ and set $x=\lambda r$, so the closed circumference
is $b=\lambda a=2\kappa+2x$. From
\eqref{eq:explicit-chord-trumpet-coefficients}, the ratio of
consecutive coefficient magnitudes approaches
$e^{-x}(1-e^{-2\kappa-x})/(1-e^{-x})$.
The coefficients grow when this ratio exceeds one and decrease
when it falls below one, so its crossing of one determines the
peak. This gives $x_*=\log(1+\sqrt{1-e^{-2\kappa}})$ and hence
\be
b_*=2\kappa+2x_*=2\operatorname{arcosh}(e^\kappa).
\label{eq:chebyshev-bottleneck-saddle}
\ee
The factor $1/a$ in the probability varies slowly across the peak,
so it does not change this leading result. A quadratic expansion
of $\log|\mathsf C_{\Delta a}|$ around the maximum gives
$|\mathsf C_{\Delta a}|\propto
\exp[-(b-b_*)^2/(4\lambda)]$ for
$b-b_*=O(\sqrt\lambda)$. Thus, the length probability concentrates
at the classical bottleneck circumference \eqref{eq:dcylinder-reconstruction}, with width
$O(\sqrt\lambda)$.

Keeping the open renormalized length
$\ell_W=2\kappa+2\log(2\lambda)$ fixed instead forces
the bottleneck to grow as $\lambda\to0$. The peak then moves to a large
circumference, where $b_*+2\log\lambda \to\ell_W$. The closed geodesic therefore becomes infinitely long, while its
renormalized length approaches the open geodesic length and the
width of its probability distribution still vanishes as
$\sqrt\lambda$. This limit reproduces the behavior found in Sec. \ref{sec:GS} and illustrated in Fig. \ref{fig:diagonal-fixed-length-geometry}.

\subsection{Baby-universe Hilbert space} 
\label{sec:BU}

Although the bulk JT Hilbert spaces are ``UV-complete,'' the interpretation is that these bulk Hilbert spaces are analogous to the perturbative bulk Hilbert spaces of higher-dimensional gravity, whose non-perturbative microscopic completion we would like to understand (cf. \cite{Iliesiu:2024cnh}). In our setting, where no non-trivial holographic Hilbert space is available, the baby-universe Hilbert space provides a candidate for this completion \cite{Marolf:2020xie}. In this subsection, we construct maps from the bulk descriptions above to the Hilbert space of wavefunctions of the couplings. These maps reproduce the wormhole amplitudes as baby-universe inner products, giving an isometric encoding of the closed-universe states constructed above.

\paragraph{Microscopic $\alpha$-states.}
Formally, the baby-universe Hilbert space is obtained by the GNS construction: functions of the
microscopic couplings are treated as states, with inner products
determined by the disorder average. 

There are $N_{c} = \binom{N}{p}$ independent couplings in the Hamiltonian.  We consider their configuration space itself as the
baby universe configuration space and use the flat Hilbert space inner product
\be
\mathcal H_{\rm BU}
=
L^2\!\left(\mathbb R^{N_c}\right),
\qquad
\langle\mathbf J|\mathbf J'\rangle
=
\delta(\mathbf J-\mathbf J').
\label{eq:flat-alpha-space}
\ee
Thus $\ket{\mathbf J}$ is the simultaneous eigenstate of all
the coupling operators, with
$\widehat {\mathbf J}\ket{\mathbf J}={\mathbf J}\ket{\mathbf J}$ and ${\mathbf J} = (J_{I_1},...,J_{I_{N_c}})$.  Every coupling-dependent function $F = F({\mathbf J})$ is promoted to an operator $\widehat{F} = F(\widehat{{\mathbf J}})$ in this Hilbert space. The coupling eigenstates are the microscopic $\alpha$-states.

Because the measure in \eqref{eq:flat-alpha-space} is flat, the Gaussian
disorder distribution must be carried by the state rather than by the
inner product.  We define the ``no-boundary'' Hartle-Hawking state $|\rm HH\rangle$ to be the normalized
ground state of $N_c$ independent harmonic oscillators in coupling space.\footnote{One may also embed $H_{\rm BU}$ in the larger Hilbert space of a free boson, retaining $N_c$ modes (one for each microscopic coupling) and leaving the remaining modes in the vacuum. The Hartle–Hawking state would then map to the Fock vacuum.}
Its wavefunction is
\be
\Psi_{\rm HH}(\mathbf J) = \langle\mathbf J|{\rm HH}\rangle
=
\frac{1}{(2\pi\sigma_J^2)^{N_c/4}}
\exp\!\left(-\frac{\mathbf J^2}{4\sigma_J^2}\right),\qquad
\sigma_J^2
=\frac{\mathcal J^2}{\lambda N_c}.
\label{eq:HH-oscillator-ground-state}
\ee
The square of the Hartle-Hawking wavefunction is
precisely the disorder probability density,
\be
\bra{{\rm HH}}\widehat F\ket{{\rm HH}}
=
\int {\rm d}^{N_c}\mathbf J\,
\big|\Psi_{\rm HH}(\mathbf J)\big|^2
F(\mathbf J)
=
\langle F\rangle_J .
\label{eq:HH-disorder-average}
\ee   

The thermal one-point function is promoted to a diagonal operator in this Hilbert space:
\be
 Z_\beta(W;\mathbf J)
=
\Tr\!\left(e^{-\beta H(\mathbf J)} \psi_W\right),
\qquad
\widehat {Z}_\beta(W)\ket{\mathbf J}
:=
Z_\beta(W;\mathbf J)\ket{\mathbf J}.
\label{eq:fixed-J-one-point-function}
\ee

Acting with one asymptotic boundary on the Hartle-Hawking state produces
the unnormalized baby universe state
\be
\ket{\Psi_{\rm BU}}
:=
\widehat Z_\beta(W)\ket{{\rm HH}}.
\label{eq:boundary-alpha-state}
\ee
The vanishing disorder average implies that this state is orthogonal to the Hartle-Hawking state, $\bra{\rm HH}\ket{\Psi_{\rm BU}} = 0$; in this representation, the vanishing follows from the $O(N)$ invariance of the Hartle-Hawking wavefunction.

The squared norm computes the coupling variance of the thermal one-point function and hence
reproduces the wormhole amplitude in the double-scaled limit
\be
\bra{\Psi_{\rm BU}}\ket{\Psi_{\rm BU}}
= \mathcal{C}_\beta(W) \sim \langle\Psi_{\rm cl}|\Psi_{\rm cl} \rangle_{\rm cl} .
\label{eq:alpha-overlap-wormhole}
\ee
Recall that the left-hand side of this equation is the norm of a state in the formal baby-universe Hilbert space, while the right-hand side is the norm in the closed geodesic Hilbert space of the bulk. For a size-smeared operator $\mathcal O$, the corresponding state is defined by $\ket{\Psi_{\rm BU}(\beta)}:=\widehat Z_\beta(\mathcal O)\ket{\rm HH}$. We now make its relation to both bulk descriptions explicit.

\paragraph{Bulk-to-BU maps.}

For a fixed operator $\mathcal O$, let $V_{\rm op}$ and $V_{\rm cl}$
map the open and closed bulk Hilbert spaces into $\mathcal H_{\rm BU}$,
\be
V_{\rm op}:\mathcal H_{\rm op}\longrightarrow\mathcal H_{\rm BU},
\qquad
V_{\rm cl}:\mathcal H_{\rm cl}\longrightarrow\mathcal H_{\rm BU}.
\ee

The open map is most easily defined through its coupling-space wavefunction:
\be
V_{\rm op}
=\int \text{d}\mathbf{J}\,\Psi_{\rm HH}(\mathbf J)
\ket{\mathbf{J}}\,{}_{\rm op}\langle\Psi_{\mathbf J,\mathcal O}|,
\label{eq:chord-to-coupling-map}
\ee
It takes the bulk overlap with the open half-wormhole wavefunction at each coupling realization
and multiplies it by the Hartle-Hawking wavefunction in coupling space.

The trumpet expansion defines a map from the closed to the open bulk Hilbert space,
\be
R:\mathcal H_{\rm cl}\longrightarrow\mathcal H_{\rm op},
\qquad
\langle K|R|a\rangle=\frac{\mathsf C_{Ka}}{a},
\qquad K,a\geq1.
\ee
The factor $1/a$ compensates for the weight $a$ in the closed completeness relation. By construction, $R$ maps the closed thermal state to the $K\geq1$ part of the open Hartle-Hawking state $R\ket{\Psi_\beta^{\rm cl}} = \sum_{K\geq1}\Psi_\beta(K)\ket K$. We set $\langle0|R=0$, so the image has no open-vacuum component. 

The closed-to-BU map first uses the trumpet expansion to pass to the open description, then applies the open-to-BU map:
\be
V_{\rm cl}=V_{\rm op} \circ R.
\ee
Thus, both maps relate the thermal Hartle-Hawking state to the same baby-universe state:
\be
\ket{\Psi_{\rm BU}(\beta)}
\sim
V_{\rm op}\ket{\Psi_\beta}
=
V_{\rm cl}\ket{\Psi_\beta^{\rm cl}}.
\label{eq:boundary-state-map}
\ee

Via this map, the baby-universe inner product induces precisely the bulk
wormhole inner products:
\be 
V_{\rm op}^\dagger V_{\rm op}=G_{\mathcal O}\,,\qquad (V_{\rm op}R)^\dagger(V_{\rm op}R)
=R^\dagger G_{\mathcal O}R=G_{\mathcal O}^{\rm cl}\,.
\ee
Making the temperature
label of $\ket{\Psi_{\rm BU}}$ explicit, we obtain the inner product relation
\be
\langle\Psi_{\rm BU}(\beta)|\Psi_{\rm BU}(\beta')\rangle_{\rm BU}
\sim\langle\Psi_\beta|G_{\mathcal O}|\Psi_{\beta'}\rangle_{\rm op} =\langle\Psi_{\rm cl}(\beta)|\Psi_{\rm cl}(\beta')\rangle_{\rm cl}.
\label{eq:wormhole-Gram}
\ee
The closed-universe states prepared by the wormhole thus have a
microscopic realization as wavefunctions of the SYK couplings.
Within the bulk approximation, their overlaps in the baby-universe
Hilbert space reproduce the closed universe inner products defined by the wormhole.

\section{Discussion}
\label{sec:discussion}

In this paper we have studied the thermal one-point function of a heavy Majorana string in SYK and derived a wormhole contribution to its variance over the couplings. In the double-scaling limit, this contribution becomes a bulk amplitude in the chord Hilbert space with an operator-dependent gluing kernel, making explicit how correlations between the boundaries give rise to the wormhole inner product. In the Schwarzian limit, we have compared the SYK wormhole with its counterpart in JT gravity coupled to a worldline matter particle. We have also described what happens at fixed couplings, where the half-wormhole is encoded in an erratic bulk wavefunction that does not self-average. The SYK couplings thus give a concrete realization of the $\alpha$-parameters for this wormhole. We now discuss some implications of these results and a few of the questions they leave open.

\subsection*{Heavy operator randomness \& the SYK wormhole}

One motivation for this work was to test the heavy-operator randomness hypothesis in a microscopic model. SYK is a useful setting for this question. Its couplings are random, but the Hamiltonian still has a definite many-body structure, and a heavy Majorana string is a specific operator built from the microscopic fermions. However, we will now explain that the SYK wormhole comes from a different source of randomness.

To explain this, we can organize the matrix elements of the heavy operator $\psi_W$ in the energy basis using an ETH-type ansatz,\footnote{The usual ETH discussion concerns few-body observables, or at least observables whose size is kept fixed in the thermodynamic limit. Our heavy strings contain a growing number of fermions, so applying the same model to them is an additional hypothesis.}
\be\label{eq:ETHansatz}
\langle E_i|\psi_W|E_j\rangle
= \psi_{W}(E_i;\mathbf J)\delta_{ij}
+e^{-S(\bar E)/2}\sqrt{f_W(E_i,E_j)}\,R_{ij},
\qquad
\bar E=\frac{E_i+E_j}{2}.
\ee
Here $f_W$ is a smooth variance envelope, with the suppression due to the thermodynamic entropy $S(E)$ factored out. The simplest version of the ansatz takes the $R_{ij}$ to be Gaussian random variables with zero mean and unit variance, with independent entries subject to Hermiticity. The term $\psi_{W}(E_i;\mathbf J)$ is the part of the diagonal that varies smoothly with energy at fixed couplings. For a low-weight or ``simple'' observable, it is set by matching to thermal expectation values. In SYK, however, this smooth energy profile can itself depend erratically on the microscopic couplings~\cite{Swingle:2026ecy}. For our strings, its disorder average vanishes, even though it is present in a typical realization. A zero averaged one-point function therefore does not mean that the whole diagonal consists of fluctuations that are erratic in energy.

For fixed couplings, the normalized thermal two-point function has two parts: an off-diagonal part, which connects different energy levels, and a diagonal part, which keeps the energy unchanged. Schematically,
\begin{equation}
\begin{aligned}
G_\beta(\tau)
&=\langle\psi_W(\tau)\psi_W(0)\rangle_\beta
=G_\beta^{\rm off}(\tau)+G_\beta^{\rm diag},\\
G_\beta^{\rm diag}
&\simeq\big\langle|\psi_W(E;\mathbf J)|^2\big\rangle_\beta.
\end{aligned}
\label{eq:smooth-diagonal-disk}
\end{equation}
The brackets denote thermal averaging, including division by the partition function. In the diagonal term, we keep the profile that is smooth in energy but erratic in the couplings, and omit the smaller diagonal ETH fluctuations. The two parts are represented by the diagrams below.

\begin{center}
\begin{tikzpicture}[line width=.65pt, every node/.style={font=\small}]
  \draw (0,0) circle (1);
  \draw[line width=1.2pt] (-1,0) -- (1,0);
  \fill (-1,0) circle (1.5pt);
  \fill (1,0) circle (1.5pt);
  \node[left] at (-1,0) {$\psi_W$};
  \node[right] at (1,0) {$\psi_W$};
  \node[anchor=west] at (1.9,0) {$\sim O(N^0)$};
  \node at (0,1.5) {$G_\beta^{\rm off}$ (self-avg)};
  \begin{scope}[xshift=6.5cm]
    \draw (0,0) circle (1);
    \draw[line width=1.2pt] (-1,0) -- (-.62,0);
    \draw[line width=1.2pt] (.62,0) -- (1,0);
    \draw[dashed] (-.54,0) -- (.54,0);
    \draw[fill=white] (-.62,0) circle (.04);
    \draw[fill=white] (.62,0) circle (.04);
    \fill (-1,0) circle (1.5pt);
    \fill (1,0) circle (1.5pt);
    \node[left] at (-1,0) {$\psi_W$};
    \node[right] at (1,0) {$\psi_W$};
    \node[above] at (0,.04) {$\mathbf 1_E$};
    \node[anchor=west] at (1.9,0) {$\sim O(e^{-S_w})$};
    \node at (0,1.5) {$G_\beta^{\rm diag}(\mathbf{J})$ (non-self-avg)};
  \end{scope}
\end{tikzpicture}
\end{center}
The first diagram represents the off-diagonal ETH contribution. The sum over many matrix elements makes this contribution self-averaging. If $f_W$ is chosen as the variance envelope extracted from the Schwarzian disk two-point function of a primary, replacing $|R_{ij}|^2$ by its unit mean reproduces that correlator and its JT gravity description~\cite{Stanford:2020wkf}.\footnote{For light primaries, correlations between different matrix elements in the erratic part of \eqref{eq:ETHansatz} are
important to correctly capture higher-point functions. They can be described by a generalized ETH ansatz \cite{Jafferis:2022uhu,Jafferis:2022wez}. Their consistency with wormhole corrections to other observables will be demonstrated in \cite{gabriele}.} At fixed double-scaling parameters and fixed thermal kinematics, this propagating contribution is of order one in $N$. Under the Lorentzian continuation $\tau={\rm i}t$, it carries phases $e^{{\rm i} t(E_i-E_j)}$. These phases lead to dephasing, so the smooth off-diagonal contribution decays with time.

The second diagram represents the contribution from the smooth diagonal part in \eqref{eq:ETHansatz}. The trajectory of the particle is broken by two vertices, indicating that the heavy operator is annihilated by interactions with the bulk fermions. The dashed line represents that the identity propagates. Each vertex supplies a coefficient that is erratic in the couplings and the resulting contribution has a non-zero average over couplings, but is non-self-averaging for individual realizations. This diagram is suppressed by the entropic cost to annihilate the heavy operator $e^{-S_w}$, coming from $\psi_{W}(E_i;\mathbf J) \sim e^{-S_w/2}$. Since $E_i=E_j$, the contribution is independent of the separation $\tau$.  Under Lorentzian continuation, the phases cancel: this term remains time-independent and gives a plateau contribution. Once the smooth off-diagonal contribution has decayed, the small diagonal term dominates the correlator. Because the diagonal term does not self-average, its plateau height remains erratic in the couplings.

For the squared thermal one-point function, the hierarchy is reversed. Each trace sees only diagonal matrix elements, so the competition is between the erratic ETH noise $R_{ii}$ and the diagonal profile that is smooth in energy but erratic in the couplings. After dividing the two-boundary amplitude by the two disk partition functions, the usual ETH/JT particle wormhole is suppressed by $e^{-2S_0}$~\cite{Stanford:2020wkf}, while the SYK wormhole carries $e^{-S_w}$, up to thermal profiles and the remaining prefactors. Since for the class of operators we considered $S_w$ is subextensive (or extensive with a vanishing coefficient) and $S_0$ is extensive, the SYK contribution dominates in the double-scaled limit. The same contribution can therefore be subleading on the disk and leading in the wormhole.

\begin{center}
\begin{tikzpicture}[
    line width=.65pt,
    every node/.style={font=\small}
]
  \draw (-1,0) ellipse (.24 and .9);
  \draw (1,0) ellipse (.24 and .9);
  \draw (-1,.9)
    .. controls (-.3,.43) and (.3,.43) .. (1,.9);
  \draw (-1,-.9)
    .. controls (-.3,-.43) and (.3,-.43) .. (1,-.9);
  \draw[line width=1.2pt] (-1.24,0) -- (1.24,0);
  \fill (-1.24,0) circle (1.5pt);
  \fill (1.24,0) circle (1.5pt);
  \node[left] at (-1.24,0) {$\psi_W$};
  \node[right] at (1.24,0) {$\psi_W$};
  \node at (0,1.3) {ETH / JT wormhole};
  \node[anchor=west] at (1.9,0) {$\sim O(e^{-2S_0})$};

  \begin{scope}[xshift=6.5cm]
    \draw (-1,0) ellipse (.24 and .9);
    \draw (1,0) ellipse (.24 and .9);
    \draw (-1,.9)
      .. controls (-.3,.43) and (.3,.43) .. (1,.9);
    \draw (-1,-.9)
      .. controls (-.3,-.43) and (.3,-.43) .. (1,-.9);
    \draw[line width=1.2pt] (-1.24,0) -- (-.62,0);
    \draw[line width=1.2pt] (.62,0) -- (1.24,0);
    \draw[dashed] (-.54,0) -- (.54,0);
    \draw[fill=white] (-.62,0) circle (.04);
    \draw[fill=white] (.62,0) circle (.04);
    \fill (-1.24,0) circle (1.5pt);
    \fill (1.24,0) circle (1.5pt);
    \node[left] at (-1.24,0) {$\psi_W$};
    \node[right] at (1.24,0) {$\psi_W$};
    \node at (0,1.3) {SYK wormhole};
    \node[anchor=west] at (1.9,0) {$\sim O(e^{-S_w})$};
  \end{scope}
\end{tikzpicture}
\end{center}
The connected JT amplitude comes from averaging over the erratic matrix elements $R_{ii}$. In the bulk, this corresponds to gluing the disk amplitude along the particle worldline to form a cylinder, with the worldline connecting the two boundaries. In SYK, averaging the shared coupling-dependent coefficient glues the two half-wormholes at the length selected by the operator. The JT kernel is diagonal in energy, so its covariance for the thermal traces depends only on $\beta+\beta'$. The SYK kernel correlates different energies and depends on $\beta$ and $\beta'$ separately.

It would be interesting to extend these considerations to higher-dimensional holographic theories, where microscopic calculations are likely to be much harder. Defect operators built from many local primaries \cite{Anous:2016kss} provide natural analogs of our Majorana strings. Deriving their wormhole amplitudes directly in the CFT would test proposed random-matrix descriptions \cite{Sasieta:2022ksu} and, in two dimensions, identity dominance in their OPE at large central charge, or the Liouville line-defect description of the wormholes \cite{Chandra:2024vhm}. The closed-universe construction of \cite{Antonini:2023hdh} provides a particularly compelling motivation for this study.

\subsection*{Chords as observer worldlines}

\begin{figure}[h]
\centering
\includegraphics[width=.38\linewidth]{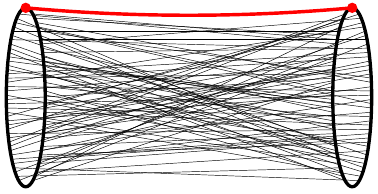}
\caption{Chords as observer worldlines, with their number determining the emergent geometry.}
\label{fig:chordsobs}
\end{figure}

The emergence of a wormhole in SYK suggests that the degrees of freedom associated with its microscopic couplings may be interpreted as a delocalized “observer” or, more broadly, as an extended quantum reference system. This is close in spirit to the observer-based descriptions of the wormhole proposed in gravity \cite{Abdalla:2025gzn,Harlow:2025pvj}. The cross-chord diagrams provide a suggestive picture: they pair microscopic coupling labels across the two halves of the wormhole, much like observer diagrams pair the observer insertions (see Fig. \ref{fig:chordsobs}). The correlations responsible for these diagrams arise from the ensemble of couplings or, equivalently, from correlations on a specific quantum state in the baby universe Hilbert space of couplings.

An interesting feature of this interpretation is that the reference system would very clearly participate in the emergence of space. Chord number provides a quantized measure of bulk geodesic length, so the putative observer would be distributed over the microscopic structure from which this very simple, gauge-invariant notion of a geometry emerges (cf. \cite{Berkooz:2018jqr,Lin:2022rbf}).\footnote{An analogy is provided by matrix integrals, where connected ribbon graphs discretize surfaces that admit a continuum description in a suitable double-scaling limit.} The microscopic description of this reference system in $\mathcal{H}_{\rm BU}$ retains information about individual couplings that chord number alone does not capture.

There is still a step between this picture and an observer that can make
measurements. To see how this system could serve as an observer, one could treat selected coupling data as records and study light-probe amplitudes conditioned on them. This would let us ask which bulk distinctions are accessible relative to those records. It would be interesting to interpret the rules of ~\cite{Abdalla:2025gzn,Harlow:2025pvj} for higher moments of these measurements from this perspective.

\subsection*{Bulk wavefunctions for half-wormholes}

At fixed couplings, the thermal one-point function is the overlap of the thermal Hartle-Hawking state with a coupling-dependent bulk state associated with the half-wormhole. For a fixed Majorana string, this state is supported at a single length and carries a random amplitude. Smearing over string sizes produces an extended wavefunction in length, with white-noise correlations between different lengths.

For an individual realization, the bulk half-wormhole wavefunction fluctuates erratically over lengths and will not admit a smooth semiclassical interpretation. It would be interesting to understand what these fluctuations mean physically. Measurements confined to a sufficiently small bulk subsystem might nevertheless agree with the semiclassical, self-averaged description (cf \cite{Nomura:2026igt}). If so, individual realizations could still encode local physics in a closed universe despite the irregularity of their global wavefunctions. 

A similar dependence on microscopic data was found in the bulk description of spectral wormholes for fixed realizations of the JT matrix integral~\cite{Blommaert:2021fob}. There, the $\alpha$-parameters are the individual eigenvalues of the Hamiltonian, so the wavefunction depends on a much larger set of microscopic parameters than in our SYK construction. The averaged wormhole, however, is directly the pure JT gravity double trumpet by construction.

Extending this microscopic description of non-averaged observables to higher-dimensional AdS/CFT remains an interesting open problem. For instance, determining the $\alpha$-parameters amounts to specifying the ensemble associated with a wormhole. In higher dimensions, this is challenging because we often lack sufficient control over multi-boundary wormholes. Once the ensemble is specified, describing its individual realizations in the bulk would require a precise dictionary relating bulk observables to microscopic CFT data.

\subsection*{Higher dimensions and non-uniqueness of $\alpha$-parameters}

In SYK, the microscopic ensemble is part of the starting point, so we know what the $\alpha$-parameters of the wormhole are. On the gravity side, we often have to work backward. We have a two-boundary wormhole, and sometimes some control over higher-boundary wormholes, but this data need not identify a unique microscopic ensemble. A covariance can be reproduced in many ways, and a few higher-boundary wormholes associated with higher moments need not fix the ensemble. Even control over all wormholes would not necessarily determine the full microscopic ensemble, since the large-$N$ expansion captured by the gravitational path integral misses information.

It is important to remark that the $\alpha$-parameters need not be simple couplings in a Lagrangian. In familiar AdS/CFT examples, such as $\mathcal N=4$ super-Yang-Mills with its gauge group and coupling fixed, we expect a single microscopic CFT. One possibility is to average over a suitable family of heavy operators within that theory~\cite{Sasieta:2022ksu,Antonini:2023hdh,Antonini:2025ioh}, with the $\alpha$-parameters labelling their microscopic data. Alternatively, one could obtain the same wormholes by averaging over the eigenbasis of the CFT Hamiltonian while keeping the operator fixed. Such an ensemble does not correspond to averaging over the couplings of simple Lagrangian terms, but it is nevertheless what gravity captures. Different choices of $\alpha$-parameters could reproduce the same two-boundary wormhole and higher moments while encoding different microscopic information. In the operator-average interpretation, this information resides in objects of a single CFT. Likewise, the information associated with the closed universe could be encoded in many different baby universe Hilbert spaces. It would be interesting to understand if this freedom has any physical significance.

\subsection*{Other directions}

A natural next step is to study thermal four-point functions of heavy Majorana strings in SYK, with the goal of identifying a contribution of a genus-$1$ saddle for the single boundary observable (cf. \cite{Stanford:2021bhl}). This would test whether our construction extends beyond the cylinder and could give a microscopic picture of ``closed-universe emission'' in a spacetime with a single asymptotic region.

Other SYK wormholes \cite{Garcia-Garcia:2020ttf,Sasieta:2025vck,Sontag:2026iiu} offer further settings in which to apply these methods. In particular, it would be interesting to derive the half-wormholes of \cite{Garcia-Garcia:2021squ} using the same approach.

\paragraph{Acknowledgements.}  We thank Stefano Antonini, Jeevan Chandra, Gabriele Di Ubaldo, Luca Iliesiu,  Kristan Jensen, Ohad Mamroud, Vladimir Narovlansky, Moshe Rozali, and Josef Seitz for useful conversations. We thank the ``Pathways to Quantum Black Holes: from Effective Theories to Exact Methods'' workshop at the Galileo Galilei Institute (INFN, Florence), and the workshop ``Observers, wormholes, and complex saddles in cosmology" organized at the Bernoulli Center for Fundamental Studies (EPFL, Lausanne), where part of this work was completed. We benefited from the assistance of OpenAI’s GPT models and Anthropic’s Claude. We acknowledge support from the U.S. Department of Energy through GeoFlow DE-SC0019380 (MS and BS) and from the Heising-Simons Foundation through grant 2024-4849 (BS). The work of VB is supported by the Quantum Horizons Alberta initiative, and the University of Calgary through the VPR Postdoctoral Excellence Award. The work of MS is supported in part by the Leinweber Institute for Theoretical Physics at UC Berkeley. The work of AVL is supported by the Natural Sciences and Engineering Research Council of Canada (NSERC) and the Simons Foundation via a Simons Investigator Award.

\appendix

\section{Details on the large-$p$ saddle}
\label{app:SD-onepoint}

In this appendix, we provide details on the replica-connected solution of Sec. \ref{sec:GS}.

\subsection{Large-$p$ equations}
\label{app:large-p-equations}

We start by deriving the coupled large-$p$ equations for the replica-symmetric saddle. Throughout this
subsection, $h$ denotes the correction to the mixed $LR$ bilocal defined
in the main text:
\be
G^{LR}(x,y)=1+\frac{h(x,y)}p+O(p^{-2}),
\qquad
\Sigma^{LR}(x,y)=\frac{\mathcal J^2}{2p}e^{h(x,y)}+O(p^{-2}).
\label{eq:app-LR-convention}
\ee

Consider the replica-diagonal correlators in each sector and retain the leading
nontrivial terms at large $p$. Factoring out the free AP propagator, we have
\begin{gather}
G_{\rm AP}^{LL}(x,y)
=
\sign(x-y)
\left[1+\frac{g(x,y)}p+O(p^{-2})\right],
\\
G_{\rm P}^{LL}(x,y)
=
P(x,y)+O(p^{-1}).
\end{gather}
with $g(x,y)=g(y,x)$ and $P(x,y) = -P(y,x)$. Note that in our normalization $g(0,0) = 0$. Although $\Sigma^{LR}(x,y)=O(p^{-1})$, its product with the enhanced periodic
zero-mode propagator is order one and therefore modifies $P(x,y)$. The
corresponding source is
\be\label{eq:app-definitionrx}
r(x):=\dfrac{1}{\beta m}\int_0^\beta {\rm d}y\,\Sigma^{LR}(x,y)
=\frac{\mathcal J^2}{2\kappa}\int_0^\beta {\rm d}y\,e^{h(x,y)}.
\ee 
The zero-mode normalization \eqref{eq:app-zero-mode-normalization} implies
$\int_0^\beta {\rm d}x\,r(x)=1$. Defining its primitive
\be 
R(x):=\int_0^x{\rm d}t\,r(t),
\ee
with $R(0)=0$ and $R(\beta)=1$, the
periodic Dyson equation \eqref{eq:app-resolvents} and its solution are
\begin{align}
\frac12\partial_xP(x,y)
&=
\delta_{\rm P}(x-y)-r(x) \nonumber \\[.2cm] & \Rightarrow \quad 
P(x,y)
=
\sign(x-y)
\left[
1-2\left|R(x)-R(y)\right|
\right].
\label{eq:app-P-solution}
\end{align}

For $0<y<x<\beta$, the total diagonal bilocal is
\begin{gather}
G^{LL}(x,y)=1+\dfrac{\Phi(x,y)}{p}+O(p^{-2})\\[.2cm]
\Phi(x,y):=g(x,y)-2\kappa[R(x)-R(y)]. \end{gather}
Hence using \eqref{eq:app-self-energy}
we get that $\Sigma^{LL}(x,y)=({\mathcal J^2}/{2p})e^{\Phi(x,y)}
+O(p^{-2})$. The AP Dyson equation then reduces to
\be
\begin{gathered}
\partial_x\partial_yg(x,y)
=
-2\mathcal J^2e^{\Phi(x,y)},
\qquad
0<y<x<\beta,\\
g(\beta,y)=g(0,y)\,,\qquad g(x,\beta)=g(x,0)\,.
\end{gathered}
\label{eq:app-diagonal-Liouville-coupled}
\ee
The rest of the domain is fixed by $g(x,y)=g(y,x)$.

The diagonal self-energy also changes the profile of the periodic zero
mode. Its row integral and primitive are
\be
\begin{aligned}
s(x)
&:=
\frac{\mathcal J^2}{2}
\left[
\int_0^x{\rm d}y\,e^{\Phi(x,y)}
-
\int_x^\beta {\rm d}y\,e^{\Phi(y,x)}
\right],
\\
S(x)
&:=
\int_0^x{\rm d}y\,s(y),
\qquad
S(0)=S(\beta)=0.
\end{aligned}
\label{eq:app-s-definition}
\ee
Here $p\int {\rm d}y\,\Sigma^{LL}(x,y)=s(x)+O(p^{-1})$, and
$S(\beta)=0$ follows from antisymmetry. Indeed, if $u(x)$ is the
dressed zero-mode profile, its equation
$(\frac12\partial_x-\Sigma^{LL})u=0$ gives
$u(x)=1+2S(x)/p+O(p^{-2})$, after choosing $u(0)=1$.

At order $p^{-1}$, the periodic zero mode contributes $1+[c+2S(x)+2S(y)]/p$ to the mixed $LR$ bilocal. Here $c$ is the correction to the zero-mode self-energy $\frac{\alpha}{p}[\int_0^\beta {\rm d}x\,{\rm d}y\,u(x)\Sigma^{LR}(x,y)u(y)]^{-1}=1+c/p+O(p^{-2})$. The term $2S(x)+2S(y)$ arises from expanding the correction to the profiles $u(x)u(y)$ with the leading self-energy. The AP sector contributes $A(x,y)/p$ to the mixed $LR$ bilocal. Thus
$h(x,y)=A(x,y)+c+2S(x)+2S(y)$. Since $A$ is antiperiodic in each
argument, it cancels when the two endpoints of the thermal interval are
added. Its Dyson equation then gives
\be
\begin{gathered}
\partial_x\partial_yh(x,y)
=
-2\mathcal J^2e^{h(x,y)},
\\
h(\beta,y)+h(0,y)
=
2c+4S(y),
\qquad
h(x,\beta)+h(x,0)
=
2c+4S(x).
\end{gathered}
\label{eq:app-offdiagonal-Liouville-coupled}
\ee
The constant $c$ is fixed by the zero-mode normalization
\eqref{eq:app-zero-mode-normalization}. The equations \eqref{eq:app-diagonal-Liouville-coupled} and \eqref{eq:app-offdiagonal-Liouville-coupled} form a closed
system for $g$ and $h$.

\subsection{General solution}
\label{app:exact-coupled-saddle}

Integrating \eqref{eq:app-diagonal-Liouville-coupled} on the two sides of $x$, and
using $g(x,x)=0$ together with the boundary conditions, gives
$s(x)=\frac12\partial_xg(x,0)$. Since $S(0)=g(0,0)=0$, it follows that
$S(x)=\frac12g(x,0)$.

We now solve the off-diagonal Liouville equation \eqref{eq:app-offdiagonal-Liouville-coupled} for an arbitrary
profile $S(x)$. Introduce
\be
L_S:=\int_0^\beta {\rm d}x\,e^{2S(x)},
\qquad
\Xi(x):=\frac{1}{L_S}\int_0^x{\rm d}t\,e^{2S(t)}.
\label{eq:app-Xi-definition}
\ee
The regular solution of the off-diagonal Liouville
equation with the boundary conditions \eqref{eq:app-offdiagonal-Liouville-coupled} and zero-mode normalization \eqref{eq:app-zero-mode-normalization} is
\be
e^{h(x,y)}
=
\frac{
4\kappa^2(e^{2\kappa}-1)e^{2S(x)+2S(y)}
e^{2\kappa[\Xi(x)+\Xi(y)]}
}{
(\mathcal J L_S)^2
\left[
(e^{2\kappa\Xi(x)}-1)(e^{2\kappa\Xi(y)}-1)
+e^{2\kappa}-1
\right]^2
}.
\label{eq:app-h-general-solution}
\ee
Its endpoint constant is
\be
c=
\log\left[
\frac{4\kappa^2e^\kappa}
{(\mathcal J L_S)^2(e^{2\kappa}-1)}
\right].
\ee

We can now determine the zero-mode density $r(x)$ using \eqref{eq:app-definitionrx}. The integral of \eqref{eq:app-h-general-solution} gives
\be
\int_0^\beta {\rm d}y\,e^{h(x,y)}
=
\frac{2\kappa}{\mathcal J^2}
\frac{e^{2S(x)}}{L_S}.
\label{eq:app-h-row-integral}
\ee
Comparing with \eqref{eq:app-definitionrx}, we obtain
\be
r(x)=\frac{e^{2S(x)}}{L_S}=\Xi'(x),
\qquad
R(x)=\Xi(x).
\ee

Using $S(x)=g(x,0)/2$, the coordinate is determined entirely by the
boundary value of $g$:
\be
\Xi(x)=
\frac{\displaystyle\int_0^x{\rm d}t\,e^{g(t,0)}}
{\displaystyle\int_0^\beta {\rm d}t\,e^{g(t,0)}}.
\ee
The diagonal equation therefore closes as
\be
\Phi(x,y)
=
g(x,y)-2\kappa[\Xi(x)-\Xi(y)],
\qquad
\partial_x\partial_yg(x,y)
=
-2\mathcal J^2e^{\Phi(x,y)}.
\label{eq:app-closed-system}
\ee
This is a single nonlocal equation for $g$. Once it is solved, $S$,
$\Xi$, and the off-diagonal field $h$ follow from the expressions above.

To solve it, set $t=\Xi(x)$ and $u=\Xi(y)$, and denote the inverse map
by $x=x(t)$. Writing $a(t):=g(x(t),0)$, the definition of $\Xi$
gives ${\rm d}x/{\rm d}t=L_Se^{-a(t)}$. The factors generated by this Jacobian are
removed by defining
$F(t,u):=g(x(t),x(u))-a(t)-a(u)-2\kappa(t-u)$, which obeys
\be
\partial_t\partial_uF=-2(\mathcal J L_S)^2e^F,
\qquad
F(t,0)=-2\kappa t,
\qquad
F(1,u)=-2\kappa(1-u).
\label{eq:app-F-Liouville}
\ee
This characteristic boundary-value problem, known as a Goursat problem, has a unique regular solution. Introducing
$\rho:=\mathcal{J}L_S/(2\kappa)$, the solution can be written as
\be
\begin{gathered}
F(t,u)=-2\kappa(t-u)-2\log D(t,u),\\
D(t,u):=
1-\rho^2
\left(e^{-2\kappa t}-e^{-2\kappa}\right)
\left(e^{2\kappa u}-1\right),\\
C(t):=D(t,t)
=
1-\rho^2
\left(1-e^{-2\kappa t}\right)
\left(1-e^{-2\kappa(1-t)}\right).
\end{gathered}
\label{eq:app-CD-definition}
\ee
On the diagonal, $F(t,t)=-2a(t)$, and therefore
$e^{a(t)}=C(t)$. Equivalently,
$e^{2S(x)}=C(\Xi(x))$, so the same function $C$ determines both the
boundary dressing and the map between $x$ and $\Xi$.

It remains only to fix the parameter $\rho$. Since
${\rm d}x/{\rm d}t=L_S/C(t)$ and $\mathcal J L_S=2\kappa\rho$,
the total length of the thermal interval gives
\be
\beta\mathcal J
=
2\kappa\rho\int_0^1\frac{{\rm d}t}{C(t)},
\qquad
0<\rho<\frac{1}{1-e^{-\kappa}}.
\label{eq:app-rho-temperature-equation}
\ee
The right-hand side increases monotonically from zero to infinity,
so this equation determines a unique regular $\rho$ for every
positive $\kappa$ and $\beta\mathcal J$.

Once $\rho$ is known, the relation between physical and zero-mode time is
\be
\mathcal Jx
=
2\kappa\rho\int_0^{\Xi(x)}\frac{{\rm d}t}{C(t)}.
\label{eq:app-Xi-parametric}
\ee
This determines $\Xi(x)$ by a monotone one-dimensional inversion.
Both integrals are elementary.

The diagonal correlator is now explicit. For $x>y$, set
$t=\Xi(x)$ and $u=\Xi(y)$. Then
\be
e^{g(x,y)}
=
\frac{C(t)C(u)}{D(t,u)^2},
\qquad
e^{\Phi(x,y)}
=
\frac{C(t)C(u)e^{-2\kappa(t-u)}}{D(t,u)^2}.
\label{eq:app-exact-diagonal-solution}
\ee
The periodic propagator is
$P(x,y)=1-2[\Xi(x)-\Xi(y)]$ for $x>y$, with the opposite ordering
fixed by antisymmetry.

The off-diagonal saddle is already given by
\eqref{eq:app-h-general-solution}. It becomes fully explicit upon using
$e^{2S(x)}=C(\Xi(x))$ and
$\mathcal J L_S=2\kappa\rho$. Its endpoint constant correspondingly
reduces to $c=-\log(2\rho^2\sinh\kappa)$. Notice that the two arguments
of $h$ range independently over the thermal interval; unlike the
diagonal fields, no ordering is required.

\begin{figure}[h]
\centering
\includegraphics[width=\textwidth]{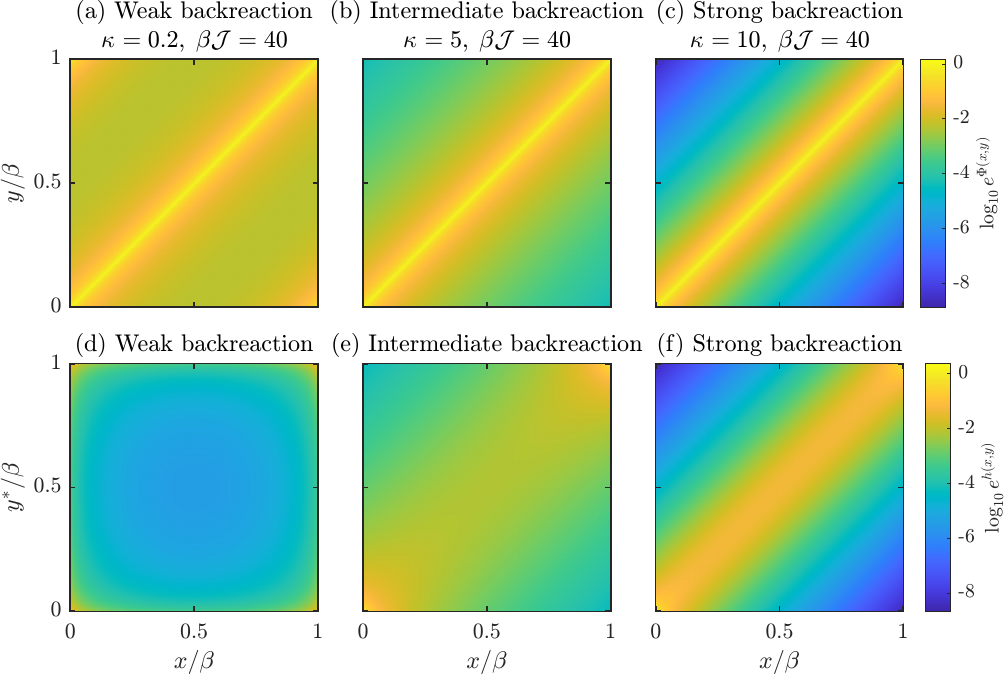}
\caption{Exact diagonal and off-diagonal solutions in three regimes of backreaction at low temperature. The diagonal field approaches the universal conformal bulk profile, while the dependence on the backreaction becomes concentrated near the thermal junctions. For strong backreaction the off-diagonal correlations become approximately translation-invariant for the backward time $y^* = \beta -y$.}
\label{fig:app-wormhole-correlation-regimes}
\end{figure}

The construction therefore involves only three steps: solve \eqref{eq:app-rho-temperature-equation} for $\rho$, determine
$\Xi(x)$ from \eqref{eq:app-Xi-parametric}, and substitute the result
into \eqref{eq:app-exact-diagonal-solution} and
\eqref{eq:app-h-general-solution}. Fig. ~\ref{fig:app-wormhole-correlation-regimes} shows the exact solution at low but finite temperature and different regimes of backreaction.

For later use, the diagonal row integrals are
\be
\int_0^x{\rm d}y\,e^{\Phi(x,y)}
=
\frac{\rho}{\mathcal J}\left(1-e^{-2\kappa t}\right),
\qquad
\int_x^\beta{\rm d}y\,e^{\Phi(y,x)}
=
\frac{\rho}{\mathcal J}
\left(1-e^{-2\kappa(1-t)}\right),
\qquad t=\Xi(x).
\label{eq:app-exact-diagonal-row-integrals}
\ee
Their difference gives $s(x)$; integrating it with $S(0)=0$ reproduces
$S(x)=\frac12\log C(\Xi(x))$, providing a final check of the solution.

\paragraph{The boundary reparametrization.}
To connect this solution directly to Sec.~\ref{sec:solwinding}, define
\be
b:=2\operatorname{arcosh}(e^\kappa),\qquad
\tanh\frac{\sigma(\mathcal Jx)}2
=\frac{\tanh\left[\kappa\left(\Xi(x)-\frac12\right)\right]}
{\tanh(b/4)}.
\label{eq:app-sigma-Xi}
\ee
Since $\tanh^2(b/4)=\tanh(\kappa/2)$, this map has precisely the endpoints
$\sigma(0)=-b/2$ and $\sigma(\beta\mathcal J)=b/2$.
It is increasing and satisfies
$\sigma(\mathcal J(\beta-x))=-\sigma(\mathcal Jx)$.
Reversing the orientation of the right time as in the main text,
$u=\mathcal Jx$ and $v=\mathcal J(\beta-y)$, reduces
\eqref{eq:app-h-general-solution} to
\be
e^{h(x,y)}=
\frac{\dot\sigma(u)\dot\sigma(v)}
{4\cosh^2[(\sigma(u)-\sigma(v))/2]}.
\label{eq:app-LR-clock-kernel}
\ee
The dot denotes a derivative with respect to dimensionless time.
This is the $LR$ kernel of Sec.~\ref{sec:solwinding}, with no change
in the definition of $h$.

Differentiating \eqref{eq:app-sigma-Xi} using
$\Xi'(x)=\mathcal J C(\Xi(x))/(2\kappa\rho)$ gives
\be
\begin{gathered}
\dot\sigma=A_0+A_1\cosh\sigma,\\
A_0=\frac{1}{\rho\sqrt{1-e^{-2\kappa}}}
-\rho\sqrt{1-e^{-2\kappa}},\qquad
A_1=e^{-\kappa}\left[
\frac{1}{\rho\sqrt{1-e^{-2\kappa}}}
+\rho\sqrt{1-e^{-2\kappa}}
\right]>0.
\end{gathered}
\label{eq:app-clock-dictionary}
\ee
In particular,
\be
e^{2\kappa}A_1^2-A_0^2=4,\qquad
\beta\mathcal J=\int_{-b/2}^{b/2}
\frac{{\rm d}\sigma}{A_0+A_1\cosh\sigma},\qquad
E:=\{e^\sigma,u\}=\frac{A_1^2-A_0^2}{2}.
\label{eq:app-clock-constraints}
\ee
These are exactly the boundary reparametrization equations and constant Schwarzian used in
Sec.~\ref{sec:solwinding}. The temperature equation here and
\eqref{eq:app-rho-temperature-equation} are related by the change of
variables \eqref{eq:app-sigma-Xi}. Note that on the positive-Schwarzian branch, these conditions imply $0<E<2$.

For completeness, the coordinate inversion can also be written
explicitly. For $A_0+A_1>0$,
\be
\tanh\frac{\sigma(u)}2=
\begin{cases}
\displaystyle\frac{A_0+A_1}{\sqrt{-2E}}
\tanh\left[\frac{\sqrt{-2E}}2
\left(u-\frac{\beta\mathcal J}2\right)\right],
& E<0,\\[6pt]
\displaystyle\frac{A_0+A_1}2
\left(u-\frac{\beta\mathcal J}2\right),
& E=0,\\[6pt]
\displaystyle\frac{A_0+A_1}{\sqrt{2E}}
\tan\left[\frac{\sqrt{2E}}2
\left(u-\frac{\beta\mathcal J}2\right)\right],
& E>0.
\end{cases}
\label{eq:app-explicit-sigma}
\ee
The endpoint condition fixes the temperature equivalently as
\be
\beta\mathcal J=
\begin{cases}
\displaystyle\frac4{\sqrt{-2E}}
\operatorname{artanh}\frac{\sqrt{-2E}\,\tanh(b/4)}{A_0+A_1},
& E<0,\\[6pt]
\displaystyle\frac{4\tanh(b/4)}{A_0+A_1},
& E=0,\\[6pt]
\displaystyle\frac4{\sqrt{2E}}
\arctan\frac{\sqrt{2E}\,\tanh(b/4)}{A_0+A_1},
& E>0.
\end{cases}
\label{eq:app-elementary-temperature}
\ee
The real principal branches give the regular boundary reparametrization on
$0\leq u\leq\beta\mathcal J$.
Together with
\be
\Xi(x)=\frac12+\frac1\kappa
\operatorname{artanh}\left[
\tanh\frac b4\,\tanh\frac{\sigma(\mathcal Jx)}2
\right],
\label{eq:app-Xi-explicit}
\ee
these expressions evaluate the integral and its inverse in
\eqref{eq:app-Xi-parametric}. The finite-temperature zero of $E$ occurs
at $\rho=(1+e^{-\kappa})^{-1}$, or $\beta\mathcal J=e^\kappa-1$;
the solution is regular across this point.

\paragraph{The diagonal correlator.}
Although $g$ and $P$ depend separately on both insertions, their mixed
combination $\Phi$ depends only on $\mathcal J(x-y)>0$.
The Schwarzian energy satisfies
\be
E=2-\frac12\left[\rho^{-1}+\rho(1-e^{-2\kappa})\right]^2
=2-\frac{A_1^2}{2}\sinh^2\frac b2.
\label{eq:app-diagonal-energy}
\ee
Eq.~\eqref{eq:app-exact-diagonal-solution} can then be written as
\be
e^{\Phi(x,y)}=
\begin{cases}
\displaystyle
\left[
\cosh\frac{\sqrt{-2E}\,\mathcal J(x-y)}2
+\sqrt{\frac{2-E}{-E}}\,
\sinh\frac{\sqrt{-2E}\,\mathcal J(x-y)}2
\right]^{-2},
& E<0,\\[6pt]
\displaystyle
\left[1+\mathcal J(x-y)\right]^{-2},
& E=0,\\[6pt]
\displaystyle
\left[
\cos\frac{\sqrt{2E}\,\mathcal J(x-y)}2
+\sqrt{\frac{2-E}{E}}\,
\sin\frac{\sqrt{2E}\,\mathcal J(x-y)}2
\right]^{-2},
& E>0.
\end{cases}
\label{eq:app-diagonal-physical-time}
\ee
The bracketed factors equal $1$ at $x=y$ and $e^\kappa$ at
$x-y=\beta$, and remain positive throughout the thermal interval.
On the positive-Schwarzian branch, $0<E<2$, so equivalently
\be
e^{\Phi(x,y)}=
\frac{E}{
2\sin^2\!\left[
\sqrt{\frac E2}\,\mathcal J(x-y)
+\arcsin\sqrt{\frac E2}
\right]}.
\label{eq:app-diagonal-shifted-kernel}
\ee
This reproduces the shifted diagonal kernel of
Sec.~\ref{sec:diagonal-correlators-low-temperature}.
The shift in the sine's argument enforces $\Phi(x,x)=0$;
it does not enter the $LR$ kernel \eqref{eq:app-LR-clock-kernel}.
The negative- and zero-Schwarzian expressions in
\eqref{eq:app-diagonal-physical-time} give its regular continuation
to the other branches.

\begin{figure}[h]
    \centering
    \includegraphics[width=.9\linewidth]{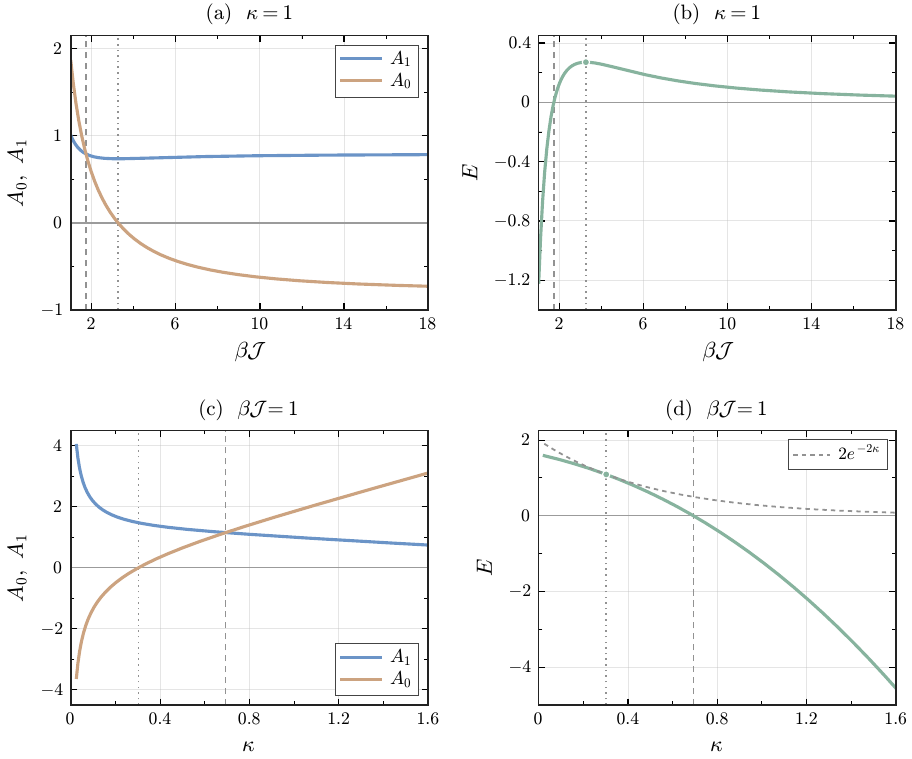}
   \caption{Coefficients and Schwarzian energy of the boundary trajectory. Panels (a,b) show the dependence on $\beta\mathcal J$
at fixed $\kappa=1$, while panels (c,d) show the dependence on $\kappa$
at fixed $\beta\mathcal J=1$. Vertical dashed
lines mark the zero of $E$, and vertical dotted lines mark $A_0=0$.
The gray dashed curve in (d) is the upper bound $2e^{-2\kappa}$.
Filled points mark saturation of this bound; in the temperature
scan, this coincides with the maximum of $E$ and the minimum of $A_1$.}
\label{fig:app-clock-scans}
\end{figure}

Fig.~\ref{fig:app-clock-scans} illustrates how the thermal boundary conditions determine the boundary trajectory coefficients and their Schwarzian energy. Although $E$ is conserved along each trajectory, its
value depends on $\beta\mathcal J$ and $\kappa$.

\subsection{Simple regimes}

It is instructive to expand the solution in different regimes where the solution simplifies considerably.

\paragraph{Low temperature at finite backreaction.}
First take $\beta\mathcal J\gg1$ at fixed $\kappa$. The parameter $\rho$ approaches
the upper endpoint of its allowed interval,
\be
\rho=
\frac{1}{1-e^{-\kappa}}
\left[
1-\frac{\pi^2e^\kappa}{2(\beta\mathcal J)^2}
+O((\beta\mathcal J)^{-3})
\right],
\qquad
L_S\longrightarrow
\frac{2\kappa}{\mathcal J(1-e^{-\kappa})}.
\label{eq:app-low-temperature-rho}
\ee
In physical time, $\Xi(x)=\frac12+O((\beta\mathcal J)^{-1})$.
Its variation from $0$ to $1/2$, and from $1/2$ to $1$, is confined
to two layers of width $O(\mathcal J^{-1})$ near the thermal junction.
Thus the density $r(x)$ becomes localized near the two ends of the interval, with algebraic tails. The dressed zero-mode wavefunction remains perturbatively close to a constant.

Away from the thermal junction and, for the diagonal field, away from coincidence,
the solution reduces to
\be
\begin{aligned}
e^{2S(x)}
&=
\frac{\pi^2e^\kappa}{(\beta\mathcal J)^2\sin^2(\pi x/\beta)}
\left[1+O((\beta\mathcal J)^{-1})\right],\\
e^{g(x,y)}=e^{\Phi(x,y)}
&=
\frac{\pi^2}{(\beta\mathcal J)^2\sin^2[\pi(x-y)/\beta]}
\left[1+O((\beta\mathcal J)^{-1})\right],\\
e^{h(x,y)}
&=
\frac{\pi^4(e^{2\kappa}-1)}
{4(\beta\mathcal J)^4\sin^2(\pi x/\beta)\sin^2(\pi y/\beta)}
\left[1+O((\beta\mathcal J)^{-1})\right].
\end{aligned}
\label{eq:app-low-temperature-bulk}
\ee
The diagonal bulk profile is therefore universal and independent of
$\kappa$ at leading order. Finite backreaction is carried by the
zero-mode map and the layers near the thermal junction. Although $e^h$ is suppressed away from the junction, those layers retain the exact normalization
$\int {\rm d}x\,{\rm d}y\,e^h=2\kappa/\mathcal J^2$. The formulas above are
accordingly nonuniform near the seams. Fig. \ref{fig:app-wormhole-correlation-regimes} shows this behavior for sufficiently weak backreaction and finite but low temperature.

\paragraph{Weak backreaction at arbitrary temperature.}
Now take $\kappa\ll1$ with $\beta\mathcal J$ fixed. At leading order,
the diagonal sector reduces to the standard translation-invariant thermal
saddle $g(x,y) \approx g_0(x-y)$ with
\be
e^{g_0(\tau)}
=
\left[
\frac{\cos\vartheta}
{\cos[2\vartheta((x-y)/\beta-1/2)]}
\right]^2,
\qquad
\beta\mathcal J=\frac{2\vartheta}{\cos\vartheta},
\qquad
0<\vartheta<\frac{\pi}{2}.
\label{eq:app-standard-thermal-saddle}
\ee
At this order, $S(x)=\frac12g_0(x)+O(\kappa)$. Defining
$L_0:=\int_0^\beta {\rm d}x\,e^{g_0(x)}
=2\sin\vartheta/\mathcal J$, the dressed zero-mode density is
$r(x)=e^{g_0(x)}/L_0+O(\kappa)$. The leading off-diagonal saddle
factorizes:
\be
e^{h(x,y)}
=
\frac{2\kappa}{\mathcal J^2L_0^2}
e^{g_0(x)+g_0(y)}
\left[1+O(\kappa)\right].
\label{eq:app-small-alpha-saddle}
\ee
Thus, to first order in $\kappa$, the off-diagonal self-energy is a
rank-one kernel constructed from the ordinary thermal saddle.

At low temperature,
$\vartheta=\frac{\pi}{2}
-\frac{\pi}{\beta\mathcal J}
+O((\beta\mathcal J)^{-2})$, and
\[
e^{g_0(\tau)}
\simeq
\left[
\frac{\pi}
{\beta\mathcal J\sin(\pi\tau/\beta)}
\right]^2.
\]
Consequently, both the density $r$ and $\Sigma^{LR}$ become localized within $O(\mathcal J^{-1})$ of the thermal junction, in agreement with the finite-$\kappa$ result above. The simultaneous
low-temperature, large-$p$ expansion remains valid provided
$\log(\beta\mathcal J)=o(\sqrt p)$.

\paragraph{High temperature at finite backreaction.}
Finally, take $\beta\mathcal J\ll1$ with $\kappa$ fixed. The diagonal dressing is small:
$S(x)=O((\beta\mathcal J)^2)$,
$\Xi(x)=x/\beta+O((\beta\mathcal J)^2)$, and, for $x>y$,
$P(x,y)=1-2(x-y)/\beta+O((\beta\mathcal J)^2)$. The off-diagonal field,
however, remains nontrivial:
\be
e^{h(x,y)}
=
\frac{
4\kappa^2(e^{2\kappa}-1)e^{2\kappa(x+y)/\beta}
}{
(\beta\mathcal J)^2
\left[
(e^{2\kappa x/\beta}-1)(e^{2\kappa y/\beta}-1)
+e^{2\kappa}-1
\right]^2
}
\left[1+O((\beta\mathcal J)^2)\right].
\label{eq:app-h-high-temperature}
\ee
This result is exact in $\kappa$ at leading high-temperature order and
is generally time dependent. 

\begin{figure}[h]
\centering
\includegraphics[width=0.92\textwidth]{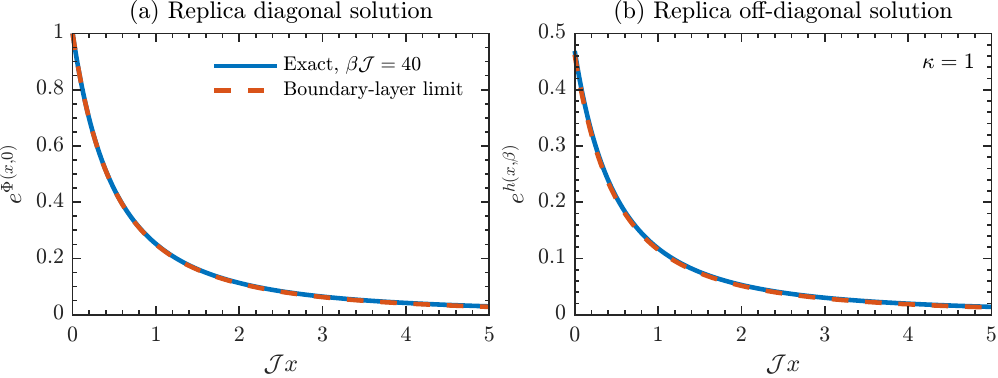}
\caption{Replica diagonal and off-diagonal kernels near the thermal
junction at $\kappa=1$ and $\beta\mathcal J=40$.
The left panel shows $e^{\Phi(x,0)}$; the right panel shows
$e^{h(x,\beta)}$, with the second insertion fixed at the opposite
junction. Solid curves show the exact finite-temperature solution;
dashed curves show the approximate solutions.
The diagonal profile is independent of $\kappa$,
and both kernels vary on the time scale $\mathcal J^{-1}$.}
\label{fig:app-wormhole-boundary-layers}
\end{figure}

\paragraph{Solution near the thermal junction.} To resolve the thermal junction at
low temperatures, we additionally take $\beta\mathcal J\to\infty$
at fixed $\kappa>0$ and fixed $\mathcal Jx$.
The leading solution near the left side of the thermal junction simplifies to
\be
e^{\Phi_{\rm BL}(x,0)}
\approx
\frac{1}{(1+\mathcal Jx)^2},
\qquad
e^{h_{\rm BL}(x,\beta)}
\approx 
\frac{\tanh(\kappa/2)}{(1+\mathcal Jx)^2}.
\ee
The diagonal boundary-layer profile is therefore universal,
independent of $\kappa$, while the $LR$ kernel retains a
$\kappa$-dependent amplitude.
This approximation applies for the left side of the thermal junction $x=O(\mathcal J^{-1})$;
the right side of the thermal junction follows from
$\Xi(\beta-x)=1-\Xi(x)$.
Fig.~\ref{fig:app-wormhole-boundary-layers} compares the exact
low-temperature kernels near the thermal junctions with this approximation.

\subsection{Contribution to the observable}

We now evaluate the large-$p$ saddle contribution to
$\mathcal C_\beta(W)$, fix its normalization by microscopic
high-temperature matching, and compare its low-temperature
expansion with the classical limit of the double-scaled result.
The calculation uses the partition threshold and the endpoint
probability $P_K(W)$ introduced in Sec.~\ref{sec:thermalonept}
and Appendix~\ref{app:ConditionedEnsemble}.

Using the self-energy equation \eqref{eq:app-self-energy}, the on-shell action becomes
\be
\begin{split}
I_{*}
={}&
-(1-\nu)\log{\rm Pf}_{\rm AP}
\left(\frac12\partial_\tau-\Sigma_{\rm AP}\right)
-\nu\log{\rm Pf}_{\rm P}
\left(\frac12\partial_\tau-\Sigma_{\rm P}\right)
\\
&+
\frac{\mathcal J^2(p-1)}{4p^2}
\sum_{a,b=L,R}
\int_0^\beta {\rm d}\tau_1\,{\rm d}\tau_2\,
(G^{ab})^p .
\end{split}
\label{eq:app-onshell-action}
\ee
The diagonal row integrals \eqref{eq:app-exact-diagonal-row-integrals} and the off-diagonal normalization \eqref{eq:app-zero-mode-normalization} give
\be
\sum_{a,b}
\int_0^\beta {\rm d}\tau_1\,{\rm d}\tau_2\,(G^{ab})^p
=
\frac{2\beta}{\mathcal J\rho}
\left[1+\rho^2(1-e^{-2\kappa})\right]
+O(p^{-1}).
\label{eq:app-full-interaction-integral}
\ee

At a stationary point, differentiating the original collective action with respect to $\mathcal J$ acts only on its explicit interaction coefficient. Therefore
\be
\frac{\partial}{\partial(\beta\mathcal J)}
\left[
\frac{p^2}{N}
\log\mathcal{C}_{\beta}(W)
\right]
=
\frac1{\rho}
+\rho(1-e^{-2\kappa})
+O(p^{-1}).
\label{eq:app-classical-action-derivative}
\ee
Together with \eqref{eq:app-rho-temperature-equation}, this determines the complete temperature dependence by a one-dimensional integral. In particular, as $\beta\mathcal J\to0$,
\be
\rho=\frac{\beta\mathcal J}{2\kappa}+O((\beta\mathcal J)^3),
\qquad
\frac{p^2}{N}\log\mathcal{C}_{\beta,{\rm saddle}}(W)
=
2\kappa\log(\beta\mathcal J)+\text{constant}
+O((\beta\mathcal J)^2),
\ee
reproducing the threshold power with $K=N\kappa/p^2=w/p$.

The remaining constant is fixed by the microscopic high-temperature expansion. This matching is necessary because the periodic Pfaffian contains the lifted zero-mode eigenvalue $\beta m\approx\frac{\kappa}{p}$ whose logarithm is multiplied by $N\nu=w$ and contributes at leading order. At the partition threshold, with $K_0=N\kappa/p^2$ an integer and $w=K_0p$, the first nonzero microscopic term is
\be
\mathcal{C}_\beta(W)
=
\frac{\beta^{2K_0}}{K_0!}
\left(\frac{\mathcal J^2}{\lambda}\right)^{K_0}
P_{K_0}(W)
+O(\beta^{2K_0+2}).
\label{eq:app-UV-normalization-general}
\ee
With the two Pfaffians defined using a common microscopic regulator, matching this expression fixes the remaining total prefactor $\mathcal N_W$.

To display the resulting temperature dependence, it is natural to compare the wormhole amplitude with the normalized thermal partition function
$Z(2\beta)=\langle\Tr(e^{-2\beta H})\rangle_J$. The period is $2\beta$, rather than $\beta$, because $\mathcal{C}_\beta(W)$ contains two Euclidean evolutions of length $\beta$. At leading order in large $p$ \cite{Maldacena:2016hyu},
\be
\frac{p^2}{N}\log Z(2\beta)
=
2\vartheta\tan\vartheta-\vartheta^2,
\qquad
\beta\mathcal J=\frac{\vartheta}{\cos\vartheta}.
\label{eq:app-largep-Z2beta}
\ee
At low temperature, the derivatives of  $\log \mathcal{C}_\beta(W)$ and $\log Z(2\beta)$ with respect to $\beta\mathcal J$ approach $2 N/p^2$. The ratio therefore remains finite and defines the infrared action of the operator
\be
F_W^{\rm IR}
:=
\lim_{\beta\mathcal J\to\infty}
\frac{\mathcal{C}_{\beta,\mathrm{class}}(W)}
{Z_{\mathrm{class}}(2\beta)}.
\label{eq:app-IR-operator-factor}
\ee
Here the subscript “class” means that the leading classical exponent is evaluated before taking the low-temperature limit.

\begin{figure}[h]
\centering
\includegraphics[width=.9\textwidth]{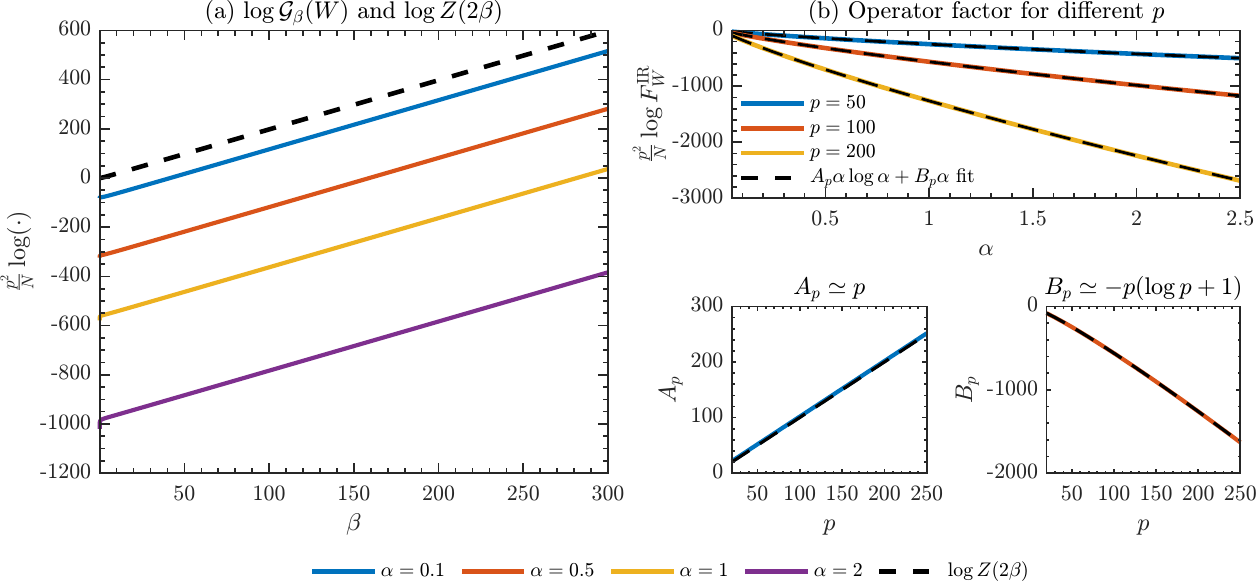}
\caption{On-shell wormhole amplitude and infrared operator factor at leading order in large $p$, with $N/p^2=100$ and normalized traces. Left: $\frac{p^2}{N}\log\mathcal{C}_\beta(W)$ for several values of $\kappa$, compared with $\frac{p^2}{N}\log Z(2\beta)$ (black dashed curve). The curves become parallel at low temperature, showing that their difference approaches a constant. Upper right: the resulting infrared factor $F_W^{\rm IR}$ as a function of $\kappa$ for $p=50,100,200$. Lower right: fitted coefficients $A_p$ and $B_p$ as functions of $p$. The black dashed curves show the large-$p$ predictions $A_p=p$ and $B_p=-p(\log p+1)$.}
\label{fig:app-wormhole-action-factor}
\end{figure}

The analytic solution also determines the low-temperature action, including its constant term. At the threshold $K_0=N\kappa/p^2$, define
\be
\gamma_W
:=
\frac{p^2}{N}
\log\left[
\frac{P_{K_0}(W)}
{K_0!\lambda^{K_0}}
\right].
\ee
The exact large-$p$ solution then gives
\be
\frac{p^2}{N}\log\mathcal{C}_\beta(W)
=
2\beta\mathcal J+\gamma_W
+2\kappa\bigl[\log(2\kappa)-1\bigr]
-\frac{\pi^2}{6}
-2\operatorname{Li}_2(e^{-\kappa})
+\frac{\pi^2}{\beta\mathcal J}
+O\!\left((\beta\mathcal J)^{-2},p^{-1}\right).
\label{eq:app-low-temperature-action}
\ee
Since
$
(p^2/N)\log Z(2\beta)
=
2\beta\mathcal J-\pi^2/4
+\pi^2/(4\beta\mathcal J)+\cdots
$,
the infrared action of the operator is
\be
\frac{p^2}{N}\log F_W^{\rm IR}
=
\gamma_W
+\frac{\pi^2}{12}
+2\kappa\bigl[\log(2\kappa)-1\bigr]
-2\operatorname{Li}_2(e^{-\kappa}).
\label{eq:app-exact-IR-operator-factor}
\ee
Fig.~\ref{fig:app-wormhole-action-factor} compares these predictions with the exact finite-temperature saddle.

\paragraph{Comparison with the double-scaled answer.} We now compare the contribution \eqref{eq:app-low-temperature-action} to the one obtained in Sec.~\ref{sec:thermalonept}. We take $\lambda\to0$ at fixed
$\kappa$ and $\beta\mathcal J$, followed by $\beta\mathcal J\to\infty$.
At the integer partition threshold $w=\Delta p$, with
$\Delta=2\kappa/\lambda$, threshold localization and
Eqs.~\eqref{eq:Gkerneldiag} and \eqref{eq:GWfinal} give
\be
\mathcal{C}_\beta(W)\simeq
\frac{\Delta!P_\Delta(W)}{[\Delta]_q!}\,|\Psi_\beta(\Delta)|^2.
\label{eq:app-DS-threshold-check}
\ee
Here $\Psi_\beta$ is the chord Hartle-Hawking wavefunction
\eqref{eq:HHchordwf}, with $\mathcal J$ restored. Recall that the heavy operator scaling here is different, by a logarithmic shift, from the one in \eqref{eq:finite-length-heavy-scaling}, which is why in the JT limit \eqref{eq:SYKwormhole} we get $\ell_W=2\kappa+2\log(2\lambda)\to-\infty$ instead of a finite length.

For the kernel, the $q$-factorial \eqref{eq:qfact}, the asymptotic
$\lambda\log(q;q)_\Delta
=\operatorname{Li}_2(e^{-2\kappa})-\pi^2/6+o(1)$,
and Stirling's formula give
\be
\frac{\lambda}{2}\log\frac{\Delta!P_\Delta(W)}{[\Delta]_q!}
=\gamma_W+2\kappa[\log(2\kappa)-1]
+\frac{\pi^2}{12}
-\frac12\operatorname{Li}_2(e^{-2\kappa})+o(1).
\ee
For the Hartle-Hawking wavefunction, set $\theta=\pi-\epsilon$ in
\eqref{eq:HHchordwf}. The $q$-binomial saddle $k=\Delta/2$ gives
\be
\lambda\log|\Phi_{\pi-\epsilon}(\Delta)|
=\frac{\pi^2}{4}
+\frac12\operatorname{Li}_2(e^{-2\kappa})
-2\operatorname{Li}_2(e^{-\kappa})
+O_\kappa(\epsilon^2)+o(1).
\ee
The spectrum \eqref{eq:chordspec} and measure
\eqref{eq:qhermitemeasure} then place the thermal saddle at
$\epsilon_*=\pi/(\beta\mathcal J)
+O_\kappa((\beta\mathcal J)^{-2})$, yielding
\be
\frac{\lambda}{2}\log|\Psi_\beta(\Delta)|^2
=2\beta\mathcal J+\frac{\pi^2}{\beta\mathcal J}
-\frac{\pi^2}{4}
-\operatorname{Li}_2(e^{-\kappa})
+\operatorname{Li}_2(-e^{-\kappa})
+O_\kappa((\beta\mathcal J)^{-2})+o(1).
\ee
Adding the two exponents and using the dilogarithm duplication
identity reproduces \eqref{eq:app-low-temperature-action},
including its constant term. Recall that we extract the operator effective action \eqref{eq:app-exact-IR-operator-factor} taking the ratio of classical contributions and then the low-temperature limit; including the one-loop fluctuations (even at finite $\lambda$) would instead give
$\mathcal{C}_\beta(W)/Z(2\beta)\propto\beta^{-3/2}\to0$.

\section{Extensive operators}
\label{sec:extensive-wormhole-windings}

In this appendix, we study the wormhole saddle for the same observable
$\mathcal C_\beta(W)$, now with a Majorana string $\psi_W$ that contains
a finite fraction of the fermions. The saddle follows from the
collective action \eqref{eq:app-correct-collective-action} in the limit:
\be
N\longrightarrow\infty,\qquad p\longrightarrow\infty,\qquad
\nu=\frac wN\in(0,1/2),\qquad
\beta\mathcal J\ \text{fixed}.
\label{eq:extensive-sd-scaling}
\ee
The periodic sector now contains an order-one fraction of the
fermions, so the fixed $\kappa=p\nu$ expansion of
Appendix~\ref{app:SD-onepoint} no longer applies.
We retain the full periodic resolvent and return to the
Schwinger-Dyson equations,
\be
\begin{gathered}
G^{ab}=(1-\nu)G_{\rm AP}^{ab}+\nu G_{\rm P}^{ab},\qquad
\Sigma^{ab}=\frac{\mathcal J^2}{2p}(G^{ab})^{p-1},\\
G_s=\left(\frac12\partial-\Sigma\right)^{-1}_s,
\qquad a,b=L,R,\qquad s={\rm AP,P}.
\end{gathered}
\label{eq:extensive-sd-equations}
\ee

\paragraph{The contact solution.}
We construct a reflection-symmetric branch whose $LR$ self-energy
forms a narrow peak along $x+y=\beta$. The contact approximation
replaces this peak by a delta function with the same integrated
strength:
\be
\Sigma^{LR}(x,y)\simeq\mu\,\delta(x+y-\beta),\qquad
\Sigma^{RL}(x,y)\simeq-\mu\,\delta(x+y-\beta).
\label{eq:extensive-sd-contact}
\ee
Here $\mu>0$ denotes the integrated strength of the peak.
At finite $p$, the peak has a smooth profile with width of order
$(p\mu)^{-1}$, so the contact approximation applies at separations
large compared with this scale. The diagonal self-energies are
negligible at leading order. Below, we resolve the peak, verify
its normalization, and determine $\mu$ by matching.

Reflecting the right coordinate, $y\mapsto\beta-y$, makes the
coupling local and reverses the sign of its kinetic derivative.
In Fourier space, the reflected kinetic operator has inverse
\be
\begin{pmatrix}
i\omega/2&-\mu\\
\mu&-i\omega/2
\end{pmatrix}^{-1}
=\frac{1}{\omega^2/4+\mu^2}
\begin{pmatrix}
-i\omega/2&\mu\\
-\mu&i\omega/2
\end{pmatrix}.
\label{eq:extensive-sd-folded-inverse}
\ee
Summing over $\omega=2\pi n/\beta$ for P and
$\omega=2\pi(n+1/2)/\beta$ for AP, then undoing the reflection,
gives the leading propagators in the contact approximation:
\be
\begin{aligned}
G_{\rm AP}^{LL}(x,y)
&\simeq\sign(x-y)
\frac{\cosh\!\left[\mu(\beta-2|x-y|)\right]}
{\cosh(\beta\mu)},\\
G_{\rm P}^{LL}(x,y)
&\simeq\sign(x-y)
\frac{\sinh\!\left[\mu(\beta-2|x-y|)\right]}
{\sinh(\beta\mu)},\\
G_{\rm AP}^{LR}(x,y)
&\simeq
\frac{\sinh\!\left[\mu(\beta-2|x+y-\beta|)\right]}
{\cosh(\beta\mu)},\\
G_{\rm P}^{LR}(x,y)
&\simeq
\frac{\cosh\!\left[\mu(\beta-2|x+y-\beta|)\right]}
{\sinh(\beta\mu)}.
\end{aligned}
\label{eq:extensive-sd-sector-resolvents}
\ee
These expressions satisfy the required boundary conditions.
The $RR$ and $RL$ components follow from replica symmetry and
antisymmetry. In the contact approximation, extrapolating the
weighted $LR$ propagator to $x+y=\beta$ gives the peak value
$(1-\nu)\tanh(\beta\mu)+\nu\coth(\beta\mu)$.
Its initial slope is $-2\mu$ as the distance $|x+y-\beta|$
from this line increases from zero.

\paragraph{Resolving the contact and fixing its strength.}
The left and right Dyson equations imply the interior identity
$\partial_x\partial_yG^{LR}
=-4\Sigma^{LR}-4(\Sigma\star G\star\Sigma)^{LR}$.
Here $\star$ denotes integration over intermediate times in
$[0,\beta]$ together with summation over intermediate replica
indices. Explicitly,
\[
\begin{aligned}
(\Sigma\star G\star\Sigma)^{LR}(x,y)
={}&\sum_{a,b=L,R}
\int_0^\beta{\rm d}\tau_1
\int_0^\beta{\rm d}\tau_2\\
&\qquad\times
\Sigma^{La}(x,\tau_1)
G^{ab}(\tau_1,\tau_2)
\Sigma^{bR}(\tau_2,y).
\end{aligned}
\]
Within the peak of width $(p\mu)^{-1}$, the local self-energy
term is of order $p\mu^2$. In the convolution, each integrated
self-energy contributes a factor of order $\mu$, while the
propagator remains of order one. The convolution is therefore
of order $\mu^2$ and is suppressed by $1/p$ relative to the
local term. Away from the endpoints of the contact line,
$G^{LR}=1+O(\log p/p)$, so the leading equation within the peak is
\be
\partial_x\partial_y
\log\!\left[\bigl(G^{LR}(x,y)\bigr)^p\right]
=-2\mathcal J^2\bigl(G^{LR}(x,y)\bigr)^p[1+o(1)].
\label{eq:extensive-sd-inner-equation}
\ee
Reflection symmetry about $x+y=\beta$ and matching to the outer
slope fix the peak:
\be
\begin{gathered}
\bigl(G^{LR}(x,y)\bigr)^p
\sim\frac{p^2\mu^2}{\mathcal J^2}
\operatorname{sech}^2\!\left[p\mu(x+y-\beta)\right],\\
\Sigma^{LR}(x,y)
\sim\frac{p\mu^2}{2}
\operatorname{sech}^2\!\left[p\mu(x+y-\beta)\right].
\end{gathered}
\label{eq:extensive-sd-inner-solution}
\ee
For $y$ away from the thermal endpoints, integrating the peak
over $x$ gives $\int_0^\beta{\rm d}x\,\Sigma^{LR}(x,y)\to\mu$,
recovering the assumed contact strength.

Matching in the overlap
$1\ll p\mu|x+y-\beta|\ll\sqrt p$
gives, to leading logarithmic accuracy,
\be
(1-\nu)\tanh(\beta\mu)+\nu\coth(\beta\mu)-1
=\frac2p\log\frac{2p\mu}{\mathcal J}
+o\!\left(\frac{\log p}{p}\right).
\label{eq:extensive-sd-matching}
\ee
The limiting contact strength therefore satisfies
$(1-\nu)\tanh(\beta\mu)+\nu\coth(\beta\mu)=1$.
Its finite positive solution is equivalent to
$\tanh(\beta\mu)=\nu/(1-\nu)$.
Using this limiting value of $\mu$, the mixed propagators become
\be
\begin{aligned}
\mu&=-\frac{1}{2\beta}\log(1-2\nu),\\
G^{LR}(x,y)&=e^{-2\mu|x+y-\beta|}+o(1),\\
G^{LL}(x,y)&=\sign(x-y)e^{-2\mu|x-y|}+o(1).
\end{aligned}
\label{eq:extensive-sd-solved}
\ee
The diagonal self-energy is confined near coincidence and has
row-integrated absolute size $O(\mathcal J^2/(p^2\mu))$.
It vanishes in the stated limit, confirming the assumed outer
scaling.

\subsection{Double-trumpet geometry and winding}

The leading propagators in each spin sector admit an image
representation on the double trumpet,
\be
{\rm d}s^2={\rm d}\rho^2+\cosh^2\rho\,{\rm d}\sigma^2,
\qquad \sigma\sim\sigma+b_{\rm w},
\qquad b_{\rm w}=2p\beta\mu.
\label{eq:extensive-winding-metric}
\ee
Here $\rho$ and $\sigma$ are the radial and angular coordinates.
Matching the solution fixes the uniform boundary reparametrizations
$\sigma_L(x)=b_{\rm w}x/\beta$ and
$\sigma_R(\beta-y)=b_{\rm w}(\beta-y)/\beta$.
The extensive limit keeps $b_{\rm w}/p$ finite.
For a fermion of scaling dimension $1/p$, the covering-space
$LR$ contribution with winding number $n$ has the leading form
\be
\left[
\frac{b_{\rm w}^2}
{4(\beta\mathcal J)^2
\cosh^2\!\left[
\frac{b_{\rm w}}2
\left(\frac{x+y}{\beta}-1+n\right)
\right]}
\right]^{1/p}
\simeq
e^{-\frac{b_{\rm w}}p
\left|\frac{x+y}{\beta}-1+n\right|}.
\label{eq:extensive-winding-geodesic-kernel}
\ee
This is the winding image construction of
Ref.~\cite{Bettaque:2026vpl}.
Raising the zero-winding covering-space kernel to the $p$th power
reproduces the peak profile and normalization in
\eqref{eq:extensive-sd-inner-solution}.
For insertions on the same boundary, the image kernel instead
contains $\sinh^2$ in the denominator, depends on the separation
$x-y+n\beta$, and carries the fermionic ordering sign.
At fixed separated points, its leading large-$p$ form is
$\sign(x-y+n\beta)
e^{-(b_{\rm w}/p)|(x-y)/\beta+n|}$.

Although the bottleneck circumference grows linearly with $p$,
the suppression per winding approaches a nonzero value:
$e^{-b_{\rm w}/p}\to1-2\nu$.
Since the image tails decay exponentially, we can interchange
the large-$p$ limit and the winding sums away from coincident
insertions on the same boundary.
Antiperiodic boundary conditions give the $n$th image a factor
$(-1)^n$, while periodic boundary conditions give every image
unit weight. The leading propagators are therefore
\be
\begin{aligned}
G_{\rm AP}^{LR}(x,y)
&\simeq\sum_{n\in\mathbb Z}(-1)^n
e^{-\frac{b_{\rm w}}p
\left|\frac{x+y}{\beta}-1+n\right|},\\
G_{\rm P}^{LR}(x,y)
&\simeq\sum_{n\in\mathbb Z}
e^{-\frac{b_{\rm w}}p
\left|\frac{x+y}{\beta}-1+n\right|},\\
G_{\rm AP}^{LL}(x,y)
&\simeq\sum_{n\in\mathbb Z}(-1)^n
\sign(x-y+n\beta)
e^{-\frac{b_{\rm w}}p
\left|\frac{x-y}{\beta}+n\right|},\\
G_{\rm P}^{LL}(x,y)
&\simeq\sum_{n\in\mathbb Z}
\sign(x-y+n\beta)
e^{-\frac{b_{\rm w}}p
\left|\frac{x-y}{\beta}+n\right|}.
\end{aligned}
\label{eq:extensive-winding-sums}
\ee
Summing these geometric series reproduces
\eqref{eq:extensive-sd-sector-resolvents} in both spin sectors.

At leading order, the saddle condition gives
$(1-\nu)/(1+e^{-b_{\rm w}/p})
=\nu/(1-e^{-b_{\rm w}/p})=1/2$.
In the weighted propagators $G^{ab}$, the coefficients of the decaying
exponentials therefore add to one, while those of the growing
exponentials cancel. This recovers
\eqref{eq:extensive-sd-solved}.
At this order, the total contribution of nonzero windings
vanishes after summing the images and combining the two spin
structures. Each spin sector separately retains its winding
contributions.

\section{Numerical analysis at finite $p$}
\label{sec:finitepnum}

We consider numerical solutions of the SYK wormhole Schwinger-Dyson equations \eqref{eq:app-resolvents} at finite $p$, and then compare them to the analytical results in the double-scaling limit. Unlike the finite-difference method employed in \cite{Bettaque:2026vpl}, we use a spectral method to compute the saddle point solution with higher accuracy. A more sophisticated initial value based on the approximate high-temperature solution is also introduced. 

\subsection{Spectral decomposition}

 To derive the spectral method, it is instructive to first rewrite the (matrix-valued) Schwinger-Dyson equations \eqref{eq:app-resolvents} as
 \be
 \begin{aligned}
    \label{eq:SD1_AP_P}
    \frac{1}{2} \partial_{\tau_1} G_\mathrm{AP}(\tau_1, \tau_2) - \int_0^\beta d \tau_3 \, \Sigma(\tau_1, \tau_3) \, G_\mathrm{AP}(\tau_3, \tau_2) &= \delta_\mathrm{AP}(\tau_1 - \tau_2) \, I_2 \\
     \frac{1}{2} \partial_{\tau_1} G_\mathrm{P}(\tau_1, \tau_2) - \int_0^\beta d \tau_3 \, \Sigma(\tau_1, \tau_3) \, G_\mathrm{P}(\tau_3, \tau_2) &= \delta_\mathrm{P}(\tau_1 - \tau_2) \, I_2,
\end{aligned}
\ee
where we identified $\Sigma \equiv \Sigma_\mathrm{AP} \equiv \Sigma_\mathrm{P}$ and introduced (anti-)periodic delta functions 
\be
    \delta_\mathrm{AP}(\tau) = \sum_{m \in \mathbb{Z}} (-1)^m \, \delta(\tau - m \beta), \quad \delta_\mathrm{P}(\tau) = \sum_{m \in \mathbb{Z}} \delta(\tau - m \beta).
\ee 
We can now Fourier-transform both equations into their respective spectral decompositions. The resulting equations will still involve the convolution of non-trivial matrices due to the lack of time translation-invariance, but iteratively solving them in Fourier space achieves a considerably higher degree of accuracy compared to ordinary finite-difference methods, at the small cost of two additional Fourier transforms.

Since  the two-point functions $G_s$ have definite (anti-)periodicity in these equations, they can be expanded in terms of modes with either odd or even Matsubara frequency $\omega_n = n \pi /\beta$:
\be
\begin{aligned}
    G_\mathrm{AP}(\tau_1, \tau_2) &= \sum_{k, l}  \widehat{G}_\mathrm{AP}(\omega_{2k+1}, \omega_{2l+1}) \, e^{-i \omega_{2k+1} \tau_1} \, e^{i \omega_{2l+1} \tau_2}, \\
     G_\mathrm{P}(\tau_1, \tau_2) &= \sum_{k, l}  \widehat{G}_\mathrm{P}(\omega_{2k}, \omega_{2l}) \, e^{-i \omega_{2k} \tau_1} \, e^{i \omega_{2l} \tau_2}.
\end{aligned}
\ee
The same is also possible for the (anti-) periodic delta functions, which become diagonal in Fourier space:
\be
\begin{aligned}
    \delta_\mathrm{AP}(\tau_1 - \tau_2) = \frac{1}{\beta} \sum_{k} e^{-i \omega_{2k+1} (\tau_1 - \tau_2)}, \quad
     \delta_\mathrm{P}(\tau_1 - \tau_2) =  \frac{1}{\beta} \sum_{k} e^{-i \omega_{2k} (\tau_1 - \tau_2)}.
\end{aligned}
\ee
The only exception is $\Sigma$, which must necessarily contain both periodic and anti-periodic sectors, and must therefore be expanded as
\be
    \Sigma(\tau_1, \tau_2) = \sum_{n, m}  \widehat{\Sigma}(\omega_n, \omega_m) \, e^{-i \omega_n \tau_1} \, e^{i \omega_m \tau_2}.
\ee
This means that both even and odd Matsubara frequencies occur in the convolution integrals in \eqref{eq:SD1_AP_P}, which leads to a nontrivial overlap over the $[0, \beta)$ interval. This overlap between individual modes is of the form
\be
    \beta \, T_{nm} \coloneqq \int_0^\beta d \tau \, e^{i (\omega_n - \omega_m) \tau} = \begin{cases}
        \beta \, \delta_{nm}, \quad &\text{$n$, $m$ both even/odd,} \\
        \frac{2 i}{\omega_n - \omega_m}, & \text{else.}
    \end{cases}
\ee
The matrix $T$ whose elements are these overlaps is therefore of the form $T \equiv I + H$, where $I$ is the identity and $H$ is a Hermitian (infinite) matrix whose non-zero entries (those for which $n$ and $m$ differ in parity) are
\be
    H_{nm} = \frac{2i}{\beta (\omega_n - \omega_m)}.
\ee
It is straightforward to show that this matrix is invertible and squares to the identity $I$. Indeed, it can be considered to be a discrete Hilbert transform which maps bosonic degrees of freedom to fermionic ones, and vice versa. On the principal interval in Euclidean time, this transformation corresponds to applying a factor of $\sign(\tau)$ to each individual Fourier mode in the expansion.

Under these considerations, and by further defining $\Omega_{nm} = \omega_n \delta_{nm}$ to be the diagonal matrix of Matsubara frequencies, the Schwinger Dyson equations \eqref{eq:SD1_AP_P} can be fully recast as (infinite) block matrix equations\footnote{Note the insertion of an additional overlap matrix $T$ on the left of $\widehat{\Sigma}$ due to the integral necessary to project onto a single spectral matrix element.}:
\be
\begin{aligned}
    \left[-\frac{1}{2} i \beta (I_2 \otimes \Omega) \widehat{G}_\mathrm{AP} - \beta^2 (I_2 \otimes T)  \widehat{\Sigma} (I_2 \otimes T) \widehat{G}_\mathrm{AP}\right](\omega_{2k+1}, \omega_{2l+1}) &= \delta_{kl} \, I_2, \\
    \left[-\frac{1}{2} i \beta (I_2 \otimes \Omega) \widehat{G}_\mathrm{P} - \beta^2 (I_2 \otimes T) \widehat{\Sigma} (I_2 \otimes T) \widehat{G}_\mathrm{P}\right](\omega_{2k}, \omega_{2l}) &= \delta_{kl} \, I_2.
\end{aligned}
\ee
Solving these equations means inverting only the purely antiperiodic or purely periodic support of the operator, but not both at once. Doing so then leaves us with
\be
\begin{aligned}
    \widehat{G}_\mathrm{AP} = \left[-\frac{1}{2} i \beta (I_2 \otimes \Omega) - \beta^2 (I_2 \otimes T)  \widehat{\Sigma} (I_2 \otimes T)\right]_\mathrm{AP}^{-1}, \\
    \widehat{G}_\mathrm{P} = \left[-\frac{1}{2} i \beta (I_2 \otimes \Omega) - \beta^2 (I_2 \otimes T)  \widehat{\Sigma} (I_2 \otimes T)\right]_\mathrm{P}^{-1}.\\
\end{aligned}
\ee
These expressions can be evaluated numerically by introducing an appropriate cutoff on the spectrum. Assuming the number of total modes to be $L$, then this involves inverting two roughly $L \times L$ block matrices, one for the $\sim L/2$ even modes and one for the $\sim L/2$ odd modes. This can be further reduced to the inversion of two $L/2 \times L/2$ matrices by imposing and then block-diagonalizing a particular choice of replica structure, as discussed in \cite{Bettaque:2026vpl}. The second Schwinger-Dyson equation \eqref{eq:app-self-energy} can then be iteratively applied in Euclidean time after performing the inverse (discrete) Fourier transform.

\subsection{Initial value from the high-temperature limit}

\begin{figure}[h]
\centering
\includegraphics[width=.9\textwidth]{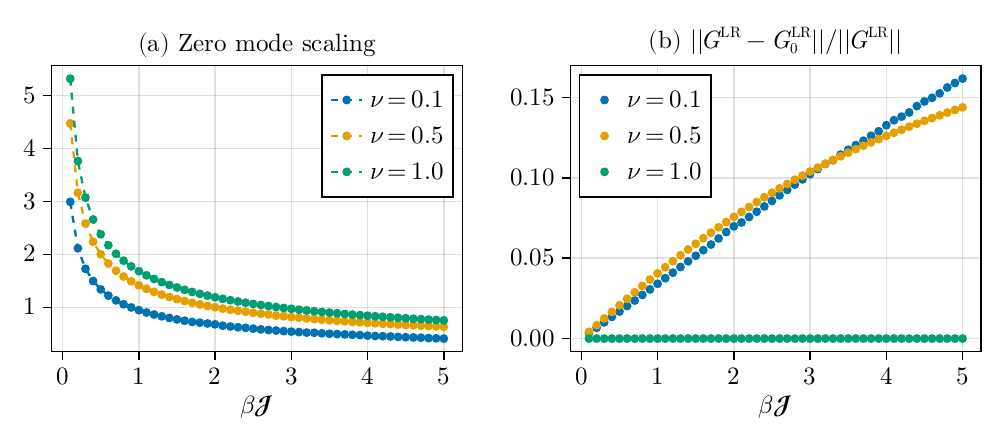}
\caption{Figure (a) depicts the zero mode of $G^{LR} = -G^{RL}$ as determined from finite-$p$ numerics ($p = 4$, scatter points), as well as its analytical high-temperature approximation $G_0^{LR} = (2 p \nu / (\beta \mathcal{J})^2)^{1/p}$ (dashed lines). Figure (b) shows the relative Frobenius-distance between the full numerical solution $G^{LR}$ and the zero-mode approximation $G_0^\mathrm{LR}$. Both figures indicate great agreement between the high-temperature ansatz \eqref{eq:G_high_temp} and actual numerics, which make it a suitable initial value. In particular for $\nu = 1$, $G^{LR}$ seems to be fully determined by the zero mode $G_0^{LR}$.}
\label{fig:G_zero_mode}
\end{figure}

The choice of initial value is essential to retrieve good results, especially for the setup considered here. Without taking the large-$p$ limit, a reasonable ansatz is to choose the approximate high-temperature solutions to the Schwinger-Dyson equations, which one might expect to be determined by
\be
    \frac{1}{2} \partial_{\tau_1} G_\mathrm{AP}(\tau_1, \tau_2) \stackrel{?}{=} \delta_{AP}(\tau_1 - \tau_2) \, I_2, \quad \frac{1}{2} \partial_{\tau_1} G_\mathrm{P}(\tau_1, \tau_2) \stackrel{?}{=} \delta_{P}(\tau_1 - \tau_2) \, I_2.
\ee
Just as in ordinary SYK, the first equation is naturally satisfied by $G^{LL}_\mathrm{AP}(\tau_1, \tau_2) = \sign(\tau_1 - \tau_2)$, and $G^{LR}_\mathrm{AP}(\tau_1, \tau_2) = 0$\footnote{The cross-replica correlations don't admit a non-zero constant due to antiperiodicity.}, but the second equation is actually not satisfiable in its current form. To see this, it suffices to consider the antiderivative of $\delta_{P}(\tau)$, which is also equal to $\sign(\tau)/2$ over the principal domain $(-\beta, \beta)$, but is not $\beta$-periodic (and is actually unbounded) due to the presence of the integrated zero-mode term $\tau / \beta$. Removing this term restores both properties, which means that
\be
    G^{LL}_\mathrm{P}(\tau_1, \tau_2) = \sign(\tau_1 - \tau_2) - \frac{2 (\tau_1 -\tau_2)}{\beta}, \quad G^{LR}_\mathrm{P}(\tau_1, \tau_2) = c
\ee
is a reasonable ansatz for the "free" periodic two-point function, and also aligns with the structure of \eqref{eq:app-LR-convention} in the large-$p$ limit. Still, this ansatz does not actually solve the equation because 
\be
    \frac{1}{2} \partial_{\tau_1} G^{LL}_\mathrm{P}(\tau_1, \tau_2) = \delta_{P}(\tau_1 - \tau_2) - \frac{1}{\beta},
    \label{eq:free_G_P}
\ee
and so one necessarily needs to consider additional contributions to the relevant Schwinger-Dyson equation. Inserting \eqref{eq:free_G_P} into the periodic equation \eqref{eq:SD1_AP_P} yields
\be
    \int_0^\beta d \tau_3 \, \Sigma(\tau_1, \tau_3) \, G_\mathrm{P}(\tau_3, \tau_2) \sim - \frac{1}{\beta} \, I_2.
\ee
While the left-hand side expression contains a few time-dependent integrals, we assume that they don't contribute significantly in the $\beta \rightarrow 0$ limit, as further suggested by the right-hand side being constant. Only the integrals over inherently constant terms are of relevance:
\be
    G^{LR} = (1-\nu) G^{LR}_\mathrm{AP} + \nu G^{LR}_\mathrm{P} = \nu c \quad  \Longrightarrow \quad \Sigma^{LR} = \frac{\mathcal{J}^2}{2p} (G^{LR})^{p-1} = \frac{\mathcal{J}^2 (\nu c)^{p-1}}{2p}.
\ee
Performing the matrix multiplication and integrating the constants then results in the following constraint equation for $G^{LR}_\mathrm{P} = c$:
\be
    -\beta \Sigma^{LR} G^{LR}_\mathrm{P} \sim - \frac{1}{\beta},
\ee
which when solved gives
\be
    G^{LR}_\mathrm{P} \sim \frac{1}{\nu} \left(\frac{2p\nu}{(\beta \mathcal{J})^2}\right)^{1/p}.
\ee
The full high-temperature ansatz for $G$ is therefore
\be
    G(\tau_1, \tau_2) = \begin{pmatrix}
        \sign(\tau_1 - \tau_2) - \frac{2 \nu}{\beta} (\tau_1 -\tau_2) & \left(\frac{2p\nu}{(\beta \mathcal{J})^2}\right)^{1/p} \\
        -\left(\frac{2p\nu}{(\beta \mathcal{J})^2}\right)^{1/p} & \sign(\tau_1 - \tau_2) - \frac{2 \nu}{\beta} (\tau_1 -\tau_2)
    \end{pmatrix},
    \label{eq:G_high_temp}
\ee
which reproduces \eqref{eq:app-LR-convention} in the limit of $p \rightarrow \infty$ with $\kappa = p \, \nu$ and $\beta \mathcal{J}$ fixed. At finite $p$, however, the off-diagonal (zero mode) entries become singular when $\beta \mathcal{J} \rightarrow 0$, which means they will grow past the diagonal entries and dominate the solution. To achieve numerical convergence in these regimes -- in particular at larger $p$ -- the initial value for $G$ must therefore account for this. Figure \ref{fig:G_zero_mode} shows how the (constant) cross-replica correlations of this ansatz match the numerical results with great precision at high temperatures for $p = 4$, but the same is true for all $p$. In fact, the off-diagonal constant 
\be
    G_0^{LR}  = \left(\frac{2p\nu}{(\beta \mathcal{J})^2}\right)^{1/p}
\ee
determines the zero-mode of $G^{LR}$ to leading order, and even all of $G^{LR}$ exactly when $\nu = 1$.

Note, however, that this constant vanishes at fixed $p$ and $\beta \mathcal{J} \to \infty$, and so it ceases to be a useful initial value. This does not mean the saddle point becomes singular, though, as it is more and more supported by higher-order modes instead. For numerical purpose, it is therefore more reasonable to choose a cross-replica initial value that captures the singular behavior of $G_0^\mathrm{LR}$ at high temperatures, but does not go below a certain threshold at low temperatures, e.g.:
\be
    G_\mathrm{init}^{LR} \sim \max(G_0^{LR}, 1).
\ee

\subsection{Results}

A numerical ansatz based on these considerations allows us to probe finite-$p$ saddles with high precision and at temperatures that manage to probe the SYK model's conformal regime. Here we focus on the structure of the saddle two-point function $G$, see \cite{Bettaque:2026vpl} for more detail on the actual distribution of thermal one-point function moments.

\begin{figure}[h!]
\centering
\includegraphics[width=\textwidth]{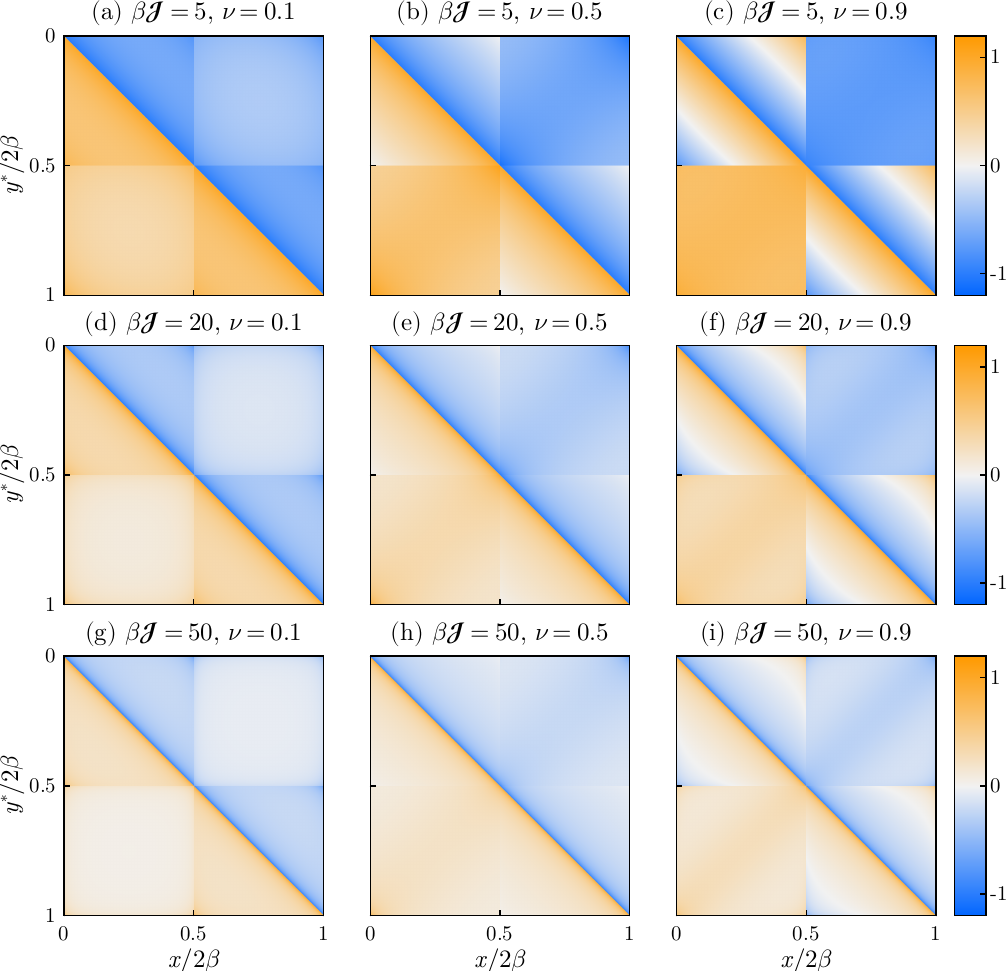}
\caption{Overview of numerical solutions for the (both same- and cross-replica) two-point function $G$, with $p = 4$ and varying values of $\beta \mathcal{J}$ and $\nu$. All three regimes of backreaction can be observed in $G^\mathrm{LR}$, even though they arise in regimes where the large-$p$ ansatz is not strictly applicable. Note that we take $G^{LR}$ to be negative here to highlight the emergence of an effective one-replica two-point function with inverse temperature $2 \beta$ at e.g. $\nu = 0.5$ and large $\beta \mathcal{J}$}
\label{fig:G_landscape}
\end{figure}

Fig. \ref{fig:G_landscape} provides an overview of the $p=4$ SD solutions for $\beta \mathcal{J} = 5, 20, 50$ and $\nu = 0.1, 0.5, 0.9$. The diagonal blocks show $G^{LL}$ and $G^{RR}$ while the off-diagonal blocks show $G^{LR}$ and $G^{RL}$. For example, panel (h) shows $\nu=0.5$ and $\beta \mathcal{J}=50$ and demonstrates an emergent time-translation symmetry in which the two separate thermal circles effectively fuse to form one thermal circle of extent $2\beta$. As discussed in \cite{Bettaque:2026vpl}, this is a manifestation of the identity $\sum_{\psi_W} \langle \Tr(e^{-\beta H} \psi_W)^2 \rangle_J \propto \langle \Tr(e^{-2 \beta H}) \rangle_J$. When performing the sum over operators on the left-hand side by saddle point, one finds that the sum is controlled by operators of a critical weight $w = \nu_* N$ with $\nu_* \to .5$ at large $\beta \mathcal{J}$. This gives $\binom{N}{\nu_* N} \langle \Tr(e^{-\beta H} \psi_{W | w = \nu_* N})^2 \rangle_J \sim \langle \Tr(e^{-2 \beta H}) \rangle_J$.

\begin{figure}[h!]
\centering
\includegraphics[width=\textwidth]{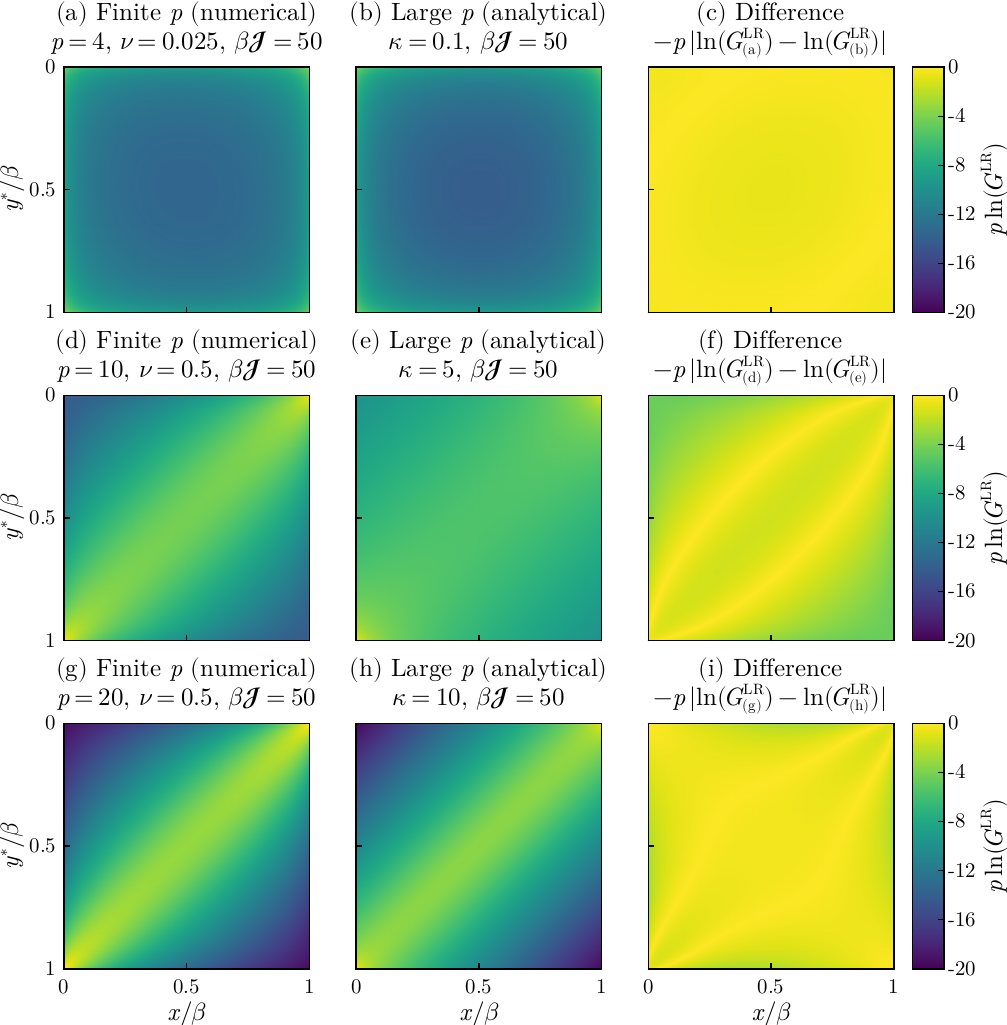}
\caption{Direct comparison between $p \ln(G^\mathrm{LR}(x, y))$ as obtained through finite-$p$ numerics, and the analytical large-$p$ solution $h(x, y)$. Both are evaluated at the same $\beta \mathcal{J}$ and for the same value of $\kappa = p \, \nu$. We observe great agreement between the two methods -- both at small and large $p$ -- for all three regimes of backreaction, although the intermediate backreaction regime exhibits the largest deviations. In general, both methods are not expected to match over the whole parameter space.}
\label{fig:G_LR_comparison}
\end{figure}

Fig. \ref{fig:G_LR_comparison} shows a comparison of $G^{LR}$ between finite $p$ numerics\footnote{$p$ was chosen to be as small as possible while still allowing one to probe the relevant values for $\kappa$ with $\nu \leq 0.5$. This was done to minimize the numerical error that comes with considering large $p$.} (left column) and large $p$ analytics (middle column) and their difference, plotted by comparing $p \ln G^{LR}$ (right column). All three regimes of backreaction as explored in Appendix \ref{app:exact-coupled-saddle} for large $p$ can also be recreated at finite $p$, with good overall agreement.

\cleardoublepage

\section{Derivation of the endpoint probability formula}
\label{sec:appKrawtchouk}

In this appendix, we provide a derivation of Eq. \eqref{eq:PW-final} used in the main text.

Let $I_1,\ldots,I_K$ be independent, uniformly chosen $p$-subsets of
$\{1,\ldots,N\}$, and let $|W|=w$. In terms of their symmetric difference,
the probability is
\be
P_K(W):=\operatorname{Prob}\!\left(
I_1\oplus\cdots\oplus I_K=W\right).
\label{eq:app-endpoint-probability}
\ee
We derive it by following the set of labels that have appeared an odd
number of times. Adding one interaction string gives a random-walk
step, and diagonalizing this walk leads to the Krawtchouk formula
quoted in the main text.

\paragraph{Single step.}
Write $S_a=I_1\oplus\cdots\oplus I_a$ for the odd support after $a$
insertions. Starting from $S_0=\varnothing$, it evolves by
$S_{a+1}=S_a\oplus I_{a+1}$. The probability of an overlap between a
random interaction set $I$ and any fixed set $W$ is hypergeometric:
\be
\operatorname{Prob}(|I\cap W|=s)
=\frac{\binom{w}{s}\binom{N-w}{p-s}}{\binom{N}{p}}.
\label{eq:hypergeom}
\ee
The numerator counts the choices of $s$ labels inside $W$ and $p-s$
outside it.

Suppose the current support has size $m$ and shares $j$ labels with
the next interaction set. Those $j$ labels are removed, while the
remaining $p-j$ are added. The new support size is therefore
\be
n=m+p-2j.
\label{eq:nmp2j}
\ee
Applying \eqref{eq:hypergeom} to the current support gives the
one-step transition matrix
\be
\mathcal T_{mn}
=\frac{\binom{m}{j}\binom{N-m}{p-j}}{\binom{N}{p}},
\qquad j=\frac{m+p-n}{2}.
\label{eq:onestepkernel}
\ee
The entry vanishes unless $j$ is an integer satisfying
$\max(0,p-N+m)\leq j\leq\min(m,p)$. Since this transition probability
depends only on $m$, the support size is itself a Markov chain.
After $K$ steps, $(\mathcal T^K)_{0w}$ is the probability of reaching
any set of size $w$. All such sets are equally likely, so
\be
P_K(W)=\frac{(\mathcal T^K)_{0w}}{\binom{N}{w}}.
\label{eq:app-size-probability}
\ee
What remains is to evaluate this more explicitly by diagonalizing $\mathcal{T}$.

\paragraph{Diagonalizing the random walk.}
The diagonalization is simplest on the full space of subsets. For a
fixed subset $A$, define the sign function
\be
\chi_A(S):=(-1)^{|A\cap S|},
\qquad
\chi_A(S\oplus I)=\chi_A(S)\chi_A(I).
\ee
This function records whether $S$ contains an even or odd number of
labels from $A$. The second identity shows that averaging over one
step simply multiplies it by a constant. If $|A|=k$, that constant is
\be
\lambda_k
=\mathbb E_I\chi_A(I)
=\frac{1}{\binom{N}{p}}
\sum_{j=0}^{p}(-1)^j\binom{k}{j}\binom{N-k}{p-j}.
\label{eq:app-eigenvalue}
\ee
Thus each $\chi_A$ is an eigenfunction of the walk, with an eigenvalue
depending only on $|A|$.

To obtain eigenvectors for the support-size chain, sum these
functions over all sets $A$ of size $k$. For $|S|=n$, the sum depends
only on $n$ and defines the Krawtchouk polynomial:
\be
K_k(n;N)
:=\sum_{r=0}^{k}(-1)^r\binom{n}{r}\binom{N-n}{k-r}
=\sum_{|A|=k}\chi_A(S).
\label{eq:Kt-def}
\ee
Here $r$ counts the labels of $A$ that lie in $S$, and binomial
coefficients outside their usual integer range are understood to
vanish. Every term in the character sum has the same eigenvalue;
consequently,
\be
\sum_{n=0}^{N}\mathcal T_{mn}K_k(n;N)
=\lambda_k K_k(m;N),
\qquad
\lambda_k=\frac{K_p(k;N)}{\binom{N}{p}}.
\label{eq:eigenvalue}
\ee
The second equality follows by comparing
\eqref{eq:app-eigenvalue} with \eqref{eq:Kt-def}.

The orthogonality relation follows from the same construction.
Summing $\chi_A(S)\chi_B(S)$ over all subsets $S$ gives $2^N$ when
$A=B$ and zero otherwise. Summing also over $A$ and $B$ at fixed sizes,
and grouping the sets $S$ by their size, gives
\be
\sum_{n=0}^{N}\binom{N}{n}K_k(n;N)K_{k'}(n;N)
=2^N\binom{N}{k}\delta_{kk'}.
\label{eq:K-orth}
\ee

\paragraph{Endpoint probability.}
The eigenvalues and orthogonality relation now determine the
$K$-step transition probability. Using
$K_k(0;N)=\binom{N}{k}$ for the empty initial support, we obtain
\be
(\mathcal T^K)_{0w}
=\frac{\binom{N}{w}}{2^N}
\sum_{k=0}^{N}
\left(\frac{K_p(k;N)}{\binom{N}{p}}\right)^K
K_k(w;N).
\label{eq:TK-0w}
\ee
Dividing by the number of possible endpoints of size $w$, as in
\eqref{eq:app-size-probability}, yields the desired result:
\be
P_K(W)
=\frac{1}{2^N}\sum_{k=0}^{N}
\left(\frac{K_p(k;N)}{\binom{N}{p}}\right)^K
K_k(w;N).
\label{eq:PW-final2}
\ee
This is the exact finite-$N$ formula used in the main text.

The same result can be read directly from the character expansion of
the fixed-endpoint constraint,
\be
\mathbf 1_{\{S=W\}}
=\frac{1}{2^N}\sum_{A\subseteq\{1,\ldots,N\}}
\chi_A(S)\chi_A(W).
\label{eq:app-delta}
\ee
Indeed, independence of the interaction sets gives
$\mathbb E\chi_A(S_K)=\lambda_{|A|}^K$. Averaging
\eqref{eq:app-delta} and grouping the sets $A$ by their size reproduces
\eqref{eq:PW-final2}.

Finally, the derivation of the formula makes the support and parity constraints explicit:
\be
P_K(W)=0\quad\text{unless}\quad
w\leq Kp,
\qquad w\equiv Kp\pmod 2.
\label{eq:app-endpoint-selection}
\ee
At most $Kp$ labels can occur an odd number of times, and their number
has the same parity as the total number of insertions of fermion
labels. These necessary conditions are implemented automatically by
cancellations in the spectral sum.

\section{Analysis in the double-scaling limit}
\label{app:ConditionedEnsemble}

In this appendix, we will introduce the techniques for computing the squared thermal one-point function of the Majorana string $\psi_W$, to leading order in the double-scaling limit. Recall that the first step is to expand in moments
\begin{equation}
\mathcal{C}_{\beta}(W) = \sum_{n,m \geq 0} \frac{(- \beta)^{n+m}}{n! \, m!}\Gamma_{n,m} \, , \qquad  \Gamma_{n,m} =  \left\langle \Tr(H^n \psi_W) \Tr(H^m \psi_W) \right\rangle_J \, .
\end{equation}
The $J$-pairings can be done in many different ways, each of them defines a chord diagram. We focus on a particular chord diagram $D$ now, for which $K$ chords connect the two traces, while $d_L=(n-K)/2$ and $d_R=(m-K)/2$ are disk-chords internal to each trace. Note that $n$, $m$ and $K$ must all have the same parity for this to make sense. Furthermore, $K p \equiv w \; ({\rm mod}\,2)$ and $K p \geq w$, since otherwise the traces just give zero as explained in the main text.

For a given diagram, within each trace, we commute disk-chords until the two endpoints are together where they square to one. Let us introduce some notation to write the result of this procedure. We label $I_a$ with $a = 1, \dots, K$ the cross-chords, $J_b$ with $b = 1, \dots, d_L$ the disk-chords in the first trace, and $L_c$ with $c = 1, \dots, d_R$ the disk-chords in the second trace. We capture which disk-chords have to be commuted (i.e., which pairs of chords intersect) by sets $E_L$ and $E_R$ that denote pairs of disk-chords that cross on the left and right boundaries, respectively. In addition, for each left disk-chord $J_b$, let $S_b^{L} \subseteq \{1, \dots, K\}$ be the set of cross-chords it crosses, and define $S_c^{R}$ analogously. Finally, if the cross-chords occur in the order $I_{\sigma(1)}, \dots, I_{\sigma(K)}$ in the right trace, let $R_{\sigma}$ be the set of pairs whose relative order is reversed with respect to the left trace. In terms of these sets, we can write the contribution of the diagram $D$ to $\Gamma_{n,m}$ as
\begin{equation}
\begin{aligned}
\mathcal{A}(D) & = \left( \frac{\mathcal{J}^2}{\lambda} \right)^{\frac{n+m}{2}} \binom{N}{p}^{- \frac{n + m}{2}} \sum_{I_a, J_b, L_c} (-1)^{Q_D} \Tr (\psi_{I_1} \dots \psi_{I_K} \psi_W) \Tr ( \psi_{I_{\sigma(1)}} \dots \psi_{I_{\sigma(K)}} \psi_W) \\
& = (-1)^{\frac{K p + w}{2}} \left( \frac{\mathcal{J}^2}{\lambda} \right)^{\frac{n+m}{2}} \binom{N}{p}^{- \frac{n + m}{2}} \sum_{I_a, J_b, L_c} (-1)^{Q_D(I,J,L)} (-1)^{X_{\sigma}(I)} \, .
\end{aligned}
\end{equation}
In this expression, the sum over $I_a$ must be restricted to meet the endpoint condition \eqref{eq:xorcondition}
\begin{equation}
I_1 \oplus I_2 \oplus \dots \oplus I_K = W \, ,
\end{equation}
and we have defined the quantities
\begin{subequations}
\begin{align}
\nonumber Q_D (I, J, L) = & \sum_{\{b,b'\} \in E_L} |J_{b} \cap J_{b'} | + \sum_{\{c,c'\} \in E_R} |L_{c} \cap L_{c'} | \\ 
 & + \sum_{b=1}^{d_L} \sum_{a \in S^L_b} |I_a \cap J_b| + \sum_{c=1}^{d_R} \sum_{a \in S^R_c} |I_a \cap L_c| \, , \\
X_{\sigma} (I) = & \sum_{\{a,a'\} \in R_{\sigma}} |I_{a} \cap I_{a'} | \, ,
\end{align}
\end{subequations}
which take care of signs from commuting fermion strings.

The idea in the following is to treat the sum over $p$-index sets as a probabilistic computation in a conditioned ensemble. For the disk-chords, we are sampling uniformly at random from all $p$-subsets, but for the cross-chords we impose the endpoint condition. So, given sets $I_a$, $J_b$, $L_c$, their probability is
\begin{equation}
\mathbb{P}_{K,W}(I,J,L) = \binom{N}{p}^{- d_L - d_R} \frac{1}{\mathcal{N}_{K,w}} \mathbf{1}_{I_1 \oplus \dots \oplus I_K = W} \, .
\end{equation}
Here, we have defined $\mathcal{N}_{K,w}$ as the number of $K$ $p$-subsets that meet the endpoint condition, and the binomial factor is just the normalization by the total number of disk-chords. Note that the probability that $K$ cross-chords meet the endpoint condition sampling from all possible sets of $K$ cross-chords is
\begin{equation}
\label{ConditionedEnsemble:PKfromNK}
P_K(W) = \binom{N}{p}^{-K}  \mathcal{N}_{K,w} \, .
\end{equation}
This quantity was introduced in the main text \eqref{eq:PKWdefinition} and computed exactly (at finite $N$) in Appendix \ref{sec:appKrawtchouk}; we will obtain it again below in the relevant limits for the current discussion. Using this probability distribution over $p$-subsets, we can write the value of the chord diagram as an expectation value weighted by $P_K(W)$:
\begin{equation}
\label{ConditionedEnsemble:GeneralDiagramForm}
\mathcal{A}(D) = (-1)^{\frac{K p + w}{2}} \left( \frac{\mathcal{J}^2}{\lambda} \right)^{\frac{n+m}{2}} P_K(W) \, \mathbb{E}_{K,W}^{(I, J, L)} \left[ (-1)^{Q_D(I, J, L)} (-1)^{X_{\sigma}(I)} \right] \, .
\end{equation}
In the following we will assume $p, w \in 4\mathbb{Z}_{>0}$; consequently, the sign $(-1)^{(Kp+w)/2}$ is trivial.

\subsection{A canonical ensemble of cross-chords}

We will now develop techniques to study the conditioned ensemble of cross-chords. We will characterize a given microscopic configuration via sets $M_i \subseteq \{1, \dots, K\}$ which indicate in which cross-chords the Majorana $\psi_i$ appears, for $i = 1, \dots, N$. Alternatively, it will occasionally be useful to refer to the binary vectors
\begin{equation}
X_i = \sum_{a \in M_i} e_a \in \{0, 1\}^K \, ,
\end{equation}
where $e_a$ is the unit Cartesian vector in the $a$-th direction --- $X_i$ is thus a vector with $1$'s indicating in which cross-chords $\psi_i$ appears. The endpoint condition is translated into the following requirement: if $i \in W$, $|M_i|$ must be odd, while if $i \notin W$, $|M_i|$ must be even. Consider the space $\mathcal{M}$ of all possible such $\{M_i\}$. For each configuration $M \in \mathcal{M}$, we can determine the total number of fermions in a given cross-chord as
\begin{equation}
d_a(M) = |\{i \; / \; a \in M_i\}| = |I_a| \, , \qquad a=1, \dots, K \, .
\end{equation}
The conditioned ensemble of cross-chords does not include all configurations in $\mathcal{M}$ because the chords are required to have length $p$, so for valid configurations $d_a(M) = p$ for all $a = 1, \dots, K$. We will see this as analogous to a microcanonical conditioning: the charges we are fixing here are the total number of fermions per chord. In this language, we can write the microcanonical partition function and identify it with the total number of valid configurations
\begin{equation}
\mathcal{N}_{K,w} = \sum_{M \in \mathcal{M}} \prod_{a=1}^K \delta_{d_a(M),p} \, .
\end{equation}

The advantage of this language is that the microcanonical conditioning is the only reason different fermion flavors are coupled. The analysis will become simpler if we go to a canonical ensemble in which we introduce chemical potentials $e^{\mu_a} := z_a$ for the charges $d_a$. The canonical partition function is now
\begin{equation}
Z(z) = \sum_{M \in \mathcal{M}} \prod_{a=1}^K z_a^{d_a(M)} = \sum_{M \in \mathcal{M}} \prod_{i=1}^N \prod_{a \in M_i} z_a = \prod_{i=1}^N \left( \sum_{M_i \subseteq \{1, \dots, K\}} \prod_{a \in M_i} z_a \right) \, .
\end{equation}
The last step shows that the fermions behave independently in the canonical ensemble: each of them can appear in any subset of cross-chords, and we introduce a factor of $z_a$ for each chord in which it appears. Note that we are implicitly imposing $|M_i|$ to be even or odd depending on whether $i \in W$ or $i \notin W$. Given this parity condition, a more compact form to write the partition function is
\begin{equation}
Z (z) = O(z)^w E(z)^{N-w} \, , 
\end{equation}
where $O(z)$ and $E(z)$ are respectively the combination of elementary symmetric polynomials of odd and even degrees, up to order $K$
\begin{subequations}
\begin{align}
O(z) & = \frac{1}{2} \left[ \prod_{a=1}^K (1 + z_a) - \prod_{a=1}^{K} (1-z_a) \right] = \sum_a z_a + \sum_{a_1 < a_2 < a_3} z_{a_1} z_{a_2} z_{a_3} + \dots \, , \\
E(z) & = \frac{1}{2} \left[ \prod_{a=1}^K (1 + z_a) + \prod_{a=1}^{K} (1-z_a) \right] = 1 + \sum_{a_1 < a_2} z_{a_1} z_{a_2} + \dots \, .
\end{align}
\end{subequations}

In the canonical ensemble with fugacities $z = (z_1, \dots, z_K)$, the probability of a given configuration $M \in \mathcal{M}$ is
\begin{equation}
\mathbb{P}_z(M) = \frac{\prod_{a=1}^K z_a^{d_a(M)}}{Z(z)} \, ,
\end{equation}
and we denote by $\mathbb{E}_z[\cdot]$ expectation values computed with respect to this probability distribution. As usual, in order to use the canonical ensemble to extract information about the relevant, microcanonical setup, we need to impose that the expectation value of the charges reproduces the microcanonical values. This gives
\begin{equation}
p = \mathbb{E}_z[d_a] = z_a \partial_{z_a} \log Z(z) = w z_a \frac{\partial_{z_a} O(z)}{O(z)} + (N - w) z_a \frac{\partial_{z_a} E(z)}{E(z)} \, .
\end{equation}
The fugacities $z_a$ are fixed by solving these $K$ equations. Since all $K$ cross-chords are equivalent, it is natural to look for a symmetric solution $z_a = z_{\star}$ for all $a$. Such solution exists and has $z_{\star} \ll 1$, so expanding $O(z)$ and $E(z)$ at small values of the arguments we can find it explicitly:
\begin{equation}
z_{\star}^2 = \frac{K p - w}{N K(K-1)} + O(p^{-2}) = \frac{K - \Delta}{K(K-1)} \frac{p}{N} + O(p^{-2}) \, .
\end{equation}
We see that $z_{\star} \sim O(p^{-1/2})$ is indeed small. In this and the following analysis, we assume $K \geq 2$ and $0 \leq \Delta < K$;  $K = 0, 1$ are elementary exceptional cases while the possible limiting case $K = \Delta$ will be analyzed separately at the end.

We will now use the canonical ensemble with $z_a = z_{\star}$ to extract information relevant for our problem, which really involves the microcanonical ensemble. There is, of course, an assumption about ensemble equivalence behind this logic; proving that this holds requires us to show that fluctuations in the canonical ensemble are under control. Given the technical nature of the argument, we postpone it until the end of the appendix, and just assume ensemble equivalence whenever relevant in what follows. The first result we can extract is the total number of microcanonical configurations $\mathcal{N}_{K,w}$ or, equivalently, the probability $P_K(W)$. In the canonical ensemble, the probability that the occupation numbers of the cross-chords are all $p$ is
\begin{equation}
\mathbb{P}_{z_{\star}}(d_1 = \dots = d_K = p) = \mathcal{N}_{K,w} \frac{z_{\star}^{Kp}}{Z(z_{\star})} \, .
\end{equation}
Taking logarithms and evaluating the canonical partition function at $z_{\star}$, we get
\begin{equation}
\label{ConditionedEnsemble:NumberConfigurationsLeadingTerms}
\begin{aligned}
\log \mathcal{N}_{K,w} & = \log Z(z_{\star}) - K p \log z_{\star} +  \log \mathbb{P}_{z_{\star}}(d_1 = \dots =d_K = p) \\
& = \frac{K - \Delta}{2} p \log p + p \left[ \Delta \log K + \frac{K - \Delta}{2} \left( 1 + \log \frac{2 K(K-1)}{\lambda (K-\Delta)} \right) \right] + O(\log p) \, ,
\end{aligned}
\end{equation}
where we used that $\mathbb{P}_{z_{\star}}(d_1 = \dots = d_K = p)$ is only polynomial in $p$, a result we will show when discussing ensemble equivalence. Using Stirling's formula we can also obtain
\begin{equation}
\label{ConditionedEnsemble:EndpointProbabilityLeadingTerms}
\begin{aligned}
\log P_K(W) & = \log \mathcal{N}_{K,w} - K \log \binom{N}{p} \\
& = - \frac{K + \Delta}{2} p \log p + p \left[ \Delta \log K + \frac{K-\Delta}{2} \log \frac{K(K-1)}{K-\Delta} \right. \\
& \left. \qquad \qquad \qquad \qquad \qquad \qquad - \frac{K + \Delta}{2} \left( 1 + \log \frac{2}{\lambda} \right) \right] + \mathcal{O} (\log p) \, .
\end{aligned}
\end{equation}

These must be supplemented with exact results in the case $\Delta \in \mathbb{Z}_{>0}$ and $K = \Delta$. In that situation, since $K p = \Delta p = w$, the cross-chords provide a partition of the Majoranas in $\psi_W$ in $K$ sets of $p$ elements. Thus
\begin{equation}
\mathcal{N}_{\Delta, \Delta p} = \frac{(\Delta p)!}{(p!)^{\Delta}} \; \Rightarrow \; \log \mathcal{N}_{\Delta, \Delta p} = \Delta p \log \Delta + O(\log p) \, ,
\end{equation}
and 
\begin{equation}
\label{ConditionedEnsemble:EndpointProbabilityAtThreshold}
P_{\Delta}(W) = \frac{(\Delta p)!}{(p!)^{\Delta} \binom{N}{p}^{\Delta}} \; \Rightarrow \; \log P_{\Delta}(W) = - \Delta p \log p + \Delta p \left[ \log \Delta - 1 - \log \frac{2}{\lambda} \right] + O(\log p) \, ,
\end{equation}
which reproduce the previous expressions for $K = \Delta$.

We can also get an idea of how the fermions are distributed between the cross-chords using this canonical ensemble. For a given flavor $i$, we have
\begin{equation}
\mathbb{P}_{z_{\star}}(|M_i| = m) = \begin{cases}
    \displaystyle \binom{K}{m} \frac{z_{\star}^{m}}{O(z_{\star})} & {\rm if}\; i \in W \; {\rm and}\; m \in 2 \mathbb{N} + 1 \, , m \leq K \, ,\\
    \displaystyle \binom{K}{m} \frac{z_{\star}^{m}}{E(z_{\star})} & {\rm if}\; i \notin W \; {\rm and}\; m \in 2 \mathbb{N} \, , m \leq K \, . \end{cases}
\end{equation}
We can define variables keeping track of how many fermions appear in a number $m$ of cross-chords:
\begin{equation}
n_m^W = |\{i \in W \; / \; |M_i| = m\}| \quad (m \;{\rm odd}) \, , \qquad n_m^{\bar{W}} = |\{i \notin W \; / \; |M_i| = m\}| \quad (m \;{\rm even})\, .
\end{equation}
Using the above probabilities, we can compute the expectation values of these variables as
\begin{equation}
\mathbb{E}_{z_{\star}} [n_m^W] = w \binom{K}{m} \frac{z_{\star}^{m}}{O(z_{\star})} \, , \qquad \mathbb{E}_{z_{\star}} [n_m^{\bar{W}}] = (N - w) \binom{K}{m} \frac{z_{\star}^{m}}{E(z_{\star})} \, ,
\end{equation}
We can expand these results for the known small value of $z_{\star} \sim O(p^{-1/2})$, but to get an idea of the distribution it is enough to read the scaling with $p$. For fermions in $W$, we get
\begin{equation}
\mathbb{E}_{z_{\star}} [n_1^W] = O(p) \, , \qquad \mathbb{E}_{z_{\star}} [n_3^W] = O(1) \, , \qquad \mathbb{E}_{z_{\star}} [n_5^W] = O(p^{-1}) \, ,
\end{equation}
so most of the fermions appear in a single cross-chord, but there is an order $1$ number of them which appear in three. Similarly, for fermions not in $W$
\begin{equation}
\mathbb{E}_{z_{\star}} [n_0^{\bar{W}}] = O(p^2) \, , \qquad \mathbb{E}_{z_{\star}} [n_2^{\bar{W}}] = O(p) \, , \qquad \mathbb{E}_{z_{\star}} [n_4^{\bar{W}}] = O(1) \, ,
\end{equation}
so most fermions outside $W$ do not appear at all, and those which appear do it overwhelmingly in pairs. To leading order, we can imagine we have $\sim w$ fermions in $W$ appearing in a single cross-chord, and $\sim (Kp-w)/2$ fermions outside $W$ appearing in two cross-chords.

Similarly, it will be useful below to define the number of fermions that appear just in a specific cross-chord $a$, or in a pair $\{a_1, a_2\}$ of cross-chords:
\begin{equation}
A_{a}^W = |\{i \in W \; / \; M_i = \{a\}\}| \, , \qquad A_{a_1 a_2}^{\bar{W}} = |\{i \notin W \; / \; M_i = \{a_1, a_2\}\}| \, .
\end{equation}
Each of the Majorana flavors is independent in the canonical ensemble, so these are just binomial variables with probability of success 
\begin{subequations}
\begin{align}
\mathbb{P}_{z_{\star}}(M_i = \{a\}) & = \frac{z_{\star}}{O(z_{\star})} = \frac{1}{K} + O(p^{-1}) \, , \\
\mathbb{P}_{z_{\star}}(M_i = \{a_1, a_2\}) & = \frac{z_{\star}^2}{E(z_{\star})} = \frac{K-\Delta}{K(K-1)} \frac{p}{N} + O(p^{-2}) \, .
\end{align}
\end{subequations}
and $w$ and $N-w$ trials, respectively. Dividing by $p$ and using standard results for the binomial distribution, we get
\begin{equation}
\label{ConditionedEnsemble:SingleChordNumberFermions}
\mathbb{E}_{z_{\star}} \left[ \frac{A_a^W}{p} \right] = \frac{\Delta}{K} + O(p^{-1}) \, , \qquad {\rm Var}_{z_{\star}}\left[ \frac{A_a^W}{p} \right] = O(p^{-1}) \, , 
\end{equation}
and
\begin{equation}
\label{ConditionedEnsemble:TwoChordsNumberFermions}
\mathbb{E}_{z_{\star}} \left[ \frac{A_{a_1 a_2}^{\bar{W}}}{p} \right] = \frac{K-\Delta}{K(K-1)} + O(p^{-1}) \, , \qquad {\rm Var}_{z_{\star}}\left[ \frac{A_{a_1 a_2}^{\bar{W}}}{p} \right] = O(p^{-1}) \, .
\end{equation}
Thus, the random variables $A_a^W/p$ and $A_{a_1 a_2}^{\bar{W}}/p$ localize in probability to their expectation values.\footnote{\label{ConditionedEnsemble:FootnoteRareEvnets}This can be made more rigorous; for example, for $A_a^W$, the binomial distribution can be used to show that
\begin{equation*}
\mathbb{P}_{z_{\star}} \left( \left| \frac{A_a^W}{p} - \frac{\Delta}{K} \right| > \epsilon \right) \lesssim e^{-c_{\epsilon} p} \, , \qquad \epsilon > 0 \, ,
\end{equation*}
up to polynomial prefactors that don't change the behavior as $p \to \infty$.
}

\subsection{Diagrams with trivial permutation between the traces}

With the technology of the previous section, we can now compute the value of the diagram $\mathcal{A}(D)$ for the simplest case in which the permutation between the traces is trivial, $\sigma = {\rm Id}$. First note the following. For any fixed disk-chord $J_b$, we have
\begin{equation}
\sum_{a \in S_b^L} |I_a \cap J_b| = |T_b^{L} \cap J_b| \pmod 2 \, , \qquad T_b^{L} = \bigoplus_{a \in S_b^L} I_a \, ,
\end{equation}
because a fermion contributing to the sum over $a$ must appear an odd number of times in the cross-chords that intersect $J_b$, which are those with indices in $S_b^L$. Since that sum over $a$ appears within $(-1)^{Q_D}$, only its sign matters. To compute the value of the diagram, we thus want to obtain the following expectation value (where we define $T_c^R$ analogously to $T_b^L$)
\begin{equation}
\mathbb{E}_{K,W}^{(I, J, L)} \left[ (-1)^{\sum_{\{b,b'\} \in E_L} |J_{b} \cap J_{b'} | + \sum_{\{c,c'\} \in E_R} |L_{c} \cap L_{c'} | + \sum_b |T_b^{L} \cap J_b| + \sum_c |T_c^{R} \cap L_c|} \right] \, .
\end{equation}

We now compute the expectation value over the disk-chord ensemble, holding fixed a specific instance of the cross-chords. The disk-chords are sampled uniformly at random from all possible $p$-subsets, so the conventional chord techniques apply. In particular, triple intersections are suppressed in the double-scaling limit as $\sim p^{-1}$, so intersections become asymptotically independent with
\begin{equation}
|J_{b} \cap J_{b'}| \to {\rm Poisson}\left( \frac{\lambda}{2} \right) \, , \qquad |T^L_{b} \cap J_{b}| \to {\rm Poisson}\left( \frac{\lambda}{2} \frac{|T^L_b|}{p} \right) \, ,
\end{equation}
and identical expressions for chords in the right trace. Note that $|T^L_b|$ is at most $O(p)$, since it is the symmetric difference of an order $1$ number of $p$-subsets, and $|T^L_b|/p$ is a slight abuse of notation to refer to the limit of this quantity as $p \to \infty$. Using these Poisson distributions for intersections, it is immediate to obtain
\begin{equation}
\label{ConditionedEnsemble:PartialExpectationValue}
\mathbb{E}_{K,W}^{(I, J, L)} \left[(-1)^{Q_D(I,J,L)} \right] = q^{C_{\rm dd}} \mathbb{E}_{K,W}^{(I)} \left[ q^{\left(\sum_b |T^{L}_b| + \sum_c |T^{R}_c| \right)/p} \right] \left(1 + o(1) \right) \, ,
\end{equation}
where we have introduced the total number of disk-disk intersections
\begin{equation}
C_{\rm dd} = |E_L| + |E_R| \, .
\end{equation}

We finally compute the expectation value over the ensemble of cross-chords; both trace sectors are identical so we focus on the first one. Consider one disk-chord $J_b$ and let the number of cross-chords it intersects be $s_b^L:=|S_b^L|$. We want to understand how $|T^{L}_b|/p$ behaves; the factor of $p$ in the denominator means that we only care about the size of $T^{L}_b$ to order $p$, and note that a given fermion flavor $i$ belongs to $T^{L}_b$ if and only if it belongs to an odd number of the cross-chords that intersect $J_b$. We can thus count fermions in $T^{L}_b$ by looking at those that appear in a single chord which intersects $J_b$, those that appear in two chords but only one intersects $J_b$, etc. In this way, we get
\begin{equation}
\frac{|T^{L}_b|}{p} = \sum_{a \in S_b^L} \frac{A_a^W}{p} + \sum_{a \in S_b^L ,\,\bar{a} \notin S_b^L} \frac{A_{a \bar{a}}^{\bar{W}}}{p} + \dots \, ,
\end{equation}
with dots denoting terms that are suppressed as $p \to \infty$, and an identical expression holding for the right trace sector. We showed that the variables $A_a^W/p$ and $A_{a \bar{a}}^{\bar{W}}/p$ localize in probability, \eqref{ConditionedEnsemble:SingleChordNumberFermions} and \eqref{ConditionedEnsemble:TwoChordsNumberFermions}. Therefore, taking into account the number of terms in each sum (which is order $1$) and the assumed ensemble equivalence which implies that the convergence in probability holds also in the microcanonical conditioned ensemble, we conclude
\begin{equation}
\frac{|T^{(1)}_b|}{p} \underset{\mathbb{P}_{K,W}}{\longrightarrow} s_b^{L} \frac{\Delta}{K} + s_b^{L} (K- s_b^{L} )\frac{K - \Delta}{K(K-1)} = s_b^{L} - s_b^{L} (s_b^{L}-1) \frac{K - \Delta}{K(K-1)} \equiv F_{K, \Delta} (s_b^{L}) \, .
\end{equation}

This then shows that the exponent in \eqref{ConditionedEnsemble:PartialExpectationValue} localizes in probability to the sum over disk-chords $b$ of this value (note that this sum has an order $1$ number of terms). What does this imply for the exponent itself? Since the exponential function is continuous, its value also localizes in probability
\begin{equation}
q^{\left(\sum_b |T^{(1)}_b| + \sum_c |T^{(2)}_c| \right)/p} \underset{\mathbb{P}_{K,W}}{\longrightarrow} q^{\sum_b F_{K, \Delta}(s_b^{L}) + \sum_c F_{K, \Delta}(s_c^{R})} \, .
\end{equation}
However, if the function itself could take large values (scaling with $p$) in the rare events, they could contribute a finite amount to the expectation value, so that we could not substitute this localized value inside it. This does not happen here because the function is bounded,
\begin{equation}
0 \leq q^{\left(\sum_b |T^{(1)}_b| + \sum_c |T^{(2)}_c| \right)/p} \leq 1 \, .
\end{equation}
Therefore, the localization in probability allows us to conclude
\begin{equation}
\mathbb{E}_{K,W}^{(I, J, L)} \left[(-1)^{Q_D(I,J,L)} \right] = q^{C_{\rm dd}} q^{\sum_b F_{K, \Delta}(s_b^{L}) + \sum_c F_{K, \Delta}(s_c^{R})} \left[ 1 + o(1) \right] \, .
\end{equation}

The full result for the value of a fixed chord diagram with a trivial permutation between the traces $X_\sigma(I) = 0$ is then
\begin{equation}
\label{ConditionedEnsemble:ValueDiagram}
\mathcal{A}(D) = \left( \frac{\mathcal{J}^2}{\lambda} \right)^{\frac{n+m}{2}} P_K(W) \, q^{C_{\rm dd}} q^{\sum_b F_{K, \Delta}(s_b^{L}) + \sum_c F_{K, \Delta}(s_c^{R})} \left[ 1 + o(1) \right] \, ,
\end{equation}
where $P_K(W)$ is the endpoint probability computed in Appendix \ref{sec:appKrawtchouk} and whose leading terms in the double-scaling limit were provided in \eqref{ConditionedEnsemble:EndpointProbabilityLeadingTerms}, $C_{\rm dd}$ is the total number of disk-disk intersections, $s_b^{L}$ and $s_c^{R}$ are the number of intersections between cross-chords and the disk-chords $J_b$ and $L_c$, and
\begin{equation}
F_{K, \Delta}(s) = s - s(s-1) \frac{K - \Delta}{K ( K-1)} \, .
\end{equation}
This result already shows one of the main features used in the main text: it is only for integer $\Delta$ and $K = \Delta$ that we recover the usual chord rules, in which each intersection produces a factor of $q$. For other regimes, the previous result has to be used in full, complicating the introduction of a chord Hamiltonian.

\subsection{Localization in the minimal cross-chord sector}

We have shown how to deal with a general chord diagram (restricted to have trivial permutation between the traces), but we can actually argue that not any value of $K$ is needed to obtain the double-scaling limit of the squared thermal one-point function: the sum over diagrams localizes in those that have the minimal number of cross-chords allowed, $K_{\rm min}$. The key fact is that diagrams with larger values of $K$ are suppressed by $P_K(W)$, which is a rapidly decreasing function. Indeed, if we compare $K_{\rm min}$ with $K = K_{\rm min} + k$
\begin{equation}
\label{ConditionedEnsemble:PKSuppression}
\log \frac{P_{K_{\rm min} + k}(W)}{P_{K_{\rm min}}(W)} = - \frac{k}{2} p \log p + O(p) \, ,
\end{equation}
which is an enormous suppression.

However, it is still possible that for larger values of $K$, the increasing number of allowed chord diagrams compensates for this. To discuss this, let us organize the computation of the squared thermal one-point function in terms with a fixed number of cross-chords:
\begin{equation}
\mathcal{C}_{\beta}(W) = \sum_{K \geq K_{\rm min}} \mathcal{C}_{\beta}^{(K)}(W) \, .
\end{equation}
A given diagram cannot contribute a large number to $\mathcal{C}_{\beta}^{(K)}(W)$, given that in \eqref{ConditionedEnsemble:GeneralDiagramForm} we are only calculating an expectation value of quantities which are just a sign:
\begin{equation}
|\mathcal{A}(D)| \leq \left( \frac{\mathcal{J}^2}{\lambda} \right)^{(n+m)/2} P_K(W) \, .
\end{equation}
How many diagrams with $n = K + 2d_L$ Hamiltonian insertions in the first trace and $K$ cross-chords are there? We have to pick $K$ insertions as cross-chords and pair the remaining $2d_L$ ones, so this number is
\begin{equation}
\binom{K+2d_L}{K} \frac{(2d_L)!}{2^{d_L} d_L!} = \frac{(K + 2 d_L)!}{2^{d_L} K! \, d_L!} \, .
\end{equation}
The same logic works for the second trace, and given a diagram with $K$ cross-chords, there are $K!$ ways to pair them between the traces (one for each $S_K$ permutation). Including the Taylor expansion coefficients, we get
\begin{equation}
\label{ConditionedEnsemble:BoundFixedCrossChords}
\left| \mathcal{C}_{\beta}^{(K)}(W) \right| \leq K! P_K(W) M_K(\beta)^2 \, ,
\end{equation}
where
\begin{equation}
M_K(\beta) = \sum_{d=0}^{\infty} \left(\frac{\beta \mathcal{J}}{\sqrt{\lambda}} \right)^{K + 2d} \frac{1}{(K+2d)!}\frac{(K + 2 d)!}{2^d K! \, d!} = \left(\frac{\beta \mathcal{J}}{\sqrt{\lambda}} \right)^{K} \frac{e^{\beta^2 \mathcal{J}^2 /(2 \lambda)}}{K!} \, .
\end{equation}
This is enough to show that, for fixed $k = 1, 2, \dots$
\begin{equation}
\frac{\left| \mathcal{C}_{\beta}^{(K_{\rm min} + k)}(W) \right|}{P_{K_{\rm min}}(W)} \leq \left(\frac{\beta \mathcal{J}}{\sqrt{\lambda}} \right)^{2(K_{\rm min}+k)} \frac{e^{\beta^2 \mathcal{J}^2 /\lambda}}{(K_{\rm min}+k)!} \frac{P_{K_{\rm min} + k}(W)}{P_{K_{\rm min}}(W)} \, .
\end{equation}
The last factor provides then the dominant $O(e^{- (k p/2) \log p + O(p)})$ suppression discussed before, while the rest of the terms are finite in the double-scaling limit.

Technically, this only proves that each sector $K = K_{\rm min} + k$ for fixed $k > 0$ (or a finite sum of them) is exponentially suppressed with respect to the $K_{\rm min}$ one, but the infinite sum with $k > 0$ could still add to something finite. The suppression is so large though that the tail of the series can also be controlled.\footnote{This can be shown for example by proving that $P_K(W) \leq P_{K_0}(\emptyset)$ for $K \geq K_0$ with $K_0$ even. This follows because for a given subset $U$, $P_{K_0}(U) = \sum_V P_{K_0/2}(V) P_{K_0/2}(U \oplus V)$, which using Cauchy-Schwarz and the fact that $V \to U \oplus V$ for fixed $U$ is just a permutation of all subsets implies $P_{K_0}(U) \leq P_{K_0}(\emptyset)$. Then, from $P_{K}(W) = \sum_U P_{K_0}(U) P_{K-K_0}(W \oplus U)$ and using the previous bound we conclude $P_K(W) \leq P_{K_0}(\emptyset)$. Using this in the bound \eqref{ConditionedEnsemble:BoundFixedCrossChords}, we can control the tail of the series as $\sum_{K \geq K_0} |\mathcal{C}_{\beta}^{(K)}(W)| \leq P_{K_0}(\emptyset) e^{2 \beta^2 \mathcal{J}^2 / \lambda}$.} In the end, we get the announced result: the sum over diagrams localizes on those with the minimum number of cross-chords allowed,
\begin{equation}
\label{ConditionedEnsemble:LocalizationOfSum}
\mathcal{C}_{\beta}(W) = \mathcal{C}_{\beta}^{(K_{\rm min})}(W) + o \left(P_{K_{\rm min}}(W)\right) \, .
\end{equation}
Provided cancellations between diagrams do not make $\mathcal{C}_{\beta}^{(K_{\rm min})}(W)$ smaller than the terms we are neglecting, we can just look at diagrams with $K = K_{\rm min}$ to compute $\mathcal{C}_{\beta}(W)$. 

\subsection{Complete result for integer $\Delta$}

The results of the previous two sections are particularly powerful when the matter operator has integer weight, $\Delta \in \mathbb{Z}_{>0}$. In that case, we know that $K_{\rm min} = \Delta$ exactly, and in all diagrams with this minimum number of cross-chords, they are non-intersecting:
\begin{equation}
I_{a_1} \cap I_{a_2} = \emptyset \, .
\end{equation}
This follows because there are $\Delta p$ fermions in all the cross-chords, and since they must match the matter operator $\psi_W$ (which also contains $\Delta p$ fermions) there cannot be non-trivial intersections: the cross-chords form a partition of $W$. So the permutation sign in any diagram is trivial,
\begin{equation}
X_{\sigma}(I) = 0 \qquad \forall\sigma \, .
\end{equation}
All diagrams have then the same value we obtained for the one with a trivial permutation, \eqref{ConditionedEnsemble:ValueDiagram}. We just have to multiply by $K_{\rm min}! = \Delta!$ to account for all the possible permutations between cross-chords.

As we emphasized before, when $K = K_{\rm min} = \Delta$ we also recover the ordinary chord rules for intersections between disk- and cross-chords (i.e., $F_{\Delta, \Delta}(r) = r$). We can then organize the leading contribution to the squared thermal one-point function as
\begin{equation}
\mathcal{C}_{\beta}(W) = \Delta! P_{\Delta}(W) \tilde{\Psi}_{\beta}(\Delta)^2 \left[1 + o(1) \right] \, ,
\end{equation}
where
\begin{equation}
\begin{aligned}
\tilde{\Psi}_{\beta}(\Delta) & = \sum_{d\geq0} \frac{1}{(\Delta+2d)!} \left( - \frac{\beta \mathcal{J}}{\sqrt{\lambda}} \right)^{\Delta+2d} \sum_{\mathcal{D}_{\Delta;d}} q^{\# \, {\rm int}(D)} \\
& = \frac{\bra{\Delta} e^{-\beta H_{\rm chord}}\ket{0}}{\sqrt{[\Delta]_q !}} \, .
\end{aligned}
\end{equation}
In the first line, we denote by $\mathcal{D}_{\Delta;d}$ the set of all chord diagrams with $\Delta$ open chords and $d$ closed disk-chords; in the second line we have rewritten this using the standard chord Hamiltonian \eqref{eq:chordH}.

\subsection{Equivalence between the canonical and microcanonical conditioned ensembles}

All the previous results used statements valid in the canonical ensemble of cross-chords, such as \eqref{ConditionedEnsemble:SingleChordNumberFermions} or \eqref{ConditionedEnsemble:TwoChordsNumberFermions}, even though the computation we are interested in must be done in the microcanonical ensemble where the number of fermions per cross-chord is fixed to $p$. We will now show that the ensembles are equivalent for all the relevant quantities, thus proving that the canonical results can indeed be used. For most of the argument, we assume $0 < \Delta < K$, we will comment on $K = \Delta$ at the end.

Start by defining a $K$-vector that keeps track of the number of fermions in each cross-chord,
\begin{equation}
D = (d_1, \dots, d_K) \, , \qquad d_a(M) = |\{ i \; / \; a \in M_i \}| = |I_a| \, ,
\end{equation}
where we recall that $M \in \mathcal{M}$ is a possible configuration, with $M_i \subseteq \{1, \dots, K\}$ indicating in which cross-chords the Majorana $\psi_i$ appears. Given the canonical measure characterized by fugacities $z = (z_1, \dots, z_K)$, $\mathbb{P}_{z}$, the microcanonical measure is obtained by conditioning on $D$ being $(p, \dots, p)$:
\begin{equation}
\mathbb{P}_{\rm mic} (M) = \mathbb{P}_{z}(M \,| \, D = p \,v_{\mathbf{1}} ) \, ,
\end{equation}
where we have defined the vector consisting of all $1$'s as $v_{\mathbf{1}} = (1, \dots, 1)$. Note that this is valid for \textit{any} positive choice of fugacities, we will of course select $z = z_{\star} v_{\mathbf{1}}$ because then the condition is satisfied at the level of the canonical mean:
\begin{equation}
\mathbb{E}_{z_{\star}}[D] = p \, v_{\mathbf{1}} \, .
\end{equation}

We now study canonical fluctuations. Start by writing
\begin{equation}
D = \sum_{i=1}^N X_i \, , 
\end{equation}
where the $X_i \in \{0,1\}^K$ are binary vectors keeping track of which cross-chords are occupied by $\psi_i$. The canonical ensemble factorizes over fermion flavors, so the $X_i$ are independent and we can obtain the covariance of the vector $D$ from that of $X_i$. As usual, let us consider $i \in W$ and $i \notin W$ separately. If $i \in W$, $X_i$ contains a single $1$ in some cross-chord with leading probability $1/K$, triple appearances are suppressed by $p^{-1}$. Thus
\begin{equation}
\mathbb{E}_{z_{\star}}[X_i] = \frac{1}{K} v_{\mathbf{1}} + O(p^{-1}) \, , \qquad \mathbb{E}_{z_{\star}}\left[ X_i X_i^T \right] = \frac{1}{K} \mathbf{1} + O(p^{-1}) \, ,
\end{equation}
where $\mathbf{1}$ is the $K \times K$ identity matrix. The covariance is
\begin{equation}
{\rm Cov}(X_i) = \mathbb{E}_{z_{\star}}\left[ X_i X_i^T \right] - \mathbb{E}_{z_{\star}}[X_i] \mathbb{E}_{z_{\star}}\left[X_i^T\right] = \frac{1}{K} \mathbf{1} - \frac{1}{K^2} J + O(p^{-1}) \equiv \frac{1}{K} P_{\perp} + O(p^{-1}) \, ,
\end{equation}
where we have defined the all-ones matrix
\begin{equation}
J = v_{\mathbf{1}} v_{\mathbf{1}}^T \, ,
\end{equation}
and $P_{\perp} = \mathbf{1} - J/K$ is a projector. Similarly, for $i \notin W$, $X_i$ is identically zero with probability $1 - O(p^{-1})$ and it contains two ones with probability $z_{\star}^2 / E(z_{\star}) = z_{\star}^2 + O(p^{-2})$ -- in this case, it is important to keep the term of order $p^{-1}$ because there are $O(N)$ fermions not in $W$, so this contributes to the same order as the $O(1)$ term for the fermions in $W$. Then, considering all possible configurations with two fermions split between the $K$ cross-chords,
\begin{equation}
\mathbb{E}[X_i] = (K-1) z_{\star}^2 v_{\mathbf{1}} + O(p^{-2}) \, , \qquad \mathbb{E}_{z_{\star}}\left[ X_i X_i^T \right] = z_{\star}^2 \left( (K-2) \mathbf{1} + J \right) + O(p^{-2}) \, ,
\end{equation}
so the covariance is
\begin{equation}
{\rm Cov}(X_i) = z_{\star}^2 \left( (K-2) \mathbf{1} + J \right) + O(p^{-2}) \equiv z_{\star}^2 \left( (K-2) P_{\perp} + 2(K-1) P_{\parallel} \right) + O(p^{-2}) \, ,
\end{equation}
where the new projector matrix is $P_{\parallel} = J/K$.

This immediately gives the covariance matrix for the random vector $D$ in the canonical ensemble,
\begin{equation}
{\rm Cov}(D) = \sum_{i=1}^N {\rm Cov}(X_i) = p \left( \lambda_{\perp} P_{\perp} + \lambda_{\parallel} P_{\parallel} \right) + O(1) \, ,
\end{equation}
with
\begin{equation}
\lambda_{\perp} := \frac{K(K-2) + \Delta}{K(K-1)} \, , \qquad \lambda_{\parallel} := 2 \frac{K - \Delta}{K} \, .
\end{equation}
The decomposition of ${\rm Cov}(D)$ using orthogonal projectors ($P_{\perp} P_{\parallel} = 0$) makes it clear that $p \lambda_{\perp}$ and $p \lambda_{\parallel}$ are the eigenvalues of the matrix (with multiplicities $K-1$ and $1$, respectively). The latter captures configurations in which all cross-chords fluctuate together, increasing or decreasing the total number of fermions, while the former captures fluctuations that preserve the total number of fermions. Equivalently,
\begin{equation}
{\rm Var}(d_a) = p \frac{K^2 - \Delta}{K^2} + O(1) \, , \qquad {\rm Cov}(d_{a_1}, d_{a_2}) = p \left( \frac{K - \Delta}{K(K-1)} - \frac{\Delta}{K^2} \right) + O(1) \, .
\end{equation}
Since $\mathbb{E}_{z_{\star}} [d_a] = p$, this means that fluctuations are suppressed
\begin{equation}
d_a = p + \delta d_a \, , \qquad \frac{\delta d_a}{p} \sim O(p^{-1/2}) \, ,
\end{equation}
and $d_a/p$ converges to $1$ in probability. Thus, the canonical ensemble is already localized at $d_a = p$, and the microcanonical constraint only removes relative fluctuations of order $p^{-1/2}$.

This analysis also provides the probability to land exactly at $D = p v_{\mathbf{1}}$ in the canonical ensemble. By the multivariate central limit theorem
\begin{equation}
\frac{D - p v_{\mathbf{1}}}{\sqrt{p}} \, ,
\end{equation}
behaves as $p \to \infty$ as a $K$-dimensional Gaussian with $O(1)$ covariance. A typical fluctuation region therefore has linear size $O(\sqrt{p})$ in each of the $K$ directions and contains $\sim O(p^{K/2})$ lattice points for $D$. We thus expect a central lattice point (such as $D = p v_{\mathbf{1}}$) to have probability
\begin{equation}
\mathbb{P}_{z_{\star}}(D = p v_{\mathbf{1}}) = p^{-K/2 + o(1)} \, .
\end{equation}
This intuition can be made precise using a lattice local central limit theorem.\footnote{The result is the following: with ${\rm Cov}(D) = p \Sigma$, since the lattice of values that $D$ takes has cell volume $2$ due to the parity restrictions
\begin{equation*}
   \mathbb{P}_{z_{\star}}(D = D_0) = \frac{2}{(2\pi p)^{K/2} \sqrt{\lambda_{\parallel} \lambda_{\perp}^{K-1}}} \exp \left[ - \frac{1}{2p} (D_0-p v_{\mathbf{1}})^T \Sigma^{-1} (D_0-p v_{\mathbf{1}}) \right] + o\left(p^{-K/2}\right) \, ,
\end{equation*}
for points $D_0$ in an $O(p^{1/2})$ window around the central point $D = p v_{\mathbf{1}}$. See Chapter 5 of \cite{BhattacharyaRao2010}.} Note that this polynomial suppression justifies dropping the last term in the first line of \eqref{ConditionedEnsemble:NumberConfigurationsLeadingTerms}, as anticipated there.

This is enough to prove that some of the results we obtained using the canonical measure apply also to the microcanonical ensemble. In particular, for \eqref{ConditionedEnsemble:SingleChordNumberFermions}, the binomial large-deviation bound quoted in footnote \ref{ConditionedEnsemble:FootnoteRareEvnets} gives exponential suppression in $p$ for rare events. Then, using
\begin{equation}
\mathbb{P}_{\rm mic} (\cdot) = \mathbb{P}_{z_{\star}}(\cdot \, | \, D = p v_{\mathbf{1}}) = \frac{\mathbb{P}_{z_{\star}}(\cdot \; {\rm and}\; D = p v_{\mathbf{1}})}{\mathbb{P}_{z_{\star}}(D = p v_{\mathbf{1}})} \leq \frac{\mathbb{P}_{z_{\star}}(\cdot )}{\mathbb{P}_{z_{\star}}(D = p v_{\mathbf{1}})} \, ,
\end{equation}
we conclude
\begin{equation}
\mathbb{P}_{\rm mic} \left( \left| \frac{A_a^W}{p} - \frac{\Delta}{K} \right| > \epsilon \right) \lesssim \frac{e^{-c_{\epsilon} p}}{p^{-K/2+o(1)}} \underset{p \to \infty}{\longrightarrow} 0 \, .
\end{equation}
Thus, $A_a^W/p$ also localizes in probability in the microcanonical ensemble, and similar results hold for $A_{a_1 a_2}^{\bar{W}}/p$. These were all the results needed to obtain the value of the chord diagram in \eqref{ConditionedEnsemble:ValueDiagram}, but similar localization results can be obtained for other variables defined, such as the number of fermions appearing in $m$ cross-chords, $n_m^W$ or $n_m^{\bar{W}}$.

Finally, a comment about the limiting case $K = \Delta$. From the perspective of the canonical ensemble, $z_{\star} \to 0$ and the longitudinal fluctuation vanishes, $\lambda_{\parallel} = 0$. This reduces the power of $p$ in $\mathbb{P}_{z_{\star}}(D = p v_{\mathbf{1}})$ from $p^{-K/2}$ to $p^{-(K-1)/2}$ -- the Gaussian is really $(K-1)$-dimensional --, but nothing significant changes. That said, as explained in previous sections, the endpoint condition for $K = \Delta$ forces the cross-chords to form a disjoint partition of $W$. This simplifies the analysis enough to obtain exact results directly in the microcanonical ensemble, without the need to invoke ensemble equivalence: one example is the result \eqref{ConditionedEnsemble:EndpointProbabilityAtThreshold}, and another is the fact that intersection rules between the chords become almost trivial when cross-chords do not intersect.

\section{Light probes on the wormhole}
\label{app:averaged-light}

In this appendix we derive the SYK wormhole amplitude for the heavy-light two-point function introduced in Sec. \ref{sec:otherobsgauss}.

\subsection{Random $p$-string}

Consider that the probe operator is an independent Gaussian random $p$-body operator,
\be
\psi_{\rm light}=\sum_{|V|=p}c_V\psi_V,
\qquad
\langle c_Vc_{V'}\rangle_{\rm light}
=\frac{\delta_{VV'}}{\binom Np},
\ee
so that $\langle\Tr\psi_{\rm light}^2\rangle_{\rm light}=1$.
We use $\mathcal J=1$ and the same real sector $p,w\in4\mathbb Z$ as in
the main text. The observable is
\be
\mathcal C^{\rm light}_{\beta,\tau}(W)
=\left\langle
\left[\Tr\!\left(e^{-(\beta-\tau)H}\psi_{\rm light}
 e^{-\tau H}\psi_W\right)\right]^2
\right\rangle_{J,\rm light},
\qquad 0\leq\tau\leq\beta.
\label{eq:light-observable-app}
\ee
We take the double-scaled limit at fixed $\lambda$, positive integer
$\Delta=w/p$, and fixed thermal times. The probe has dimension one,
which stays finite when the heavy-operator limit is subsequently taken.

\paragraph{The trace constraint and the gluing.}
Averaging the light coefficients first gives the exact identity
\be
\mathcal C^{\rm light}_{\beta,\tau}(W)
=\frac1{\binom Np}\sum_{|V|=p}
\left\langle
\left[\Tr\!\left(e^{-(\beta-\tau)H}\psi_V
 e^{-\tau H}\psi_W\right)\right]^2
\right\rangle_J.
\label{eq:matter-average-fixed-strings}
\ee
Each term contains a light chord joining the two boundaries. If there
are $K$ Hamiltonian cross-chords, their labels obey
\be
I_1\oplus\cdots\oplus I_K\oplus V=W.
\label{eq:matter-endpoint-xor}
\ee
Since $V$ is also a uniform $p$-set, the endpoint probability is simply
$P_{K+1}(W)$, already defined in the main text.

The smallest allowed value is $K=\Delta-1$. At this threshold, the light
support and the Hamiltonian supports are disjoint and partition $W$.
Their endpoint weight is $P_\Delta(W)$. The light chord therefore
supplies one of the $\Delta$ blocks required by the heavy insertion.
The remaining Hamiltonian endpoints can be paired between the two
boundaries in $(\Delta-1)!$ ways, all with the same weight.

Crossings between these disjoint blocks have sign one. Whenever a Hamiltonian
disk-chord crosses another disk-chord or a prescribed block,
the crossing has weight $q=e^{-\lambda}$. These weights
factor in the double-scaled limit because the prescribed blocks are
disjoint and intersections involving two disk supports and a prescribed
block vanish. This is why the light chord dresses the partial disks
while also changing the number of chords used in the gluing.

\paragraph{The Hartle-Hawking wavefunction with a light insertion.}
We work on the disjoint threshold background.
Cut a boundary at $W$ and at the light insertion. The two arcs have
lengths $\beta-\tau$ and $\tau$ and prepare the ordinary Hartle-Hawking
wavefunctions. Suppose $K$ Hamiltonian endpoints remain open: $j$ come
from the first disk and $K-j$ from the second. If $n$ Hamiltonian chords
join the two arcs, their weighted choices and matchings contribute
\be
\frac{[n+j]_q!}{[n]_q![j]_q!}\,
\frac{[n+K-j]_q!}{[n]_q![K-j]_q!}\,[n]_q!
=\frac{[n+j]_q![n+K-j]_q!}
{[n]_q![j]_q![K-j]_q!}.
\label{eq:light-arc-matching}
\ee
Each of these $n$ chords crosses the light chord and adds a factor $q$.
In this threshold sector, the surviving endpoints become cross-chords
and acquire no additional light-crossing weight.

Multiplying this count by the two disk amplitudes
$\Psi_{\beta-\tau}(n+j)/\sqrt{[n+j]_q!}$ and
$\Psi_\tau(n+K-j)/\sqrt{[n+K-j]_q!}$, then summing over $n$ and $j$,
gives the partial-disk amplitude. As for the ordinary Hartle-Hawking
state, multiplying by $\sqrt{[K]_q!}$ converts it to the orthonormal
chord basis. Denote the resulting wavefunction by
$\Psi^{\rm light}_{\beta,\tau}(K)$. Its threshold component is the physical partial-disk response.
Defining the other components by the same formal expression, it can
be written entirely in the existing chord Hilbert space:
\be
\Psi^{\rm light}_{\beta,\tau}(K)
=\sqrt{[K]_q!}\sum_{j=0}^{K}
\frac{\langle\Psi_\tau|
(a^\dagger)^{K-j}q^{\widehat n}a^j
|\Psi_{\beta-\tau}\rangle}
{[j]_q![K-j]_q!}.
\label{eq:light-HH-wavefunction}
\ee
Here $a$ and $a^\dagger$ are the chord operators of the main text, with
$a|n\rangle=\sqrt{[n]_q}|n-1\rangle$, and
$\widehat n|n\rangle=n|n\rangle$. Inserting chord-number states in
\eqref{eq:light-HH-wavefunction} reproduces the preceding count:
$a^j$ and $(a^\dagger)^{K-j}$ attach the surviving endpoints, while
$q^{\widehat n}$ supplies the weight $q^n$ of the chords crossing the
probe.

\paragraph{The bulk amplitude.}
Gluing the two partial disks at $K=\Delta-1$ gives
\be
\mathcal C^{\rm light}_{\beta,\tau}(W)
\sim \frac{(\Delta-1)!P_\Delta(W)}{[\Delta-1]_q!}
\left|\Psi^{\rm light}_{\beta,\tau}(\Delta-1)\right|^2.
\label{eq:matter-leading-covariance}
\ee
The factorial counts the Hamiltonian cross-chord pairings, and the
$q$-factorial converts the two boundary wavefunctions to the
orthonormal basis. Using the main-text kernel
$G_W=g_W|\Delta\rangle\langle\Delta|$, the same result is the bulk overlap
\be
\mathcal C^{\rm light}_{\beta,\tau}(W)
\sim \frac1\Delta
\langle\Psi^{\rm light}_{\beta,\tau}|
 aG_Wa^\dagger
|\Psi^{\rm light}_{\beta,\tau}\rangle,
\label{eq:light-bulk-amplitude}
\ee
where $|\Psi^{\rm light}_{\beta,\tau}\rangle
=\sum_{K\geq0}\Psi^{\rm light}_{\beta,\tau}(K)|K\rangle$.
Indeed, $aG_Wa^\dagger/\Delta$ projects onto $|\Delta-1\rangle$ with
coefficient $[\Delta]_q g_W/\Delta$.
The probe changes both the boundary wavefunction and the gluing:
one block of the heavy string is now carried by the light chord.

Two limits make this interpretation explicit. Moving the light
insertion to either end of a boundary gives
\be
\Psi^{\rm light}_{\beta,0}(K)
=\Psi^{\rm light}_{\beta,\beta}(K)=\Psi_\beta(K).
\label{eq:light-endpoint-check}
\ee
With no surviving Hamiltonian cross-chords, the wavefunction instead
reduces to the ordinary disk matter two-point function,
\be
\Psi^{\rm light}_{\beta,\tau}(0)
=\langle\Psi_\tau|q^{\widehat n}|\Psi_{\beta-\tau}\rangle.
\label{eq:light-disk-check}
\ee
Reflection of the two arcs also gives invariance under
$\tau\leftrightarrow\beta-\tau$.

\paragraph{Control of the thermal sum.}
The asymptotic expansion \eqref{eq:matter-leading-covariance} assumes that the thermal evolutions do not modify the dominance of the threshold configuration $K = \Delta - 1$. The argument for this is analogous to the one presented in Appendix \ref{app:ConditionedEnsemble}. Repeating the absolute value bound \eqref{ConditionedEnsemble:BoundFixedCrossChords} for diagrams with $K$ Hamiltonian cross-chords, with the light chord providing one additional set to the endpoint condition, gives now the bound
\be
e^{\beta^2/\lambda}P_{K+1}(W)
\frac{(\beta^2/\lambda)^K}{K!}.
\label{eq:matter-thermal-majorant}
\ee
Splitting each boundary at $\tau$ does not change this bound: summing over
the two arc orders restores the Taylor coefficient for total length
$\beta$. The remaining argument is then the same leading to the localization result \eqref{ConditionedEnsemble:LocalizationOfSum}, with the replacement $K \to K+1$ in the endpoint probability. Thus all fixed sectors $K > \Delta - 1$ are suppressed relative to $K = \Delta - 1$.


\subsection{A fixed light string}
\label{app:fixed-light-string}

For a fixed $p$-string $\psi_V$, the trace constraint is
\be
I_1\oplus\cdots\oplus I_K=W\oplus V.
\label{eq:fixed-light-endpoint}
\ee
A simple bulk formula follows when $V\subset W$. Then $W \oplus V = W \setminus V$, and the threshold again occurs at $K = \Delta - 1$. The Hamiltonian cross-chords partition $W \setminus V$; they are in particular disjoint from $V$. As a consequence, the crossing rules and the wavefunction
$\Psi^{\rm light}_{\beta,\tau}$ are unchanged. Only the endpoint weight
changes, giving
\be
\left\langle
\left[\Tr\!\left(e^{-(\beta-\tau)H}\psi_V
 e^{-\tau H}\psi_W\right)\right]^2
\right\rangle_J \sim 
\frac{(\Delta-1)!P_{\Delta-1}(W\setminus V)}{[\Delta-1]_q!}
\left|\Psi^{\rm light}_{\beta,\tau}(\Delta-1)\right|^2.
\label{eq:fixed-light-bulk-amplitude}
\ee
Here $P_0(\varnothing)=1$, so the result also includes $\Delta=1$.
Compared with the independently averaged probe, the coefficient is
larger by
\be
\frac{P_{\Delta-1}(W\setminus V)}{P_\Delta(W)}
=\frac{\binom Np}{\binom wp}.
\ee
The inverse factor is the probability that the independent probe
selects a support contained in $W$. Thus the light average includes
an additional selection cost, while the bulk propagation is the same.

For other fixed supports $V$, this simplification no longer holds. The Hamiltonian cross-chords solving \eqref{eq:fixed-light-endpoint} can have non-trivial intersections with $V$, and the associated crossing signs must be kept. In particular,
\eqref{eq:fixed-light-bulk-amplitude} cannot be extended by changing
only the endpoint probability.

\paragraph{A disjoint fixed probe.}
The case $W\cap V=\varnothing$ is also explicit. Put
\be
U=W\cup V,\qquad k=\Delta+1.
\label{eq:fixed-disjoint-endpoint}
\ee
At threshold $K = k$, the cross-chords partition $U$. Although $V$ is disjoint from $W$, it intersects the partition blocks $I_a$. In fact, if we look at a single $I_a$, this is taken uniformly at random from the $p$-subsets of $U$, so $|I_a \cap V|$ follows a hypergeometric distribution with mean $p/k$ and variance $O(p)$. Thus, $|I_a \cap V|/p$ converges to $1/k$ as $p \to \infty$. The signs $\varepsilon_a = (-1)^{|I_a \cap V|}$ satisfy the exact constraint $\prod_a \varepsilon_a =(-1)^p = 1$, but other than that they become asymptotically uniform. Indeed, for fixed overlaps $x_a = |I_a \cap V|$, the number of threshold partitions is proportional to $\prod_a\binom{p}{x_a}$. Changing $x_a\to x_a-1$ and $x_b\to x_b+1$ flips the two signs $\varepsilon_a,\varepsilon_b$ while preserving $\sum_a x_a=p$, and changes the number of partitions by the factor
\be
\frac{x_a}{p-x_a+1}\frac{p-x_b}{x_b+1} = 1+O(p^{-1/2}),
\ee
where we used $x_a,x_b=p/k+O(\sqrt p)$. Since the overlap distribution is concentrated in this region, pairwise sign flips leave the probability asymptotically unchanged. Since any two sign assignments with $\prod_a\varepsilon_a=1$ are related by such flips, all $2^{k-1}$ allowed assignments become equally likely.

Suppose now a disk-chord $J$ crosses $V$ and a collection of the threshold blocks $I_a$ with $a \in S$ (and let $s = |S|$). The product of the fermionic signs is $(-1)^{|J \cap T_S|}$, with $T_S = V \oplus (\oplus_{a \in S} I_a)$. The previous result $|I_a \cap V|/p \to 1/k$ together with the fact that the $I_a$ are disjoint implies that
\be
\frac{|T_S|}{p} \to 1 + s - \frac{2 s}{k} .
\ee
Using the double-scaled intersection rules, the disk-chord produces a weight $q r^{s}$, with $r := q^{1-2/k}$. This means that a disk-chord that crosses $V$ contributes $q$, while additional intersections of that disk-chord with the cross-chords produce factors of $r$ instead.

We now analyze the preparation on each boundary like we did for random $p$-strings. Cut at the heavy and light insertions, producing two thermal arcs of lengths $\beta - \tau$ and $\tau$, and suppose $j$ of the $k$ cross-chords survive on the first arc, $k-j$ on the second, while $n$ disk-chords join the two arcs. Within each arc, the amplitude remains the ordinary Hartle--Hawking wavefunction. Each of the $n$ joining chords crosses $V$, producing the usual factor of $q$, while every additional crossing with a cross-chord contributes $r = q^{1-2/k}$. The only modification to \eqref{eq:light-arc-matching} is thus that the base of the Gaussian binomials changes from $q$ to $r$, producing the following amplitude with $j$ surviving endpoints on the first arc:
\be
\begin{aligned}
\mathcal A_j(\beta,\tau;k)
&:=\sum_{n\geq0}q^n
\binom{n+j}{j}_{r}\binom{n+k-j}{k-j}_{r}
\frac{[n]_q!\,\Psi_{\beta - \tau}(n+j)\Psi_{\tau}(n+k-j)}
{\sqrt{[n+j]_q![n+k-j]_q!}}.
\end{aligned}
\label{eq:fixed-disjoint-disk-amplitude}
\ee
Here, $\binom mj_r=[m]_r!/([j]_r![m-j]_r!)$; for $k=2$, $r=1$ and these are ordinary binomials.

At this point, we have taken care of all disk-chords and we have to glue both boundaries. Some cross-chords might appear in different sides of the light insertion, and these must be commuted so that both traces have the same form. This produces signs $\prod_{a \in \mathcal{S}} \varepsilon_a$, where $\mathcal{S} \subseteq \{1,...,k\}$ collects the cross-chords whose endpoints lie on opposite sides of the light insertion $V$ on the two boundaries. If $\mathcal{S}$ is empty or is the whole $\{1, ..., k\}$, the product of signs give one. In any other case, it averages to zero, given our previous result about the distribution of the signs. For the channel in which $\mathcal{S} = \emptyset$, we have $j!(k-j)!$ ways to pair the cross-chords (these do not produce any sign because cross-chords are disjoint); the same number of pairings appears if $\mathcal{S} = \{1, \dots ,k\}$. Including the endpoint probability, we conclude
\be
\mathcal C^{\rm disj}_{\beta,\tau}(W,V)
\sim \frac{k!P_k(U)}{2}\sum_{j=0}^{k}
\frac{\bigl[\mathcal A_j(\beta,\tau;k)+\mathcal A_{k-j}(\beta,\tau;k)\bigr]^2}
{\binom kj}.
\label{eq:fixed-disjoint-bulk-amplitude}
\ee
where $\mathcal C^{\rm disj}$ denotes the squared fixed-string trace
in \eqref{eq:fixed-light-bulk-amplitude}, and we have manifestly symmetrized the sum under $j \leftrightarrow k-j$. We have also used that the thermal sum is dominated by the threshold sector $K = k$ by the same localization argument as in Appendix \ref{app:ConditionedEnsemble}: the endpoint probability $P_K(U)$ produces the same parametric suppression as before, while the thermal combinatorics cannot compensate for it. As a check, at $\tau = 0$ or $\tau = \beta$, only one $\mathcal A_j$ survives, recovering the ordinary one-point gluing for $U$.


\subsection{A light probe of finite dimension}
\label{app:finite-light-dimension}

Keeping the same integer heavy dimension $\Delta=w/p$, the
dimension-one calculation extends to an independent Gaussian
$v$-body probe with $v=\delta p$,
\be
\psi_{\rm light}^{(v)}=\sum_{|V|=v}c_V\psi_V,
\qquad
\langle c_Vc_{V'}\rangle_{\rm light}
=\frac{\delta_{VV'}}{\binom Nv},
\qquad
\delta=\Delta_{\rm light}\in\{1,\ldots,\Delta\}.
\label{eq:finite-light-probe}
\ee
We retain $p,v,w\in4\mathbb Z$ and keep $\delta$, $\lambda>0$, and
the thermal times fixed in the first double-scaled limit.
The probe remains light when the heavy dimension is subsequently taken large. With respect to a random $p$-string, only the endpoint weight and the light-crossing factor change, all the other steps in the derivation are analogous so we will not repeat them.

After the light average, the endpoint probability is
\be
Q_K(W;v):=\frac1{\binom Nv}\sum_{|V|=v}P_K(W\oplus V).
\label{eq:finite-light-endpoint-probability}
\ee
Equivalently, $Q_K(W;v)$ is the probability that $K$ independent Hamiltonian $p$-sets together with one independent $v$-set have symmetric difference $W$. For
\be
k=\Delta-\delta\in\mathbb Z_{\geq0},
\qquad w=kp+v,
\label{eq:finite-light-compatible-sizes}
\ee
we see that the threshold number of Hamiltonian cross-chords is $K = k$, where the light support and the $k$ Hamiltonian supports partition $W = V \sqcup I_1 \sqcup \dots \sqcup I_k$. Hence, simple counting gives
\be
\begin{aligned}
Q_k(W;v)
&=\frac{w!}{v!(p!)^k\binom Nv\binom Np^{\,k}},
\\
\binom Nw Q_k(W;v)
&\longrightarrow
\exp\!\left[-\frac\lambda4\bigl(k(k-1)+2k\delta\bigr)\right].
\end{aligned}
\label{eq:finite-light-threshold-weight}
\ee

A disk-chord crossing the light support now carries $q^\delta$.
Thus define $\Psi^{(\delta)}_{\beta,\tau}(K)$ by the substitution
\be
q^{\widehat n}\ \longmapsto\ q^{\delta\widehat n}
\label{eq:finite-light-HH-wavefunction}
\ee
in \eqref{eq:light-HH-wavefunction}, leaving its other factors
unchanged. Its $K=k$ component gives the threshold boundary amplitude.
The same $k!$ boundary pairings then yield
\be
\begin{aligned}
G_W^{(\delta)}
&:=\frac{k!Q_k(W;v)}{[k]_q!}|k\rangle\langle k|,
\\
\mathcal C^{(v)}_{\beta,\tau}(W)
&\sim \frac{k!Q_k(W;v)}{[k]_q!}
\left|\Psi^{(\delta)}_{\beta,\tau}(k)\right|^2.
\end{aligned}
\label{eq:finite-light-bulk-amplitude}
\ee
Here $\mathcal C^{(v)}$ is \eqref{eq:light-observable-app} with the
$v$-body probe. For $v=p$, we have $\delta=1$, $Q_k=P_{k+1}$, and
$G_W^{(1)}=aG_Wa^\dagger/\Delta$, recovering the worked example.

For a fixed $V\subset W$, the fixed-string result extends by replacing
$Q_k(W;v)$ with $P_k(W\setminus V)$. This coefficient is larger by
\be
\frac{P_k(W\setminus V)}{Q_k(W;v)}
=\frac{\binom Nv}{\binom wv}.
\label{eq:finite-light-selection-cost}
\ee
The omitted higher-$K$ sectors contribute a correction
$o(\binom N{w-v}^{-1})$. We emphasize that this is valid for $V \subset W$; other general fixed supports still require dealing with conditioned intersection signs.

The thermal-sum argument bound \eqref{eq:matter-thermal-majorant}
applies with $P_{K+1}$ replaced by $Q_K$: essentially, every fixed $K>k$ requires overlaps of order $p$, but this is exponentially suppressed because the expected total overlap is finite. This proves the stated
remainder at fixed $k$, $\delta$, $\lambda$, and thermal times.
If the original integer $\Delta$ is retained while $\delta$ is
noninteger, no disjoint threshold exists and the averaged amplitude
is instead $o(\binom Nw^{-1})$. The formula above therefore requires
the exact support relation \eqref{eq:finite-light-compatible-sizes}.

\section{Higher moments of the thermal one-point function}
\label{app:higher-moments}

We study the disorder statistics of the single-instance observable
constructed in Sec.~\ref{sec:alphacouplings},
\be
Z_\beta(W,\mathbf J)
=\Tr\!\left(e^{-\beta H[\mathbf J]}\psi_W\right).
\ee
Its leading bulk representation is
\be
Z_\beta(W,\mathbf J)
\sim\frac{\lambda^{\Delta/2}}{\sqrt{[\Delta]_q!}}\,
\Psi_\beta(\Delta)\,\xi_W(\mathbf J).
\label{eq:single-instance-moments-app}
\ee
Thus the temperature dependence is carried by the Hartle-Hawking
wavefunction, while the same threshold coefficient $\xi_W(\mathbf J)$
contains the disorder dependence. We show that its normalized moments
become Gaussian and that the omitted thermal sectors do not change this
conclusion.

As in the main text, $w=\Delta p$ with positive integer $\Delta$ and
$\mathcal J=1$. We take $N,p\to\infty$ at fixed $\lambda=2p^2/N>0$,
$\Delta$, moment order, and positive inverse temperatures. Reflecting
any Majorana in $W$ changes the sign of $Z_\beta$ while preserving the
disorder measure, so all odd moments vanish exactly.

\paragraph{Threshold statistics.}
Recall that $\xi_W(\mathbf J)=\Delta!\operatorname{Pf}(\mathbf J|_W)$
is a sum over partitions of $W$ into $\Delta$ disjoint interaction sets.
Its fourth moment is obtained by assigning each fermion label to one
of the three pairings of four boundaries. If the three supports have
sizes $ap$, $bp$, and $cp$, their partition count gives
\be
\frac{\langle\xi_W^4\rangle_J}{\langle\xi_W^2\rangle_J^2}
=\sum_{\substack{a,b,c\geq0\\a+b+c=\Delta}}
\left(\frac{\Delta!}{a!b!c!}\right)^2
\frac{(ap)!(bp)!(cp)!}{(\Delta p)!}
\longrightarrow3.
\label{eq:threshold-fourth-exact}
\ee
Each Clifford sign occurs twice and cancels. The three terms with only
one nonzero entry give $3$, and every mixed term vanishes as $p$ grows.
Note that $\xi_W$ is a polynomial of the couplings of degree $\Delta$. The fourth-moment
theorem then ensures that the random variable is Gaussian after normalization
\cite{NualartPeccati2005}. 

\paragraph{Control of the threshold approximation.}
The previous analysis assumes that the threshold contribution dominates the higher moments. We now show that this is true for the $n$-th moment if $n$ is not allowed to scale with $N$. To see this, expand $Z_\beta$ in Hermite polynomials of the couplings, and denote
its component of total degree $K$ by $Z_\beta^{(K)}$. This degree counts
cross-chords in a covariance. The trace constraint excludes $K<\Delta$.
At $K=\Delta$, the interaction sets partition $W$, so permutation
symmetry makes $Z_\beta^{(\Delta)}$ proportional to $\xi_W$ exactly.
Internal contractions change only the deterministic coefficient, whose
double-scaled limit is given in \eqref{eq:single-instance-moments-app}.

The endpoint probability $P_K(W)$ defined in the main text also controls
the remaining sectors. Counting Wick pairings and bounding the
normalized Clifford traces by one gives
\be
\bigl\|Z_\beta^{(K)}\bigr\|_2^2
\leq e^{\beta^2/\lambda}P_K(W)
\frac{(\beta^2/\lambda)^K}{K!}.
\label{eq:thermal-chaos-bound}
\ee
For $K>\Delta$, an endpoint of size $\Delta p$ requires an excess of
$(K-\Delta)p$ input appearances over the endpoint size. This excess is
bounded by twice the total pairwise overlap of the $K$ sets, whose expectation is
$\binom K2p^2/N$. Exchangeability and Markov's inequality therefore give
\be
\binom Nw P_K(W)
\leq\frac{K(K-1)p}{N(K-\Delta)},
\qquad K>\Delta.
\label{eq:thermal-endpoint-bound}
\ee
In contrast, $\binom Nw P_\Delta(W)$ tends to the nonzero constant
$e^{-\lambda\Delta(\Delta-1)/4}$. Using Gaussian hypercontractivity,
$\|Z_\beta^{(K)}\|_s\leq(s-1)^{K/2}\|Z_\beta^{(K)}\|_2$,
the factorial bound makes the sum over all $K$ convergent and yields
\be
\frac{\|Z_\beta-Z_\beta^{(\Delta)}\|_s}{\sqrt{P_\Delta(W)}}
=O\!\left(\sqrt{\frac pN}\right)\longrightarrow0,
\qquad s\geq2\text{ fixed}.
\label{eq:thermal-tail-Ls}
\ee
The main-text two-point function is proportional to $P_\Delta(W)$ with
a finite positive coefficient at every fixed $\beta>0$. Hence this
also controls the remainder relative to the standard deviation.

Combining the threshold moments with this bound gives
\be
\left\langle\prod_{a=1}^{2n}Z_{\beta_a}(W,\mathbf J)\right\rangle_J
\sim \sum_{\pi\in\mathcal P_2(2n)}
\prod_{(a,b)\in\pi}\mathcal{C}_{\beta_a,\beta_b}(W).
\label{eq:thermal-higher-moment-Wick}
\ee
Here $\mathcal P_2(2n)$ denotes the pairings of the boundaries, and
$\mathcal{C}_{\beta,\beta'}(W)$ is the exact microscopic covariance.
The leading diagrams are therefore disjoint unions of two-boundary
wormholes. Connected higher cumulants vanish after normalization.
This statement applies at fixed $\Delta$ in the first double-scaled
limit; no uniformity in the subsequent heavy-operator limit is assumed.

\bibliographystyle{ourbst}
\bibliography{biblio}

@article{Marolf:2020xie,
  author       = {Marolf, Donald and Maxfield, Henry},
  title        = {Transcending the ensemble: baby universes, spacetime wormholes, and the order and disorder of black hole information},
  journal      = {JHEP},
  volume       = {08},
  pages        = {044},
  year         = {2020},
  doi          = {10.1007/JHEP08(2020)044},
  eprint       = {2002.08950},
  archivePrefix= {arXiv},
  primaryClass = {hep-th}
}

@article{Abdalla:2025gzn,
    author = "Abdalla, Ahmed I. and Antonini, Stefano and Iliesiu, Luca V. and Levine, Adam",
    title = "{The gravitational path integral from an observer{\textquoteright}s point of view}",
    eprint = "2501.02632",
    archivePrefix = "arXiv",
    primaryClass = "hep-th",
    doi = "10.1007/JHEP05(2025)059",
    journal = "JHEP",
    volume = "05",
    pages = "059",
    year = "2025"
}

@article{Bettaque:2026vpl,
    author = "Bettaque, Val{\'e}rie and Swingle, Brian",
    title = "{Magic and Wormholes in the Sachdev-Ye-Kitaev Model}",
    eprint = "2602.12339",
    archivePrefix = "arXiv",
    primaryClass = "hep-th",
    month = "2",
    year = "2026"
}

@article{Berkooz:2018jqr,
    author = "Berkooz, Micha and Isachenkov, Mikhail and Narovlansky, Vladimir and Torrents, Genis",
    title = "{Towards a full solution of the large N double-scaled SYK model}",
    eprint = "1811.02584",
    archivePrefix = "arXiv",
    primaryClass = "hep-th",
    doi = "10.1007/JHEP03(2019)079",
    journal = "JHEP",
    volume = "03",
    pages = "079",
    year = "2019"
}

@article{Saad:2021rcu,
    author = "Saad, Phil and Shenker, Stephen H. and Stanford, Douglas and Yao, Shunyu",
    title = "{Wormholes without averaging}",
    eprint = "2103.16754",
    archivePrefix = "arXiv",
    primaryClass = "hep-th",
    doi = "10.1007/JHEP09(2024)133",
    journal = "JHEP",
    volume = "09",
    pages = "133",
    year = "2024"
}

@article{Cui:2025sgy,
    author = "Cui, Chuanxin and Rozali, Moshe",
    title = "{Splitting and gluing in sine-dilaton gravity: matter correlators and the wormhole Hilbert space}",
    eprint = "2509.01680",
    archivePrefix = "arXiv",
    primaryClass = "hep-th",
    doi = "10.1007/JHEP02(2026)160",
    journal = "JHEP",
    volume = "02",
    pages = "160",
    year = "2026"
}

@article{Sontag:2026iiu,
    author = "Sontag, Andrew and Verlinde, Herman",
    title = "{Baby Universe in a Coupled SYK Model}",
    eprint = "2605.05291",
    archivePrefix = "arXiv",
    primaryClass = "hep-th",
    month = "5",
    year = "2026"
}

@article{DiUbaldo:2026rly,
    author = "Di Ubaldo, Gabriele and Iliesiu, Luca V. and Lin, Henry W. and Yan, Cynthia",
    title = "{Positivity of the gravitational path integral implies the axionic weak gravity conjecture}",
    eprint = "2605.05305",
    archivePrefix = "arXiv",
    primaryClass = "hep-th",
    reportNumber = "RIKEN-iTHEMS-Report-26",
    month = "5",
    year = "2026"
}

@article{VanRaamsdonk:2026tnv,
    author = "Van Raamsdonk, Mark and Vilar L{\'o}pez, Alejandro",
    title = "{Menagerie of Euclidean constructions for 3D holographic cosmologies}",
    eprint = "2601.10906",
    archivePrefix = "arXiv",
    primaryClass = "hep-th",
    doi = "10.1103/qlnl-rpm2",
    journal = "Phys. Rev. D",
    volume = "113",
    number = "10",
    pages = "106010",
    year = "2026"
}

@article{Belin:2025ako,
    author = "Belin, Alexandre and de Boer, Jan",
    title = "{Baby Universes in AdS$_3$}",
    eprint = "2512.02098",
    archivePrefix = "arXiv",
    primaryClass = "hep-th",
    doi = "10.21468/SciPostPhys.21.2.038",
    journal = "SciPost Phys.",
    volume = "21",
    pages = "038",
    year = "2026"
}

@article{Kudler-Flam:2025cki,
    author = "Kudler-Flam, Jonah and Witten, Edward",
    title = "{Emergent mixed states for baby universes and black holes}",
    eprint = "2510.06376",
    archivePrefix = "arXiv",
    primaryClass = "hep-th",
    doi = "10.1007/JHEP05(2026)090",
    journal = "JHEP",
    volume = "05",
    pages = "090",
    year = "2026"
}

@article{Zhao:2026mpl,
    author = "Zhao, Ying",
    title = "{''It from Bit'': the Hartle-Hawking state and quantum mechanics for de Sitter observers}",
    eprint = "2602.05939",
    archivePrefix = "arXiv",
    primaryClass = "hep-th",
    reportNumber = "MIT-CTP/6000",
    doi = "10.1007/JHEP09(2026)252",
    journal = "JHEP",
    volume = "09",
    pages = "252",
    year = "2026"
}

@article{Stanford:2017thb,
    author = "Stanford, Douglas and Witten, Edward",
    title = "{Fermionic Localization of the Schwarzian Theory}",
    eprint = "1703.04612",
    archivePrefix = "arXiv",
    primaryClass = "hep-th",
    doi = "10.1007/JHEP10(2017)008",
    journal = "JHEP",
    volume = "10",
    pages = "008",
    year = "2017"
}

@article{Bagrets:2016cdf,
    author = "Bagrets, Dmitry and Altland, Alexander and Kamenev, Alex",
    editor = "Unno, Yoshinobu and Ohsugi, Takashi and Hou, Suen and Sadrozinski, Hartmut F. -W. and Lou, Xinchou and Zhu, Hongbo and Ouyang, Qun",
    title = "{Sachdev{\textendash}Ye{\textendash}Kitaev model as Liouville quantum mechanics}",
    eprint = "1607.00694",
    archivePrefix = "arXiv",
    primaryClass = "cond-mat.str-el",
    doi = "10.1016/j.nuclphysb.2016.08.002",
    journal = "Nucl. Phys. B",
    volume = "911",
    pages = "191--205",
    year = "2016"
}

@article{Raz:2025wjw,
    author = "Raz, Amir and Youssef, Merna",
    title = "{The late time ramp from chord diagrams in the double-scaled SYK model}",
    eprint = "2507.12543",
    archivePrefix = "arXiv",
    primaryClass = "hep-th",
    reportNumber = "UT-WI-10-2025",
    doi = "10.1007/JHEP06(2026)215",
    journal = "JHEP",
    volume = "06",
    pages = "215",
    year = "2026"
}

@article{Yang:2018gdb,
    author = "Yang, Zhenbin",
    title = "{The Quantum Gravity Dynamics of Near Extremal Black Holes}",
    eprint = "1809.08647",
    archivePrefix = "arXiv",
    primaryClass = "hep-th",
    doi = "10.1007/JHEP05(2019)205",
    journal = "JHEP",
    volume = "05",
    pages = "205",
    year = "2019"
}

@article{Garcia-Garcia:2020ttf,
    author = "Garc{\'\i}a-Garc{\'\i}a, Antonio M. and Godet, Victor",
    title = "{Euclidean wormhole in the Sachdev-Ye-Kitaev model}",
    eprint = "2010.11633",
    archivePrefix = "arXiv",
    primaryClass = "hep-th",
    doi = "10.1103/PhysRevD.103.046014",
    journal = "Phys. Rev. D",
    volume = "103",
    number = "4",
    pages = "046014",
    year = "2021"
}

@article{Mertens:2017mtv,
    author = "Mertens, Thomas G. and Turiaci, Gustavo J. and Verlinde, Herman L.",
    title = "{Solving the Schwarzian via the Conformal Bootstrap}",
    eprint = "1705.08408",
    archivePrefix = "arXiv",
    primaryClass = "hep-th",
    doi = "10.1007/JHEP08(2017)136",
    journal = "JHEP",
    volume = "08",
    pages = "136",
    year = "2017"
}

@article{Berkooz:2020fvm,
    author = "Berkooz, Micha and Brukner, Nadav and Narovlansky, Vladimir and Raz, Amir",
    title = "{Multi-trace correlators in the SYK model and non-geometric wormholes}",
    eprint = "2104.03336",
    archivePrefix = "arXiv",
    primaryClass = "hep-th",
    doi = "10.1007/JHEP09(2021)196",
    journal = "JHEP",
    volume = "21",
    pages = "196",
    year = "2020"
}

@article{Berkooz:2018qkz,
    author = "Berkooz, Micha and Narayan, Prithvi and Simon, Joan",
    title = "{Chord diagrams, exact correlators in spin glasses and black hole bulk reconstruction}",
    eprint = "1806.04380",
    archivePrefix = "arXiv",
    primaryClass = "hep-th",
    doi = "10.1007/JHEP08(2018)192",
    journal = "JHEP",
    volume = "08",
    pages = "192",
    year = "2018"
}

@article{Lin:2022rbf,
    author = "Lin, Henry W.",
    title = "{The bulk Hilbert space of double scaled SYK}",
    eprint = "2208.07032",
    archivePrefix = "arXiv",
    primaryClass = "hep-th",
    doi = "10.1007/JHEP11(2022)060",
    journal = "JHEP",
    volume = "11",
    pages = "060",
    year = "2022"
}

@article{Altland:2017eao,
    author = "Altland, Alexander and Bagrets, Dmitry",
    title = "{Quantum ergodicity in the SYK model}",
    eprint = "1712.05073",
    archivePrefix = "arXiv",
    primaryClass = "cond-mat.str-el",
    doi = "10.1016/j.nuclphysb.2018.02.015",
    journal = "Nucl. Phys. B",
    volume = "930",
    pages = "45--68",
    year = "2018"
}

@article{Usatyuk:2024mzs,
    author = "Usatyuk, Mykhaylo and Wang, Zi-Yue and Zhao, Ying",
    title = "{Closed universes in two dimensional gravity}",
    eprint = "2402.00098",
    archivePrefix = "arXiv",
    primaryClass = "hep-th",
    doi = "10.21468/SciPostPhys.17.2.051",
    journal = "SciPost Phys.",
    volume = "17",
    number = "2",
    pages = "051",
    year = "2024"
}

@article{Stanford:2020wkf,
    author = "Stanford, Douglas",
    title = "{More quantum noise from wormholes}",
    eprint = "2008.08570",
    archivePrefix = "arXiv",
    primaryClass = "hep-th",
    month = "8",
    year = "2020"
}

@article{gabriele,
    author = "Abdalla, Ahmed and Di Ubaldo, Gabriele and Tabor, Elisa",
    title = "{to appear}"
}

@article{Kudler-Flam:2026nzz,
    author = "Kudler-Flam, Jonah and Witten, Edward",
    title = "{Wormholes and Averaging over N}",
    eprint = "2605.15180",
    archivePrefix = "arXiv",
    primaryClass = "hep-th",
    month = "5",
    year = "2026"
}

@article{Anous:2016kss,
    author = "Anous, Tarek and Hartman, Thomas and Rovai, Antonin and Sonner, Julian",
    title = "{Black Hole Collapse in the 1/c Expansion}",
    eprint = "1603.04856",
    archivePrefix = "arXiv",
    primaryClass = "hep-th",
    doi = "10.1007/JHEP07(2016)123",
    journal = "JHEP",
    volume = "07",
    pages = "123",
    year = "2016"
}

@article{Okuyama:2024BabyUniverse,
  author = {Okuyama, Kazumi},
  title = {Baby universe operators in the {ETH} matrix model of double-scaled {SYK}},
  journal = {JHEP},
  volume = {10},
  pages = {249},
  year = {2024},
  doi = {10.1007/JHEP10(2024)249},
  eprint = {2408.03726},
  archivePrefix = {arXiv},
  primaryClass = {hep-th}
}

@article{Jafferis:2022uhu,
    author = "Jafferis, Daniel Louis and Kolchmeyer, David K. and Mukhametzhanov, Baur and Sonner, Julian",
    title = "{Matrix Models for Eigenstate Thermalization}",
    eprint = "2209.02130",
    archivePrefix = "arXiv",
    primaryClass = "hep-th",
    doi = "10.1103/PhysRevX.13.031033",
    journal = "Phys. Rev. X",
    volume = "13",
    number = "3",
    pages = "031033",
    year = "2023"
}

@article{Blommaert:2023wad,
    author = "Blommaert, Andreas and Mertens, Thomas G. and Yao, Shunyu",
    title = "{The q-Schwarzian and Liouville gravity}",
    eprint = "2312.00871",
    archivePrefix = "arXiv",
    primaryClass = "hep-th",
    doi = "10.1007/JHEP11(2024)054",
    journal = "JHEP",
    volume = "11",
    pages = "054",
    year = "2024"
}

@article{Goel:2023svz,
    author = "Goel, Akash and Narovlansky, Vladimir and Verlinde, Herman",
    title = "{Semiclassical geometry in double-scaled SYK}",
    eprint = "2301.05732",
    archivePrefix = "arXiv",
    primaryClass = "hep-th",
    doi = "10.1007/JHEP11(2023)093",
    journal = "JHEP",
    volume = "11",
    pages = "093",
    year = "2023"
}

@article{Klinger:2026kqj,
    author = "Klinger, Marc S.",
    title = "{How to have your wormholes and factorize, too}",
    eprint = "2602.15120",
    archivePrefix = "arXiv",
    primaryClass = "hep-th",
    month = "2",
    year = "2026"
}

@article{Blommaert:2024Hologram,
  author = {Blommaert, Andreas and Mertens, Thomas G. and Papalini, Jacopo},
  title = {The dilaton gravity hologram of double-scaled {SYK}},
  journal = {JHEP},
  volume = {06},
  pages = {050},
  year = {2025},
  doi = {10.1007/JHEP06(2025)050},
  eprint = {2404.03535},
  archivePrefix = {arXiv},
  primaryClass = {hep-th}
}

@article{Blommaert:2025Wormholes,
  author = {Blommaert, Andreas and Levine, Adam and Mertens, Thomas G.
            and Papalini, Jacopo and Parmentier, Klaas},
  title = {Wormholes, branes and finite matrices in sine dilaton gravity},
  journal = {JHEP},
  volume = {09},
  pages = {123},
  year = {2025},
  doi = {10.1007/JHEP09(2025)123},
  eprint = {2501.17091},
  archivePrefix = {arXiv},
  primaryClass = {hep-th}
}

@article{Garcia-Garcia:2021squ,
    author = "Garc{\'\i}a-Garc{\'\i}a, Antonio M. and Godet, Victor",
    title = "{Half-wormholes in nearly AdS$_2$ holography}",
    eprint = "2107.07720",
    archivePrefix = "arXiv",
    primaryClass = "hep-th",
    doi = "10.21468/SciPostPhys.12.4.135",
    journal = "SciPost Phys.",
    volume = "12",
    number = "4",
    pages = "135",
    year = "2022"
}

@article{CarberyWright2001,
  author  = {Carbery, Anthony and Wright, James},
  title   = {Distributional and {$L^q$} norm inequalities
             for polynomials over convex bodies in {$\mathbb{R}^n$}},
  journal = {Mathematical Research Letters},
  volume  = {8},
  number  = {3},
  pages   = {233--248},
  year    = {2001},
  doi     = {10.4310/MRL.2001.v8.n3.a1}
}

@article{Miyata:2026myd,
    author = "Miyata, Akihiro and Horikoshi, Keigo and Tajima, Hiromasa and Ugajin, Tomonori",
    title = "{Emergent Closed Universes in Symmetric Orbifold CFTs}",
    eprint = "2606.04575",
    archivePrefix = "arXiv",
    primaryClass = "hep-th",
    month = "6",
    year = "2026"
}

@article{Swingle:2026ecy,
    author = "Swingle, Brian",
    title = "{Single-Instance Observables in the Sachdev-Ye-Kitaev Model}",
    eprint = "2609.16130",
    archivePrefix = "arXiv",
    primaryClass = "hep-th",
    month = "9",
    year = "2026"
}

@article{NualartPeccati2005,
  author = {Nualart, David and Peccati, Giovanni},
  title = {Central limit theorems for sequences of multiple stochastic integrals},
  journal = {Annals of Probability}, volume = {33}, number = {1},
  pages = {177--193}, year = {2005},
  doi = {10.1214/009117904000000621},
  eprint = {math/0503598}, archivePrefix = {arXiv}, primaryClass = {math.PR}
}

@article{DiUbaldo:2026tkx,
    author = "Di Ubaldo, Gabriele and Iliesiu, Luca V. and Lin, Henry W. and Yan, Cynthia",
    title = "{When the gravity path integral describes a statistical average}",
    eprint = "2609.16113",
    archivePrefix = "arXiv",
    primaryClass = "hep-th",
    reportNumber = "RIKEN-iTHEMS-Report-26",
    month = "9",
    year = "2026"
}

@article{Lin:2023trc,
    author = "Lin, Henry W. and Stanford, Douglas",
    title = "{A symmetry algebra in double-scaled SYK}",
    eprint = "2307.15725",
    archivePrefix = "arXiv",
    primaryClass = "hep-th",
    doi = "10.21468/SciPostPhys.15.6.234",
    journal = "SciPost Phys.",
    volume = "15",
    number = "6",
    pages = "234",
    year = "2023"
}

@article{Engelhardt:2026blp,
    author = "Engelhardt, Netta and Gesteau, Elliott",
    title = "{A Semiclassical Diagnostic for Spacetime Emergence}",
    eprint = "2605.06780",
    archivePrefix = "arXiv",
    primaryClass = "hep-th",
    reportNumber = "MIT-CTP/6035",
    month = "5",
    year = "2026"
}

@article{Akers:2025ahe,
    author = "Akers, Chris and Bueller, Gracemarie and DeWolfe, Oliver and Higginbotham, Kenneth and Reinking, Johannes and Rodriguez, Rudolph",
    title = "{On observers in holographic maps}",
    eprint = "2503.09681",
    archivePrefix = "arXiv",
    primaryClass = "hep-th",
    doi = "10.1007/JHEP05(2025)201",
    journal = "JHEP",
    volume = "05",
    pages = "201",
    year = "2025"
}

@article{Antonini:2025ioh,
    author = "Antonini, Stefano and Rath, Pratik and Sasieta, Martin and Swingle, Brian and Vilar L{\'o}pez, Alejandro",
    title = "{The baby universe is fine and the CFT knows it: on holography for closed universes}",
    eprint = "2507.10649",
    archivePrefix = "arXiv",
    primaryClass = "hep-th",
    doi = "10.1007/JHEP12(2025)159",
    journal = "JHEP",
    volume = "12",
    pages = "159",
    year = "2025"
}

@article{Qi:2021oni,
    author = "Qi, Xiao-Liang and Shangnan, Zhou and Yang, Zhenbin",
    title = "{Holevo information and ensemble theory of gravity}",
    eprint = "2111.05355",
    archivePrefix = "arXiv",
    primaryClass = "hep-th",
    doi = "10.1007/JHEP02(2022)056",
    journal = "JHEP",
    volume = "02",
    pages = "056",
    year = "2022"
}

@article{Mukhametzhanov:2021nea,
    author = "Mukhametzhanov, Baur",
    title = "{Half-wormholes in SYK with one time point}",
    eprint = "2105.08207",
    archivePrefix = "arXiv",
    primaryClass = "hep-th",
    doi = "10.21468/SciPostPhys.12.1.029",
    journal = "SciPost Phys.",
    volume = "12",
    number = "1",
    pages = "029",
    year = "2022"
}

@article{Almheiri:2021jwq,
    author = "Almheiri, Ahmed and Lin, Henry W.",
    title = "{The entanglement wedge of unknown couplings}",
    eprint = "2111.06298",
    archivePrefix = "arXiv",
    primaryClass = "hep-th",
    doi = "10.1007/JHEP08(2022)062",
    journal = "JHEP",
    volume = "08",
    pages = "062",
    year = "2022"
}

@article{Stanford:2021bhl,
    author = "Stanford, Douglas and Yang, Zhenbin and Yao, Shunyu",
    title = "{Subleading Weingartens}",
    eprint = "2107.10252",
    archivePrefix = "arXiv",
    primaryClass = "hep-th",
    doi = "10.1007/JHEP02(2022)200",
    journal = "JHEP",
    volume = "02",
    pages = "200",
    year = "2022"
}

@article{Abdalla:2026wdx,
    author = "Abdalla, Ahmed I. and Chandra, Jeevan and Wang, Yifan",
    title = "{Line Defects in Liouville Conformal Field Theory: Localized Cosmological Constants and Decohered Hyperbolic Geometries}",
    eprint = "2603.02166",
    archivePrefix = "arXiv",
    primaryClass = "hep-th",
    month = "3",
    year = "2026"
}

@article{Harlow:2026hky,
    author = "Harlow, Daniel",
    title = "{Observers, $α$-parameters, and the Hartle-Hawking state}",
    eprint = "2602.03835",
    archivePrefix = "arXiv",
    primaryClass = "hep-th",
    reportNumber = "MIT-CTP/5900",
    month = "2",
    year = "2026"
}

@article{Chen:2025fwp,
    author = "Chen, Hong Zhe",
    title = "{Observers seeing gravitational Hilbert spaces: abstract sources for an abstract path integral}",
    eprint = "2505.15892",
    archivePrefix = "arXiv",
    primaryClass = "hep-th",
    doi = "10.1007/JHEP10(2025)139",
    journal = "JHEP",
    volume = "10",
    pages = "139",
    year = "2025"
}

@article{Antonini:2022blk,
    author = "Antonini, Stefano and Simidzija, Petar and Swingle, Brian and Van Raamsdonk, Mark",
    title = "{Cosmology from the vacuum}",
    eprint = "2203.11220",
    archivePrefix = "arXiv",
    primaryClass = "hep-th",
    doi = "10.1088/1361-6382/ad1d46",
    journal = "Class. Quant. Grav.",
    volume = "41",
    number = "4",
    pages = "045008",
    year = "2024"
}

@article{Blommaert:2021fob,
    author = "Blommaert, Andreas and Iliesiu, Luca V. and Kruthoff, Jorrit",
    title = "{Gravity factorized}",
    eprint = "2111.07863",
    archivePrefix = "arXiv",
    primaryClass = "hep-th",
    doi = "10.1007/JHEP09(2022)080",
    journal = "JHEP",
    volume = "09",
    pages = "080",
    year = "2022"
}

@article{Erdos:2014zgc,
    author = {Erd{\H{o}}s, L{\'a}szl{\'o} and Schr{\"o}der, Dominik},
    title = "{Phase Transition in the Density of States of Quantum Spin Glasses}",
    eprint = "1407.1552",
    archivePrefix = "arXiv",
    primaryClass = "math-ph",
    doi = "10.1007/s11040-014-9164-3",
    journal = "Math. Phys. Anal. Geom.",
    volume = "17",
    number = "3-4",
    pages = "441--464",
    year = "2014"
}

@article{Chandra:2026hfj,
    author = "Chandra, Jeevan and Post, Boris",
    title = "{Elastic stiffness of three-dimensional black holes and wormholes from Liouville line defects}",
    eprint = "2607.16155",
    archivePrefix = "arXiv",
    primaryClass = "hep-th",
    month = "7",
    year = "2026"
}

@article{Cotler:2016fpe,
    author = "Cotler, Jordan S. and Gur-Ari, Guy and Hanada, Masanori and Polchinski, Joseph and Saad, Phil and Shenker, Stephen H. and Stanford, Douglas and Streicher, Alexandre and Tezuka, Masaki",
    title = "{Black Holes and Random Matrices}",
    eprint = "1611.04650",
    archivePrefix = "arXiv",
    primaryClass = "hep-th",
    reportNumber = "SU-ITP-16-19, SU-ITP-16/19, YITP-16-124",
    doi = "10.1007/JHEP05(2017)118",
    journal = "JHEP",
    volume = "05",
    pages = "118",
    year = "2017",
    note = "[Erratum: JHEP 09, 002 (2018)]"
}

@article{Okuyama:2023yat,
    author = "Okuyama, Kazumi",
    title = "{Matter correlators through a wormhole in double-scaled SYK}",
    eprint = "2312.00880",
    archivePrefix = "arXiv",
    primaryClass = "hep-th",
    doi = "10.1007/JHEP02(2024)147",
    journal = "JHEP",
    volume = "02",
    pages = "147",
    year = "2024"
}

@article{Berkooz:2024lgq,
    author = "Berkooz, Micha and Mamroud, Ohad",
    title = "{A cordial introduction to double scaled SYK}",
    eprint = "2407.09396",
    archivePrefix = "arXiv",
    primaryClass = "hep-th",
    doi = "10.1088/1361-6633/ada889",
    journal = "Rept. Prog. Phys.",
    volume = "88",
    number = "3",
    pages = "036001",
    year = "2025"
}

@article{Coleman:1988cy,
    author = "Coleman, Sidney R.",
    title = "{Black holes as red herrings: Topological fluctuations and the loss of quantum coherence}",
    reportNumber = "HUTP-88/A008",
    doi = "10.1016/0550-3213(88)90110-1",
    journal = "Nucl. Phys. B",
    volume = "307",
    pages = "867--882",
    year = "1988"
}

@article{Harlow:2025pvj,
    author = "Harlow, Daniel and Usatyuk, Mykhaylo and Zhao, Ying",
    title = "{Quantum mechanics and observers for gravity in a closed universe}",
    eprint = "2501.02359",
    archivePrefix = "arXiv",
    primaryClass = "hep-th",
    reportNumber = "MIT-CTP/5824",
    doi = "10.1007/JHEP02(2026)108",
    journal = "JHEP",
    volume = "02",
    pages = "108",
    year = "2026"
}

@article{Maldacena:2004rf,
    author = "Maldacena, Juan Martin and Maoz, Liat",
    title = "{Wormholes in AdS}",
    eprint = "hep-th/0401024",
    archivePrefix = "arXiv",
    reportNumber = "ITFA-2003-57",
    doi = "10.1088/1126-6708/2004/02/053",
    journal = "JHEP",
    volume = "02",
    pages = "053",
    year = "2004"
}

@article{Antonini:2023hdh,
    author = "Antonini, Stefano and Sasieta, Martin and Swingle, Brian",
    title = "{Cosmology from random entanglement}",
    eprint = "2307.14416",
    archivePrefix = "arXiv",
    primaryClass = "hep-th",
    doi = "10.1007/JHEP11(2023)188",
    journal = "JHEP",
    volume = "11",
    pages = "188",
    year = "2023"
}

@article{VanRaamsdonk:2020tlr,
    author = "Van Raamsdonk, Mark",
    title = "{Comments on wormholes, ensembles, and cosmology}",
    eprint = "2008.02259",
    archivePrefix = "arXiv",
    primaryClass = "hep-th",
    doi = "10.1007/JHEP12(2021)156",
    journal = "JHEP",
    volume = "12",
    pages = "156",
    year = "2021"
}

@article{Goel:2018ubv,
    author = "Goel, Akash and Lam, Ho Tat and Turiaci, Gustavo J. and Verlinde, Herman",
    title = "{Expanding the Black Hole Interior: Partially Entangled Thermal States in SYK}",
    eprint = "1807.03916",
    archivePrefix = "arXiv",
    primaryClass = "hep-th",
    doi = "10.1007/JHEP02(2019)156",
    journal = "JHEP",
    volume = "02",
    pages = "156",
    year = "2019"
}

@article{Liu:2025cml,
    author = "Liu, Hong",
    title = "{Towards a holographic description of closed universes}",
    eprint = "2509.14327",
    archivePrefix = "arXiv",
    primaryClass = "hep-th",
    reportNumber = "MIT-CTP/5923",
    month = "9",
    year = "2025"
}

@article{Antonini:2024mci,
    author = "Antonini, Stefano and Rath, Pratik",
    title = "{Do holographic CFT states have unique semiclassical bulk duals?}",
    eprint = "2408.02720",
    archivePrefix = "arXiv",
    primaryClass = "hep-th",
    doi = "10.1142/S0218271825440250",
    journal = "Int. J. Mod. Phys. D",
    volume = "34",
    number = "16",
    pages = "2544025",
    year = "2025"
}

@article{Maldacena:2018lmt,
    author = "Maldacena, Juan and Qi, Xiao-Liang",
    title = "{Eternal traversable wormhole}",
    eprint = "1804.00491",
    archivePrefix = "arXiv",
    primaryClass = "hep-th",
    month = "4",
    year = "2018"
}

@article{Sasieta:2022ksu,
    author = "Sasieta, Martin",
    title = "{Wormholes from heavy operator statistics in AdS/CFT}",
    eprint = "2211.11794",
    archivePrefix = "arXiv",
    primaryClass = "hep-th",
    doi = "10.1007/JHEP03(2023)158",
    journal = "JHEP",
    volume = "03",
    pages = "158",
    year = "2023"
}

@article{Maldacena:2016hyu,
    author = "Maldacena, Juan and Stanford, Douglas",
    title = "{Remarks on the Sachdev-Ye-Kitaev model}",
    eprint = "1604.07818",
    archivePrefix = "arXiv",
    primaryClass = "hep-th",
    doi = "10.1103/PhysRevD.94.106002",
    journal = "Phys. Rev. D",
    volume = "94",
    number = "10",
    pages = "106002",
    year = "2016"
}

@misc{Kitaev:2015SYKTalks,
  author       = {Kitaev, Alexei},
  title        = {A simple model of quantum holography},
  year         = {2015},
  note         = {Talks at KITP, University of California, Santa Barbara, April and May 2015},
  howpublished = {\url{http://online.kitp.ucsb.edu/online/entangled15/kitaev/} and \url{http://online.kitp.ucsb.edu/online/entangled15/kitaev2/}}
}

@article{Chen:2020tes,
    author = "Chen, Yiming and Gorbenko, Victor and Maldacena, Juan",
    title = "{Bra-ket wormholes in gravitationally prepared states}",
    eprint = "2007.16091",
    archivePrefix = "arXiv",
    primaryClass = "hep-th",
    doi = "10.1007/JHEP02(2021)009",
    journal = "JHEP",
    volume = "02",
    pages = "009",
    year = "2021"
}

@article{Saad:2019lba,
    author = "Saad, Phil and Shenker, Stephen H. and Stanford, Douglas",
    title = "{JT gravity as a matrix integral}",
    eprint = "1903.11115",
    archivePrefix = "arXiv",
    primaryClass = "hep-th",
    month = "3",
    year = "2019"
}

@article{Saad:2018bqo,
    author = "Saad, Phil and Shenker, Stephen H. and Stanford, Douglas",
    title = "{A semiclassical ramp in SYK and in gravity}",
    eprint = "1806.06840",
    archivePrefix = "arXiv",
    primaryClass = "hep-th",
    month = "6",
    year = "2018"
}

@article{Iliesiu:2024cnh,
    author = "Iliesiu, Luca V. and Levine, Adam and Lin, Henry W. and Maxfield, Henry and Mezei, M{\'a}rk",
    title = "{On the non-perturbative bulk Hilbert space of JT gravity}",
    eprint = "2403.08696",
    archivePrefix = "arXiv",
    primaryClass = "hep-th",
    doi = "10.1007/JHEP10(2024)220",
    journal = "JHEP",
    volume = "10",
    pages = "220",
    year = "2024"
}

@article{Blommaert:2020seb,
    author = "Blommaert, Andreas",
    title = "{Dissecting the ensemble in JT gravity}",
    eprint = "2006.13971",
    archivePrefix = "arXiv",
    primaryClass = "hep-th",
    doi = "10.1007/JHEP09(2022)075",
    journal = "JHEP",
    volume = "09",
    pages = "075",
    year = "2022"
}

@article{Bintanja:2026vkr,
    author = "Bintanja, Suzanne and Gesteau, Elliott",
    title = "{The $α$-states of a string worldsheet}",
    eprint = "2607.15336",
    archivePrefix = "arXiv",
    primaryClass = "hep-th",
    reportNumber = "MIT-CTP/6069",
    month = "7",
    year = "2026"
}

@article{Nomura:2026igt,
    author = "Nomura, Yasunori and Ugajin, Tomonori",
    title = "{Physical Predictions in Closed Quantum Gravity}",
    eprint = "2602.13387",
    archivePrefix = "arXiv",
    primaryClass = "hep-th",
    reportNumber = "RIKEN-iTHEMS-Report-26",
    month = "2",
    year = "2026"
}

@article{Jafferis:2022wez,
    author = "Jafferis, Daniel Louis and Kolchmeyer, David K. and Mukhametzhanov, Baur and Sonner, Julian",
    title = "{Jackiw-Teitelboim gravity with matter, generalized eigenstate thermalization hypothesis, and random matrices}",
    eprint = "2209.02131",
    archivePrefix = "arXiv",
    primaryClass = "hep-th",
    doi = "10.1103/PhysRevD.108.066015",
    journal = "Phys. Rev. D",
    volume = "108",
    number = "6",
    pages = "066015",
    year = "2023"
}

@article{Higginbotham:2025dvf,
    author = "Higginbotham, Kenneth",
    title = "{On tests for baby universes in AdS/CFT}",
    eprint = "2507.05337",
    archivePrefix = "arXiv",
    primaryClass = "hep-th",
    doi = "10.1007/JHEP09(2025)038",
    journal = "JHEP",
    volume = "09",
    pages = "038",
    year = "2025"
}

@article{Harlow:2018tqv,
    author = "Harlow, Daniel and Jafferis, Daniel",
    title = "{The Factorization Problem in Jackiw-Teitelboim Gravity}",
    eprint = "1804.01081",
    archivePrefix = "arXiv",
    primaryClass = "hep-th",
    doi = "10.1007/JHEP02(2020)177",
    journal = "JHEP",
    volume = "02",
    pages = "177",
    year = "2020"
}

@article{Engelhardt:2025azi,
    author = "Engelhardt, Netta and Gesteau, Elliott and Harlow, Daniel",
    title = "{Observer complementarity for black holes and holography}",
    eprint = "2507.06046",
    archivePrefix = "arXiv",
    primaryClass = "hep-th",
    reportNumber = "MIT-CTP/5884",
    month = "7",
    year = "2025"
}

@article{Saad:2021uzi,
    author = "Saad, Phil and Shenker, Stephen H. and Yao, Shunyu",
    title = "{Comments on wormholes and factorization}",
    eprint = "2107.13130",
    archivePrefix = "arXiv",
    primaryClass = "hep-th",
    doi = "10.1007/JHEP10(2024)076",
    journal = "JHEP",
    volume = "10",
    pages = "076",
    year = "2024"
}

@article{Saad:2019pqd,
    author = "Saad, Phil",
    title = "{Late Time Correlation Functions, Baby Universes, and ETH in JT Gravity}",
    eprint = "1910.10311",
    archivePrefix = "arXiv",
    primaryClass = "hep-th",
    month = "10",
    year = "2019"
}

@article{Giddings:1988cx,
    author = "Giddings, Steven B. and Strominger, Andrew",
    title = "{Loss of incoherence and determination of coupling constants in quantum gravity}",
    reportNumber = "HUTP-88/A006",
    doi = "10.1016/0550-3213(88)90109-5",
    journal = "Nucl. Phys. B",
    volume = "307",
    pages = "854--866",
    year = "1988"
}

@article{Sasieta:2025vck,
    author = "Sasieta, Martin and Swingle, Brian and Vilar L{\'o}pez, Alejandro",
    title = "{Baby Universes from Thermal Pure States in the Sachdev-Ye-Kitaev Model}",
    eprint = "2512.00149",
    archivePrefix = "arXiv",
    primaryClass = "hep-th",
    doi = "10.1103/8p6x-w737",
    journal = "Phys. Rev. Lett.",
    volume = "137",
    number = "4",
    pages = "041501",
    year = "2026"
}

@article{Chandra:2024vhm,
    author = "Chandra, Jeevan and Hartman, Thomas and Meruliya, Viraj",
    title = "{Statistics of three-dimensional black holes from Liouville line defects}",
    eprint = "2404.15183",
    archivePrefix = "arXiv",
    primaryClass = "hep-th",
    doi = "10.1007/JHEP11(2024)090",
    journal = "JHEP",
    volume = "11",
    pages = "090",
    year = "2024"
}

@book{BhattacharyaRao2010,
  author    = {Bhattacharya, Rabi N. and Ranga Rao, R.},
  title     = {Normal Approximation and Asymptotic Expansions},
  series    = {Classics in Applied Mathematics},
  volume    = {64},
  publisher = {Society for Industrial and Applied Mathematics},
  address   = {Philadelphia},
  year      = {2010},
  isbn      = {978-0-898718-97-3},
  doi = {10.1137/1.9780898719895.fm}
}

\end{document}